# EQUATION-INFORMED TIME REPARAMETERIZATION FOR PHYSICS-INFORMED NEURAL NETWORKS

**Orestis Belokas**
INSANE Research Lab
Institute of Informatics and Telecommunications
NCSR "Demokritos", Athens, Greece
obelokas@gmail.com

Christoforos Rekatsinas
INSANE Research Lab
Institute of Informatics and Telecommunications
NCSR "Demokritos", Athens, Greece
crek@iit.demokritos.gr

## Abstract

Physics-Informed Neural Networks (PINNs) offer a mesh-free approach to solving differential equations, but their application to time-dependent problems remains challenging, particularly when the governing dynamics involve widely different time scales. With collocation points spread uniformly in time, short intervals of rapid change are under-resolved. This work introduces an equation-informed time reparameterization framework that modifies the temporal coordinate used during PINN training based on the local dynamics of the governing equation. The proposed method follows a two-stage procedure. First, a pilot PINN training provides an approximate solution, which is used to linearize the governing differential operator and estimate its local response to small perturbations. These responses are combined into a time-dependent rate function, large where the dynamics are fast, that defines a strictly increasing transformation from physical time to a new computational time coordinate. A second PINN is then trained on the transformed governing equation, with uniformly sampled collocation points in the new coordinate producing a nonuniform distribution in physical time. Different combination rules and spatial reduction strategies are considered. With the cost of the pilot training included, the method reduces the error by a factor of 4 to 29 relative to the best alternative method on the Kachanov–Rabotnov creep-damage model near rupture, and by a factor of 5 to 16 relative to every alternative method on the Frank–Kamenetskii equation near blow-up.

## 1. Introduction

Machine Learning (ML) is reshaping how problems in engineering and the physical sciences are solved, offering effective alternatives to classical numerical methods precisely where those are most limited [1]. Differential equations are among the areas where this shift has been most noticeable [1]. The classical approach to solving such equations relies on grid- or mesh-based discretization, with finite difference [2], finite element [3] and spectral methods [4] chief among them, and it has significant limitations: mesh generation grows complex for irregular or high-dimensional domains, noisy or sparse observational data are rarely easy to incorporate, and inverse problems demand a substantially different formulation [1,3]. ML has increasingly been applied in this context, used to solve both Ordinary Differential Equations (ODEs) and Partial Differential Equations (PDEs) by replacing or augmenting classical grid- or mesh-based discretization with a learned, mesh-free representation of the solution [1,5–9].

Our scope is Physics-Informed Neural Networks (PINNs) [6], a class of neural networks designed to solve differential equations directly. Unlike conventional supervised models [10], PINNs do not rely on labeled data with verified solutions. They are trained by driving the equation's own residual towards zero, which removes the need for experimental data in forward problems [7,11,12]. The residual comes directly from the network's output: since that output is a differentiable function of the input coordinates, its derivatives are obtained by automatic differentiation [11,13] and substituted into the governing equation, giving a quantity that vanishes only where the equation is satisfied [6,11]. This construction applies to any governing equation [11], but its behavior changes markedly once the equation is time-dependent.
Time-dependent differential equations have proved to be more difficult than time-independent ones, and classic PINNs may struggle to solve them [14–16]. This difficulty stems from three inherent characteristics of time-dependent problems. Of these three, our method addresses stiffness, and it does so by redefining the time coordinate itself. A pilot PINN gives a rough solution, the linearization of the governing equation about it shows where a small error would grow fastest, and a final PINN is trained on a time axis stretched there.

The first factor that contributes to the difficulty of dealing with time-dependent differential equations is causality. Causality refers to time's nature: later events depend on prior ones [14]. This is not something that standard PINNs can account for, since a PINN's residual loss is, by default, a single scalar averaged over the whole space-time domain [6,14] — nothing in that objective enforces that early-time error gets resolved before the network is credited for matching later-time behavior; in fact, PINNs trained by gradient descent tend to fit later times first [14].

Existing solutions to causality target the model's training, reordering training emphasis placed on different time instants [14,17–20]. More specifically, causal weighting reformulates the training objective, defining a weighted residual loss, so that the residual at each time enters the loss only once the residuals at all earlier times are minimized properly [14]. Turinici studies why such early-weighted time sampling works, for prototypical examples and under standard assumptions on network convergence, showing that the optimal time-sampling density is a truncated exponential distribution on a bounded interval, whose decay rate governs how strongly early times are favored [21].

The second large factor is cost over long time horizons. As the time domain grows, so does the space-time collocation grid needed to cover it, and the optimization problem the network must solve becomes substantially harder [15,16,22]. Classical solvers avoid this by marching forward in time: they advance the solution step by step, as in the central-difference [23] and Newmark [24] schemes, so that each step involves only a small problem rather than the whole time interval at once. A PINN, on the other hand, treats time as one more input coordinate and fits the entire space-time trajectory as a single optimization problem [6,16], which is why space-time problems are typically harder to solve than time-marching [16,25]. Meng et al. identify this as a limitation of PINNs: training is effective over short time horizons but becomes computationally prohibitive once the space-time domain is large [22]. This limitation differs from causality, since it concerns intractable training cost and not an incorrect ordering of the solution.

The third large issue is stiffness. A system is stiff when its governing equation includes components that differ substantially in magnitude [26,27]. In PINNs, this appears as an imbalance between the terms of the loss, whose gradients differ greatly in size, so that training favors some terms over others [28]; one remedy is to rescale each term of the governing equation by a characteristic quantity before training, so that all terms operate at comparable magnitudes [29]. For time-dependent equations, stiffness takes a more specific form: components evolve on widely different time scales, some rapidly and others slowly [30,31]. Standard PINNs do not distinguish between them. The residual loss aggregates their contributions into a single scalar, and because the fast-varying terms produce much larger residuals, the loss gradient is dominated by them, biasing training toward the fast dynamics at the expense of the slow ones [28,31,32], even when the latter govern the solution over the time scales of interest.

Existing responses to stiffness operate on the equation or optimization problem. At the equation level Ji et al. reduce stiffness through a quasi-steady-state assumption that eliminates the fast variables [31], and Ko and Park rescale the independent variables by a constant factor to mitigate stiff behavior [33]. At the loss level, Bento et al. normalize the residuals by the Jacobian to solve stiff dark matter equations [32], and Wang, Teng and Perdikaris balance the loss terms through gradient statistics [28]. A related method targets the ill-conditioning of the residual minimization itself: Cao and Zhang's time stepping oriented neural network (TSONN) integrates a time stepping scheme into training, turning one ill-conditioned problem into a sequence of well-conditioned sub-problems over discrete pseudo-time intervals — an auxiliary temporal axis introduced purely as an optimization device, distinct from the governing equation's own physical time coordinate [34].

An underlying common factor of the above responses to time-dependent equations, despite addressing three different difficulties, is that each operates within the physical time domain: reordering emphasis [14,17–20], partitioning the domain [22,35], or adjusting the equation and residual [28,29,31–34], but none redefines the time coordinate itself. A separate, less explored strategy in the PINN realm is to transform the time coordinate directly: rather than acting within physical time, redefine the domain the network is trained on so that the equation is better conditioned in the new coordinate. Our approach builds on a known technique for this: time reparameterization [36–38].

In its simplest form, time reparameterization replaces the physical time coordinate of a differential equation with a new coordinate, related to the original by a transformation that need not be uniform [37]. It is a century-old technique that stems from physics, not machine learning, and goes back to Sundman's work on the three-body problem [36]. There the physical clock was rescaled by a function of the mutual distances between the bodies, which makes the binary-collision singularity removable and lets the solution continue through collision [36]. The same transformation was later generalized for the numerical integration of orbits, where it concentrates integration effort near close approaches instead of spreading steps uniformly in physical time [37,39].

More recently, with the rapid development of Machine Learning, this method has been repurposed to serve ML methods: it has been applied to reduce stiffness in neural-ODE-based reduced-order models [38] and to restore the accuracy of Gaussian process surrogates on stiff ODE solutions [40].

SEE-PINN [29], an equation-level treatment that rescales the terms of the governing equation before training to address disparities in the magnitudes of its terms, was the inspiration for our contribution. Carrying over this principle of acting on the governing equation itself, this paper applies it to the time coordinate and introduces an equation-informed time reparameterization for PINNs, in which the new time coordinate is derived algebraically from the linearization of the governing operator about the solution of a pilot training, through a quantity we call the rate. Section 2 fixes the two pieces of machinery this construction depends on: the PINN formulation used throughout this paper, and the general mechanics of reparameterizing time in an initial value problem. Section 3 introduces our contribution to apply time reparameterization to a PINN's residual, together with the construction behind the proposed technique. Section 4 covers the practical considerations: the strength of the warp, the weighting of the residual, the bootstrapping of the rate and the implementation. Section 5 derives the rate for four case studies, the Kachanov-Rabotnov creep-damage model, Burgers' equation, the Allen-Cahn equation and the Frank-Kamenetskii thermal explosion, and Section 6 compares the method with six existing methods on them and discusses the results. The Appendix gives the nomenclature, the hyperparameters, the computational cost and further results for each case study.

# 2. Theoretical Background

This section introduces the two ingredients on which our method is based: PINNs and time reparameterization, as presented in the literature [6,36,37]. Section 2.1 covers the PINN formulation used throughout this paper. Section 2.2 shows how a change of time coordinate transforms a generic initial value problem and derives the reparameterized equation on which Section 3 builds.

## 2.1 Physics-Informed Neural Networks

Consider a generic time-dependent differential equation, with initial and boundary conditions:

$$\frac{\partial u}{\partial t} = \mathcal{N}[u] + f(\boldsymbol{x}, t), \quad u(\boldsymbol{x}, 0) = u_0(\boldsymbol{x}), \quad \mathcal{B}[u] = g(\boldsymbol{x}, t) \; on \; \partial\Omega \times [0, T] \tag{1}$$

where $\mathcal{N}$ is the equation's governing operator, $f$ is an external forcing term, and $\mathcal{B}$ an operator encoding the boundary conditions. The solution is represented by a neural network $u_\theta(\boldsymbol{x}, t)$ *[6]*, where $\boldsymbol{x} = (x_1, \ldots, x_d) \in \Omega \subset \mathbb{R}^d$ is the spatial coordinate, $t$ is the time coordinate and $\theta$ are the network's trainable parameters. In contrast with classical numerical approaches, no mesh is required [7]: the network trains directly against the governing equation itself. This is done by evaluating $u_\theta$ and its derivatives at a set of collocation points, producing the residual [6,11] given by Eq. (2):

$$r(\boldsymbol{x}, t; \theta) = \frac{\partial u_\theta}{\partial t} - \mathcal{N}[u_\theta] - f(\boldsymbol{x}, t), \tag{2}$$

that would be identically zero if $u_\theta$ solved the equation exactly. The training minimizes this residual loss, along with terms enforcing the boundary and initial conditions [6], giving the composite loss of Eq. (3):

$$\mathcal{L}(\theta) = \mathcal{L}_r(\theta) + \mathcal{L}_{BC}(\theta) + \mathcal{L}_{IC}(\theta), \mathcal{L}_r(\theta) = \frac{1}{N_r}\sum_{i=1}^{N_r} |r(\boldsymbol{x_i}, t_i; \theta)|^2, \tag{3}$$

where $\mathcal{L}_{BC}$, $\mathcal{L}_{IC}$ are defined analogously over sampled boundary and initial condition points.

An alternative to this soft constraint formulation enforces the boundary and initial conditions exactly, by constraining the network's output, rather than penalizing violations of them in the loss function. Lagaris, Likas & Fotiadis [5] constructed such trial solutions in the general form of Eq. (4):

$$u_\theta(\boldsymbol{x}, t) = A(\boldsymbol{x}, t) + D(\boldsymbol{x}, t)N_\theta(\boldsymbol{x}, t), \tag{4}$$

where $A$ satisfies the initial and boundary conditions and contains no trainable parameters, and $D$ is constructed to vanish the term $D(\boldsymbol{x}, t)N_\theta(\boldsymbol{x}, t)$ wherever those conditions are imposed, so the trainable term $N_\theta$ never disturbs them, regardless of its output. This removes $\mathcal{L}_{IC}$ and $\mathcal{L}_{BC}$ from the loss of Eq. (3) entirely, leaving only the residual loss $\mathcal{L}_r$ to be minimized. It is the formulation mainly used throughout this paper (Figure 1).

Boundary conditions that are periodic can also be enforced exactly on the network output. Composing any smooth function with a periodic function gives an output that is periodic, together with all its derivatives [41]. For a network $u_\theta(x, t)$ on a domain of period $P$, it is therefore enough to feed it a periodic embedding of $x$ in place of $x$. Following Wang, Sankaran and Perdikaris [14], who also include a constant component, which the bias of the first layer already provides, we use

$$\gamma_m(x) = [cos(2\pi x/P), \dots, cos(2\pi m x/P), sin(2\pi x/P), \dots, sin(2\pi m x/P)]. \tag{5}$$

so that the input is $(\gamma_m(x), t)$, of dimension $2m + 1$. Every component of $\gamma_m$ has period $P$, so for any weights, $u_\theta$ and all its $x$-derivatives take the same value at both ends of the domain. The boundary term of the composite loss (Eq. (3)) is then identically zero and is dropped.

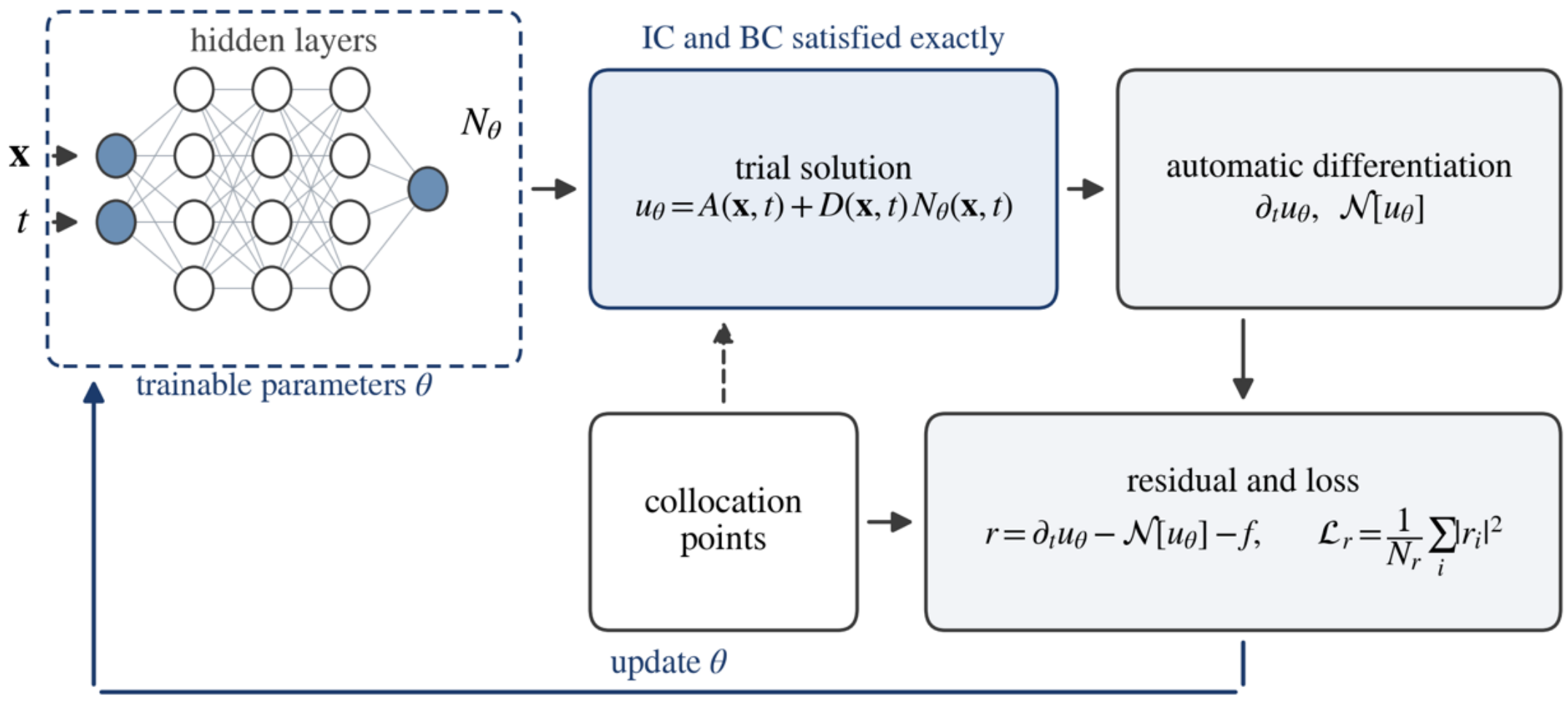

*Figure 1: A PINN with a hard-constraint trial solution: initial and boundary conditions are satisfied identically, leaving the equation residual as the only loss term.*

## 2.2 Time Reparameterization

Section 2.1 fixed how a PINN represents a governing equation and trains against its residual. Reparameterizing time changes the coordinate that equation is expressed in, not the equation itself: the physical clock is replaced by a new coordinate, related to the original by a transformation that can advance faster through some parts of the domain than others [36,37]. To state this precisely, let $u(\boldsymbol{x}, t)$ solve the initial value problem of Eq. (1).
A time reparameterization introduces a strictly increasing map $s: [0, T] \rightarrow [0, S]$ related to the physical clock by a warp density function $\rho(t) > 0$, through the equation:

$$\frac{ds}{dt} = \rho(t), \Leftrightarrow s(t) = \int_0^t \rho(t')dt', s(0) = 0. \tag{6}$$

Since $\rho(t) > 0$ for all $t \in [0, T]$, the map $s(t) = \int_0^t \rho(t')\, dt'$ is strictly increasing and therefore invertible; we write $t(s)$ for this inverse, a strictly increasing map $t: [0, S] \rightarrow [0, T]$, and $S := s(T)$ for the endpoint of the transformed domain.

Let $\tilde{u}(\boldsymbol{x}, s) := u(\boldsymbol{x}, t(s))$ denote the solution expressed in the new time coordinate s. We are looking for an equation of $\tilde{u}$ in terms of $s$, so we differentiate $\tilde{u}(\boldsymbol{x}, s) = u(\boldsymbol{x}, t(s))$ with respect to $s$. Since $\tilde{u}$ depends on $s$ only through $t(s)$, the chain rule applies:

$$\frac{\partial \tilde{u}}{\partial s} = \left.\frac{\partial u}{\partial t}\right|_{t=t(s)} \cdot \frac{dt}{ds} \tag{7}$$

The second factor, $\frac{dt}{ds}$, follows from inverting $\frac{ds}{dt} = \rho(t)$. Thus, by the inverse function rule applied to Eq. (6):

$$\frac{dt}{ds} = \frac{1}{\rho(t(s))} \tag{8}$$

The first factor, $\left.\frac{\partial u}{\partial t}\right|_{t=t(s)}$, follows directly from the original equation $\frac{\partial u}{\partial t} = \mathcal{N}[u] + f(\boldsymbol{x}, t)$, evaluated at $t = t(s)$ and using $u(\boldsymbol{x}, t(s)) = \tilde{u}(\boldsymbol{x}, s)$:

$$\left.\frac{\partial u}{\partial t}\right|_{t=t(s)} = \mathcal{N}[\tilde{u}(\boldsymbol{x}, s)] + f(\boldsymbol{x}, t(s)) \tag{9}$$

Substituting Eq. (8) and Eq. (9) back into the chain rule, Eq. (7), gives the reparameterized problem, Eq. (10):

$$\frac{\partial \tilde{u}}{\partial s} = \frac{\mathcal{N}[\tilde{u}(\boldsymbol{x}, s)] + f(\boldsymbol{x}, t(s))}{\rho(t(s))}, \tilde{u}(\boldsymbol{x}, 0) = u_0(\boldsymbol{x}), \mathcal{B}[\tilde{u}(\boldsymbol{x}, s)] = g(\boldsymbol{x}, t(s)) \; on \; \partial\Omega \times [0, S], s \in [0, S], \boldsymbol{x} \in \Omega. \tag{10}$$

The boundary operator $\mathcal{B}$ acts on $\boldsymbol{x}$ alone and contains no time derivative, so it carries through to Eq. (10) unchanged. Only the boundary data $g(\boldsymbol{x}, t)$ is evaluated at the physical time $t(s)$.

Each of the constructions listed in Section 1 [36–40] is a choice of the function $\rho(t)$ within the equation just derived. Some are tied to one class of problems, such as Sundman's transformation for gravitational singularities, and others follow the kinematics of the solution, such as arc length. Moving-mesh methods call the warp density the monitor function and likewise choose it from the solution (its arc length or curvature), from an estimate of its error, or from the scaling structure of a particular problem class, such as blow-up [43]. None builds it from the response of the governing operator to a perturbation, which is the step our method adds.

# 3. Equation-Informed Time Reparameterization for PINNs

## 3.1 Time Reparameterization for PINNs

In the context of this research, we introduce the concept of time reparameterization to the PINN's loss for a generic warp-density function $\rho(t) > 0$ as defined in Eq. (6).

A PINN trained on the reparameterized time domain s is a network $u_\theta(\boldsymbol{x}, s)$.

Collocation points are sampled uniformly over $s \in [0, S]$. Consider a short interval of physical time $[t, t + \Delta t]$. Since $s(t) = \int_0^t \rho(t')dt'$ (Eq. (6)), this interval corresponds to an interval in $s$ of length:

$$s(t + \Delta t) - s(t) = \int_t^{t+\Delta t} \rho(t')dt'. \tag{11}$$

Because points are sampled uniformly in s, the fraction of sampled points expected to land within this interval is exactly:

$$\frac{1}{S}\int_t^{t+\Delta t} \rho(t')dt' \tag{12}$$

Wherever $\rho(t)$ is large over $[t, t + \Delta t]$, this integral is large, and a proportionally larger share of collocation points lands within this interval of physical time; wherever $\rho(t)$ is small, proportionally fewer do. Sampling uniformly in $s$ therefore concentrates collocation points in physical time in proportion to $\rho(t)$ [42–44].

The residual is formed against the transformed equation derived in Section 2.2, Eq. (10):

$$r(\boldsymbol{x}, s; \theta) = \frac{\partial u_\theta}{\partial s} - \frac{\mathcal{N}[u_\theta(\boldsymbol{x}, s)] + f(\boldsymbol{x}, t(s))}{\rho(t(s))} \tag{13}$$

The loss $\mathcal{L}_r(\theta)$ is the mean square of the residual of Eq. (13) over the collocation points $(\boldsymbol{x}_i, s_i)$ [6], exactly as in Eq. (3). When the initial and boundary conditions are enforced exactly (Section 2.1), the network takes the form of Eq. (4) with the time argument transformed, $u_\theta(\boldsymbol{x}, s) = A(\boldsymbol{x}, t(s)) + D(\boldsymbol{x}, t(s))N_\theta(\boldsymbol{x}, s)$. The conditions then hold for every $\theta$, because $D$ vanishes at $s = 0$ and on $\partial\Omega$, and $\mathcal{L}_{BC}$ and $\mathcal{L}_{IC}$ do not appear. When a condition is enforced softly instead, its loss term is kept as in Eq. (3). Neither $\mathcal{L}_{BC}$ nor $\mathcal{L}_{IC}$ contains a time derivative, so $\mathcal{L}_{BC}$ is evaluated at the physical time $t(s)$ on the boundary and $\mathcal{L}_{IC}$ at $s = 0$.

However, the warp density function is a function of the solution trajectory itself, and cannot be evaluated without first having at least a rough estimate of the solution. For that purpose we use a plain PINN $u_{\theta_0}(\boldsymbol{x}, t)$, trained on the physical domain t, following the standard construction of Section 2.1 but with a reduced computational budget. We call this the **pilot training.** Thus, the warp density $\rho(t)$ can now be evaluated on the pilot-training solution and used in the full training as described above.

*Remark: An alternative approach reconstructs* $\frac{\partial u_\theta}{\partial t}$ *from* $\frac{\partial u_\theta}{\partial s}$ *via the chain rule and checks it against the original equation instead. This gives*

$$r_{alt} = \rho\big(t(s)\big) \cdot r, \tag{14}$$

sharing the same zero set as $r$ since $\rho(t) > 0$ everywhere. The two residuals weight the collocation points differently, and the weighting has to be read together with the sampling. By Eq. (12), the density of the points in physical time is proportional to $\rho(t)$. Each point contributes the square of its residual to the loss, which is ${r_{alt}}^2$ for the alternative form and $r^2 = \frac{{r_{alt}}^2}{\rho(t)^2}$ for Eq. (13). The loss contributed per unit of physical time is therefore proportional to $\rho(t)$ for $r_{alt}$ and to $\rho(t) \cdot \rho(t)^{-2} = 1/\rho(t)$ for $r$. The residual of Eq. (13) thus works against the resampling and gives the high-rate times less weight than the others, whereas $r_{alt}$ keeps the emphasis that the sampling gives them. Section 4.2 generalizes both with an explicit weight.

## 3.2 Defining an Equation-Informed Warp Density Function

The previous section established that the warp density function $\rho(t)$ directly controls where collocation points concentrate in physical time; large $\rho(t)$ draws points densely, small $\rho(t)$ draws points sparsely. This paper introduces a warp density function with a specific purpose: to concentrate the PINN's effort at the times at which the equation reacts most strongly to a small perturbation of its current state, by amplifying it or by damping it. This follows the classical analysis of stiffness, which also examines the response of the linearized equation to a perturbation [26,27,45]. We call the resulting measure the rate. It is an equation-informed measure of local dynamics, not a general measure of stiffness; the properties at the end of this section state what it does and does not capture. Consider the generic initial value problem of Eq. (1).

Let $U(\boldsymbol{x}, t)$ denote the current state of the equation, acquired in practice through the pilot training (Section 3.1) and decompose the governing operator into its constituent terms $\mathcal{N} = \sum_i \mathcal{N}_i$ (advective, diffusive, reactive, ...). Consider perturbing this state $U \to U + w$ and expanding as a Taylor series in $w$ about $w = 0$ [46,47]:

$$\mathcal{N}_i(U + w) = \mathcal{N}_i(U) + L_i[U(t)]w + R_i(w) \tag{15}$$

retaining the term linear in $w$ and discarding the remainder $R_i(w)$. $L_i[U(t)]$, being the coefficient of the single power $w$, is the Fréchet derivative of $\mathcal{N}_i$ at $U(t)$ [46]. The remainder vanishes faster than the linear term, $R_i(w) = o(\|w\|)$ as $w \to 0$ [47], so the linearization is accurate for sufficiently small perturbations; the rate is a linearized sensitivity of the equation, not a prediction of nonlinear error growth. To first order, at each fixed $t$, a perturbation to the state evolves under the equation's own dynamics as

$$\frac{\partial w}{\partial t} = L[U(t)]w = \sum_i L_i[U(t)]\, w, \tag{16}$$

where $i$ ranges over the same fixed decomposition of $\mathcal{N}$ introduced above — only the value of each $L_i[U(t)]$ changes with $t$, not which terms exist. Each summand states what one mechanism alone would do to a perturbation, that is, how strongly it would push back against or amplify it. A growth rate follows from the action of $L_i[U(t)]$ on a single spatial wave $w = e^{ikx}$ ($e^{ik\cdot x}$ in $d$ dimensions), the standard device of Fourier (von Neumann) stability analysis for discretized equations [48,49] and of normal-mode analysis for continuous ones [50,51]. On a periodic or unbounded domain a perturbation is a superposition of such waves and, for constant coefficients, each wave evolves independently [52]. Where the coefficients vary, they are frozen at the point $(x, t)$ [53], a local approximation that neglects the coupling between waves that this variation causes. The wave is then multiplied by a complex number, $L_i[U]e^{ikx} = \mu_i(x, t)e^{ikx}$, with $\mu = c$ for a term $cw$ and $\mu = c(ik)^q$ for a term $c\frac{\partial^q w}{\partial x^q}$ [49,54], with $k$ fixed by the closure given in the note*. For odd $q$ and real $c$, $\mu$ is purely imaginary: the wave is shifted without changing its amplitude and $\lambda_i = 0$ (see Pure transport below). For one mechanism acting alone, $\frac{\partial w}{\partial t} = \mu_i w$, so the amplitude changes like $e^{Re(\mu_i)t}$ while $Im(\mu_i)$ only shifts the phase. The growth coefficient is the real part, $\lambda_i(x, t) = Re(\mu_i)$: the rate, in inverse time, at which mechanism $i$ alone amplifies ($\lambda_i > 0$), damps ($\lambda_i < 0$) or only displaces ($\lambda_i = 0$) a small perturbation of the solution. With the growth coefficients $\lambda_i(x, t)$ defined, the pointwise rate field and its reduction to a function of time are then

$$|L[U(\boldsymbol{x}, t)]| \coloneqq \Phi\big(\lambda_1(\boldsymbol{x}, t), \dots, \lambda_I(\boldsymbol{x}, t)\big) \tag{17}$$

$$\boxed{rate(t) \coloneqq Q[|L[U(\cdot, t)]|]} \tag{18}$$

where the combination rule $\Phi$ and the spatial reduction $Q$ over $\Omega$ are fixed before the full training and stated with each experiment. A choice of $\Phi$ and $Q$ is admissible only when it satisfies the following conditions:

i. $\Phi$ is continuous and nonnegative, and $\Phi(0, \dots, 0) = 0$: a state to which none of the mechanisms responds has a $rate$ of zero.
ii. $\Phi$ is defined for any number of entries, and inserting an entry $\lambda_i = 0$ leaves its value unchanged: a mechanism that neither amplifies nor damps contributes nothing.
iii. $Q$ maps the field $|L[U(\boldsymbol{x}, t)]|$ over $\Omega$ at fixed t to a single number in $[0, \infty)$, and satisfies two requirements. It returns $a$ for a field equal to the constant $a$: if the field takes the same value at every point of $\Omega$, the reduction leaves it unchanged. It is monotone: if one field is pointwise no smaller than another, its reduction is no smaller, so a larger response anywhere in $\Omega$ can never lower $rate(t)$. The mean, maximum and median all satisfy both, and are the main reduction methods used in this context.

We take $\rho(t) = rate(t)$** as the choice for the warp density function, completing the specialization promised above. With this choice, $s(t) = \int_0^t rate(t')dt'$ is the accumulated response of the equation, and collocation levels equally spaced in $s$ are separated by equal amounts of it. This is the equidistribution principle of moving-mesh methods, with the rate as the monitor function [43,44]. For $\Phi_+$ and the mean, $s(t)$ is the linearized growth, in e-folds, accumulated by time $t$ and averaged over $\Omega$, with decay ignored.

*For an even-order term the response is real, but it depends on $k$, because such terms carry no intrinsic length scale. Classical explicit stability limits take the scale from the grid, $\Delta t \lesssim \Delta x^2/\nu$ [2,49]. Lacking a grid, we take it from the pilot-training state, $k(x,t) = |\nabla U(x,t)|/U_{ref}$, with $U_{ref}$ a characteristic amplitude of $U$ ($U_{ref} = 1$ in the nondimensional variables of the equation), and use the same $k$ in every even-order term. In one spatial dimension with $U_{ref} = 1$ this reads $k(x,t) = |\frac{\partial U(x,t)}{\partial x}|$, so that the response $-\nu k^2$ of a diffusion term $\nu \frac{\partial^2 w}{\partial x^2}$ becomes $-\nu \left|\frac{\partial U}{\partial x}\right|^2$, the local rate at which diffusion removes the energy of $U$ itself; it assumes that a perturbation varies on the same spatial scale as the solution. The symbol is exact for terms without derivatives and for terms with constant coefficients; for derivative terms with variable coefficients it is accurate when the wave varies faster than the coefficients, which the closure scale satisfies only to order of magnitude. An even term of order $q$ responds as $k^q$, so a relative error in $|\nabla U|$ changes it by about $q$ times that error, while terms without derivatives do not depend on $k$. The closure sees only the scale of the solution: it does not capture damping at finer scales, and where $\nabla U$ vanishes it gives no diffusive damping.

**As established in Section 2.2, a valid warp density function requires $\rho(t) > 0$ for all $t$. By construction $rate(t) = Q[|L[U(\cdot,t)]|] \geq 0$, but strict positivity is not granted automatically for all equations. In those where strict positivity cannot be guaranteed by construction, we introduce a small, fixed constant $\varepsilon$ (e.g., $\varepsilon = 10^{-6}$), for which

$$\rho(t) = rate(t) + \varepsilon, \tag{19}$$

thus ensuring positivity unconditionally.

*Definition Properties:*

1. *An additive forcing term, if existent in the equation, is invisible to rate. If an equation carries an additive forcing term of the form $f(\boldsymbol{x}, t)$, independent of u, the same expansion gives $\frac{\partial f}{\partial w} \equiv 0$, so f contributes nothing to the sum: $rate(t)$ measures how the equation reacts to a perturbation of its own state, not how hard it is being driven. This decouples where the network's effort goes from how large the forcing happens to be at any given moment: a strong push does not by itself inflate $rate(t)$, and a weak one does not by itself suppress it — only the equation's own tendency to amplify or damp a deviation does. $rate(t)$ is answering a narrower, more specific question than 'where is the state changing fastest' — it asks 'where would a small deviation from the current state grow or shrink quickly,' independent of whatever is currently pushing the state around.*
2. *Pure transport contributes nothing. Any piece of $L_i[U(t)]w$ of the form $c(x,t)\frac{\partial w}{\partial x}$, with coefficient $c$ independent of $w$, acts on a perturbation purely by translating it: at fixed $c$, $\frac{\partial w}{\partial t} = c\frac{\partial w}{\partial x}$ solves exactly via $w(x,t) = w_0(x + ct)$, an exact translation under which the amplitude of $w$ is conserved. Such a piece neither grows nor damps a perturbation, so it contributes zero to $rate(t)$.*
3. *The rate is not a general measure of stiffness.* It follows the growth and damping of a perturbation under the linearized equation at the closure scale, so it identifies stiff behavior when stiffness appears as strong local growth or damping, such as reaction-driven growth, thin layers with steep gradients and finite-time blow-up. Stiffness of other kinds is not captured. Transport and other oscillatory dynamics respond with an imaginary part and contribute zero by construction. The damping of a diffusion term is reported at the scale of the solution and not at the finer scales that limit explicit time stepping. Finally, eigenvalue-based measures describe asymptotic behavior, and when the linearized operator is non-normal the transient amplification can be larger than they indicate [27].

### 3.3 Combination rules and spatial reductions

Section 3.2 leaves the combination rule $\Phi$ and the spatial reduction $Q$ unspecified. This section describes some common ones that are used in our case studies.

The rules are built from the growth coefficient $\lambda_i$.

Each $\lambda_i$ describes one mechanism acting alone, but in the equation all mechanisms act on a perturbation at the same time. Since $\frac{\partial w}{\partial t} = \sum_i L_i[U(t)]w$, their responses to the test wave of note * add, and so do the real parts that set the change in size. We write the sum as

$$\Lambda(\boldsymbol{x}, t) \coloneqq \sum_i \lambda_i(\boldsymbol{x}, t) \tag{20}$$

$\Lambda$ is the net growth coefficient: the rate at which the equation, with all of its mechanisms acting together, changes the size of a perturbation at the closure scale. A perturbation grows where $\Lambda > 0$, decays where $\Lambda < 0$, and keeps its size where $\Lambda = 0$. $\Lambda$ cannot serve as the field directly, because a warp density function $\rho(t)$ must be nonnegative and $\Lambda$ is negative wherever damping prevails. A combination rule $\Phi$ is a way of turning the coefficients $\lambda_i$ into a nonnegative number. We describe three below, which differ in how they treat the sign of the coefficients:

- $\Phi_\Sigma \coloneqq \sum_i |\lambda_i|$, the sum of the magnitudes of the individual coefficients;
- $\Phi_{net} \coloneqq |\Lambda|$, the magnitude of the net coefficient;
- $\Phi_+ \coloneqq \max(0, \Lambda)$, the net coefficient where it is positive and zero elsewhere.

At every point they are ordered as:

$$\Phi_+ \leq \Phi_{net} \leq \Phi_\Sigma \tag{21}$$

In Eq. (21) the first inequality holds because $\Phi_+$ equals $\Phi_{\text{net}}$ where $\Lambda \geq 0$ and is zero, hence smaller, where $\Lambda < 0$. The second holds because mechanisms of opposite sign cancel in $\Lambda$ but not in the sum of magnitudes. When all nonzero $\lambda_i$ share a sign, nothing cancels, so $\Phi_{\text{net}} = \Phi_\Sigma$, and $\Phi_+$ equals both if that sign is positive and is zero if it is negative.

Sum of magnitudes. $\Phi_\Sigma$ counts each mechanism by its strength in isolation, so no mechanism that is stiff on its own is cancelled by an opposing one. Since it sums over mechanisms, it depends on the decomposition of $\mathcal{N}$ (merging two terms of opposite sign lowers it), which is why the decomposition is fixed by mechanism.

Net magnitude. $\Phi_{\text{net}}$ is the size of the net response of the frozen operator at the closure scale, the eigenvalue magnitude of classical stiffness analysis [26,27]. It is independent of the decomposition and allows opposing mechanisms to cancel.

Growth. $\Phi_+$ is the net amplification rate, zero where the frozen operator damps. Its meaning follows from the error of an approximate solution $\hat{u}$ with residual $r = \frac{\partial \hat{u}}{\partial t} - \mathcal{N}[\hat{u}] - f$. To first order, the error $\eta = \hat{u} - u$ satisfies

$$\frac{\partial \eta}{\partial t} = L[U(t)]\,\eta + r \tag{22}$$

so error introduced by the residual grows where $\Lambda > 0$ and decays where $\Lambda < 0$ [25,56,57]. $\Phi_+$ is therefore the local rate at which error is amplified at time t. It is independent of the decomposition, and it vanishes identically for an equation whose mechanisms all damp.

The first two rules follow the size of the equation's response wherever a mechanism acts, whether or not the state is evolving there. The third follows the accumulation of error. Which rule an equation calls for is decided by what happens to an error where the equation damps, $\Lambda < 0$. By Eq. (22), an error there decays at the local rate $|\Lambda|$. If the damped state relaxes towards a stable equilibrium of the equation, this decay removes the error and the state reached does not depend on it; the damped regions then need no collocation effort, and $\Phi_+$, which ignores them, is the appropriate rule. If instead the damped regions have no equilibrium to relax to and remain coupled to the regions where $\Lambda > 0$, an error in them is carried into the growing regions, and strong damping is a fast time scale that the collocation points must resolve like strong growth; $\Phi_{\text{net}}$, which measures the response in both, is then the appropriate rule. $\Phi_\Sigma$

is the conservative choice when mechanisms of opposite sign should not cancel. The rule is chosen from the sign structure of $\Lambda$ on the pilot-training state, before the full training.

Spatial reductions $Q$ turn the field $|L[U(\boldsymbol{x},t)]|$ into one number per time point $t$. We introduce three spatial reduction methods used in this paper:

- $Q_{mean}[g] := \frac{1}{|\Omega|}\int_\Omega g\ dx$, weights every point of the domain $\Omega$ equally;
- $Q_{max}[g] := \max_{x\in\Omega}[g]$, is the response at the stiffest point and an upper bound for every admissible reduction;
- $Q_{med}[g]$ is the median of $g$ over $\Omega$, is insensitive to localized features [58], and it is zero whenever the field vanishes on more than half of $\Omega$.

Both choices follow the structure of the equation. The combination rule follows the sign structure of $\Lambda$. Where competing mechanisms should each be resolved, $\Phi_\Sigma$ keeps both and lets neither cancel. Where the aim is the amplification of error and $\Lambda$ changes sign with the state, a rule that ignores damped states, such as $\Phi_+$, is the natural one. The reduction follows the spatial structure of the field. The median gives the scale of a typical point, which a thin layer barely changes; the mean weights each point by its share of the domain when nothing is known about where the error lies; and the maximum follows the single most amplifying point. Section 5 derives these choices for the benchmark equations. They are design criteria supported by those derivations, not guarantees.

## 3.4 The Arc-length Warp Density Function

In Section 3.2 we defined a new warp density function that depends on the underlying operator's structure. There exist other constructions in the literature for choosing a warp density that do not rely on this structure, one being arc-length, built instead from the pilot-training solution's own kinematics.

Arc-length reparameterizes a curve by its own path length: stretching computational time in proportion to how fast the curve $[t, U(t)]^\top$ is actually moving concentrates resolution wherever the solution changes quickly, regardless of why it is changing. This is a classical construction [37], recently used to make stiff ODE solutions more stationary for Gaussian-process (GP) surrogates [40], where a solution $y(t,\boldsymbol{p})$ is reparameterized by the arc length of $[t, y(t,\boldsymbol{p})]^\top$ purely as a preprocessing step before a stationary-kernel GP is fit.

Following Cortes Garcia et al. [40], who take the arc length of the curve $[t, y(t)]^\top$ of a solution component, we take the arc length of the curve $[t, U(\cdot,t)]^\top$, with the speed $\partial U/\partial t$ measured in the $L^2(\Omega)$ norm, which reduces the field to a function of time only. From the pilot-training state $U$ of Section 3.1, it is defined as:

$$arc(t) = \sqrt{1 + \int_\Omega \left|\frac{\partial U(\boldsymbol{x},t)}{\partial t}\right|^2 dx} \tag{23}$$

and take $\rho(t) = arc(t)$ as the warp density function. $\frac{\partial U}{\partial t}$ is evaluated directly from the pilot-training network's own output, not from evaluating $\mathcal{N}[u]$, keeping $arc(t)$ independent of the equation's structure, in contrast to $rate(t)$.

Two consequences follow directly from the definition of arc-length time reparameterization, mirroring the two properties of $rate(t)$ in Section 3.2 from the opposite side:

*$arc(t)$ is not blind to forcing.* An additive term $f(\boldsymbol{x},t)$ contributes nothing to $rate(t)$, but it enters $\partial_t U$ directly the moment the pilot-training solution responds to it — a strong forcing pulse inflates $\mathrm{arc}(t)$ immediately, while $rate(t)$ stays unmoved until the equation's own dynamics engage. This is the forcing-lag condition under which the two warps are constructed to diverge.

*$arc(t)$ does not ignore pure transport.* A solution translating at constant speed with unchanging shape has $\partial_t U \neq 0$ wherever it moves, so $\mathrm{arc}(t)$ concentrates resolution there — exactly what $rate(t)$'s transport property excludes by construction.

*Remark: Throughout the paper, we use arc(t) as the comparison method for time reparameterization on PINNs, precisely because its nature is the opposite of rate(t)'s — this comparison is not a state-of-the-art benchmark, as arc-length time reparameterization has not been implemented on PINNs before to the best of our knowledge [38,40,59], but a way to determine whether Equation-Informed Time Reparameterization offers an advantage over standard, kinematic time reparameterization for PINNs.*

# 4. Practical Considerations

Before continuing with our case studies, we find it useful to share some practical considerations that we used when applying our method to them.

## 4.1 Adjusting the warp strength

Let $\rho(t)$ denote the warp density function, derived as specified in Section 3. Collocation points are sampled uniformly in $s$, so by Eq. (12) their density in physical time is proportional to $\rho(t)$. Thus, the region of physical time densest in collocation points receives $\Gamma = \frac{\max \rho(t)}{\min\, \rho(t)}$ times more collocation points than the sparsest one. When $\Gamma$ is large, the concentration can be excessive, leaving quiet intervals with very few or no collocation points. There, the residual cannot constrain the solution. When $\Gamma$ is small, the concentration can be too weak, and the high-rate region does not receive enough collocation points to be resolved. To solve this problem, we adjust the warp density function with an exponent $\beta > 0$:

$$\rho_\beta(t) = \rho(t)^\beta \tag{24}$$

The strict positivity requirement still holds for this formulation, and the procedure in Section 3.1 applies with $\rho(t)$ replaced by $\rho(t)^\beta$. The density ratio becomes $\Gamma_\beta = \Gamma^\beta$. For example, if $\Gamma = 100$, then $\beta = 0.5$ gives a ratio of $10$, and $\beta = 1.5$ gives a ratio of $1000$. Therefore, the exponent trades concentration in the high-rate region against coverage of the rest of the domain.

The size of this trade-off depends largely on the number $N_t$ of collocation time levels. A sharp warp pulls most of the levels toward the high-rate region, and the quiet intervals keep only a small share of them. When $N_t$ is small, that share can amount to one level or none, so the residual does not constrain the solution there, and $\beta$ must be chosen with care to distribute collocation points adequately to all regions (often $\beta < 1$). When $N_t$ is large, even a small share still contains several levels. The warp can then focus on the high-rate region and still cover the rest of the time domain.

$\beta$ is a tunable hyperparameter, acquired through a sweep of different values for each problem.

## 4.2 Weighting the Residual

In Section 3.1 we introduced two residuals for the reparameterized problem: $r$ of Eq. (13), and $r_{alt}$ of Eq. (14). Both follow from the reparameterization and vanish on the same set, but they differ by a factor of $\rho(t)$ at each collocation point. They therefore distribute the gradient signal differently across time, as specified in the Remark of Section 3.1. Beyond these two, one can add an explicit weight. A pointwise weight on the residual is closely related to a change in the density of the collocation points: with many uniformly sampled points, weighting by a function of position amounts, in expectation, to sampling with a density proportional to that function [60]. Building on $r_{alt}$ we consider the family

$$r_\zeta(\boldsymbol{x}, s; \theta) = \left(\frac{\rho(t(s))}{\bar{\rho}}\right)^{\frac{\zeta}{2}} r_{alt}(\boldsymbol{x}, s; \theta) \tag{25}$$

Here, $\bar{\rho}$ is the geometric mean of $\rho$ over the collocation points. This normalization gives the weights a geometric mean of one for every $\zeta$, so changing $\zeta$ reshapes the weights across time, without changing their overall scale.

The choice $\zeta = 0$ gives the residual of Eq. (14), and $\zeta = -2$ gives $\bar{\rho}$ times the residual of Eq. (13). With the sampling density proportional to $\rho(t)$, the loss contributed per unit of physical time is proportional to $\rho(t)^{1+\zeta}$. This is $\rho(t)$ for $\zeta = 0$ and $1/\rho(t)$ for $\zeta = -2$, and a choice $\zeta = -1$ would weight all physical times equally. A choice $\zeta > 0$ gives the largest weight to the intervals the warp marked as high-rate. The optimizer is then pushed to resolve the residual most tightly where a small error would be amplified fastest and spread into the later solution. The value of $\zeta$ is stated with each case study.

## 4.3 Bootstrapping the rate

The warp of Section 3 is only as good as the state it is evaluated on. On difficult equations, the pilot training, a plain PINN trained on a reduced budget, can be so inaccurate that $rate(t)$ evaluated on it misplaces the warp. Bootstrapping refines the warp using the method's own output (Figure 2). We run a full training on the warp of the pilot training, evaluate $rate(t)$ again on its output, and use the result to define the warp of the next pass. One such step is a bootstrap round. We repeat it for $n$ rounds, and a final full training, run on the warp of the last round, gives the reported solution. The case $n = 0$ is the pilot-training-only method of Section 3.1.

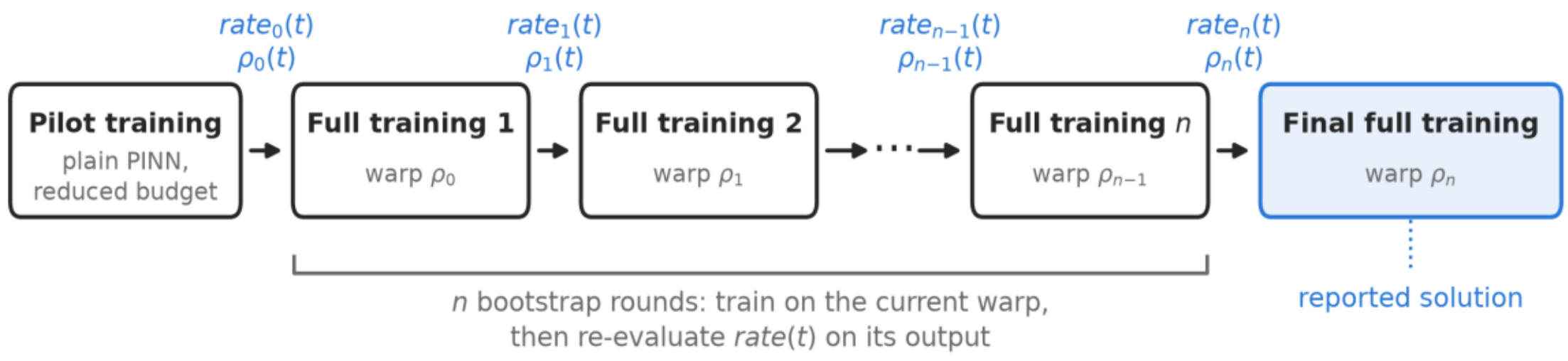

*Figure 2: Bootstrapping the rate with n rounds. The pilot training, a plain PINN trained on a reduced budget, gives a first estimate $rate_0(t)$ and the warp $\rho_0(t)$. Each bootstrap round runs a full training on the current warp and re-evaluates rate(t) on its output. After n rounds, the final full training is run on the last warp $\rho_n(t)$ and gives the reported solution. The case n = 0 is the pilot-training-only method.*

A round helps only if its full training is a more accurate state than the one before it. Additional rounds are therefore worth their training cost only while the rate estimate is still poor, and in some problems, they give little or no improvement.

## 4.4 Implementation details

**Pilot training.** The pilot training is a plain PINN trained on the physical time domain for a fixed number of iterations, without a stopping criterion. It uses fewer iterations than the full training and, in most case studies, a smaller network. The budget of each case study is given in the Appendix, and its cost is credited to the competing methods as extra training (Appendix C).

**Evaluating the rate.** After the pilot training, the network and its derivatives are evaluated by automatic differentiation on a fine grid of the domain. At every time of the grid, the growth coefficients $\lambda_i$ of Section 5 are combined by $\Phi$ and reduced over space by $Q$, which gives $rate(t)$ on the time grid. A cubic spline through these values defines $rate(t)$, which is raised to the power $\beta$ (Section 4.1) and checked for strict positivity. The rate is not updated while a training runs. It is recomputed only between trainings, by the bootstrap rounds of Section 4.3.

**Inverse map.** The map $s(t) = \int_0^t \rho(t')dt'$ is the antiderivative of the spline $\rho(t)$, with end value $S = s(T)$. The network input is the normalized coordinate $\tau = s/S$, and the collocation levels $\tau_j$ are equally spaced in [0,1]. The physical time $t_j$ of each level solves $s(t_j) = \tau_j S$, which has a unique solution because $s' = \rho > 0$. We obtain it by interpolation on a dense table of pairs $(t, s(t))$; in the Kachanov-Rabotnov and Burgers case studies the result is refined with Newton steps, $t_j \leftarrow t_j - (s(t_j) - \tau_j S)/\rho(t_j)$. Errors are evaluated at physical times through $\tau(t) = s(t)/S$.

# 5. Case Studies

Section 3 defined $rate(t)$ at the level of a general operator and left two things open by design: how the linearization is carried out and, where a spatial domain is present, how a pointwise quantity over $x$ is aggregated into a single number per $t$. Our goal in the case studies is to demonstrate that warping time improves the accuracy of PINN training on equations where uniform collocation in time under-resolves a region of fast error growth, using the same recipe for very different equations. This section prepares those experiments by carrying out the derivation explicitly for the paper's four benchmark equations, ordered so each introduces exactly one new complication. The Kachanov-Rabotnov creep-damage ODE has no spatial domain, so it isolates the linearization step on its own. Burgers' equation, Allen-Cahn, and the Frank-Kamenetskii equation then bring in a spatial domain, with increasing demands on the aggregation choice: in Burgers' equation the growth coefficients compete and the structure is localized in a thin layer, in Allen-Cahn the combined coefficient changes sign with the state, and in the Frank-Kamenetskii equation a diffusion term damps what a source term amplifies, up to a finite-time blow-up.

## 5.1 Kachanov-Rabotnov

Kachanov-Rabotnov is a continuum damage mechanics model for creep rupture [61–64], predicting when a structural component under sustained load and high temperature will fail, not through sudden fracture but through the slow accumulation of internal microvoids and microcracks that progressively erode the material's load-bearing cross-section. It underlies remaining-life assessment for components such as turbine blades, high-temperature piping, and pressure vessels [65], where a material is held under constant stress for extended periods and the question is how long it survives before rupture. The state variable $\omega(t) \in [0,1)$ is the damage fraction — the proportion of the cross-section's original load-bearing capacity already lost — with $\omega = 0$ representing pristine material and $\omega \to 1$ representing imminent rupture. The model is standard in this setting precisely because it captures the experimentally observed creep signature [66]: damage accumulates slowly at first, then accelerates sharply as the remaining intact material is forced to carry increasing effective stress, diverging in finite time.

The governing equation is:

$$\frac{d\omega}{dt} = \mathcal{N}(\omega) \coloneqq M\sigma^{\chi}(1-\omega)^{-\varphi}, \;\; \omega(0) = 0, \;\; t \in [0, t_{end}], \;\; t_{end} = (1-\psi)t_f, \tag{26}$$

where $\sigma$ is the applied stress, held fixed from $t = 0$, $M, \chi, \varphi$ are material constants fit from experimental creep curves [67], $t_f$ is the closed-form rupture time, and $\psi > 0$ is a small truncation margin keeping $\omega$ away from its singularity at 1.

The case studied in this paper is DS GTD-111, a directionally solidified Ni-base superalloy used in first-stage gas-turbine buckets, under the L-orientation, 816°C, $\sigma = 455$ MPa test condition reported by Stewart, Gordon, Hogan & Saxena [68]: $M = 6.4127 \times 10^{-10}$ $\text{MPa}^{-2.257}\text{h}^{-1}$, $\chi = 2.257$, $\varphi = 3.792$, giving $t_f \approx 326.1$ h.

To derive $rate(t)$ per Section 3.2, the operator is decomposed as $\mathcal{N} = \sum_i \mathcal{N}_i$. However, the equation has a single term, $\mathcal{N}_1 = \mathcal{N}$, so the decomposition is trivial. Let $U(t)$ represent the state of the pilot-training estimate of $\omega(t)$, perturbing it by $\omega \to \omega + w$:

$$\mathcal{N}(\omega + w) = M\sigma^{\chi}(1-\omega-w)^{-\varphi} \tag{27}$$

Let $\xi \coloneqq 1 - \omega$, then:

$$\mathcal{N}(\omega + w) = M\sigma^{\chi}(\xi - w)^{-\varphi} \tag{28}$$

Write $F(w) = (\xi - w)^{-\varphi}$ and differentiate repeatedly with respect to $w$:

$$F'(w) = \varphi(\xi - w)^{-\varphi-1}, F''(w) = \varphi(\varphi+1)(\xi - w)^{-\varphi-2},$$
$$F'''(w) = \varphi(\varphi+1)(\varphi+2)(\xi - w)^{-\varphi-3} \tag{29}$$

and by induction the $l$ -th derivative, for any fixed order $l$ is:

$$F^{(l)}(w) = \varphi(\varphi+1)(\varphi+2)\dots(\varphi+l-1)(\xi - w)^{-\varphi-l} \tag{30}$$

Evaluating this at $w = 0$ gives the Taylor coefficients $F^{(l)}(0) = \varphi(\varphi+1)\dots(\varphi+l-1)\xi^{-\varphi-l}$ [69], and substituting them into the Maclaurin series for $F$ itself [47]:

$$F(w) = \sum_{l=0}^{\infty} \frac{F^{(l)}(0)}{l!} w^l = \sum_{l=0}^{\infty} \frac{\varphi(\varphi+1)\dots(\varphi+l-1)}{l!} \xi^{-\varphi-l} w^l =$$
$$= \xi^{-\varphi} + \varphi\xi^{-\varphi-1} w + O(w^2) \tag{31}$$

The $l = 0$ term recovers $F(0) = \xi^{-\varphi}$, i.e. $\mathcal{N}(\omega)$ itself once the prefactor is restored; the $l = 1$ term is the linear-in-$w$ piece that we are looking for per Section 3.2. Restoring $M\sigma^{\chi}$:

$$\mathcal{N}(\omega + w) = M\sigma^{\chi}\xi^{-\varphi} + M\sigma^{\chi}\varphi\xi^{-\varphi-1}w + O(w^2)$$
$$\xrightarrow{\xi=1-\omega} \underbrace{M\sigma^{\chi}(1-\omega)^{-\varphi}}_{\mathcal{N}(\omega)} + M\sigma^{\chi}\varphi(1-\omega)^{-\varphi-1}w + O(w^2) \tag{32}$$

so the linear-in-$w$ coefficient is $L_1[\omega(t)] = M\sigma^{\chi}\varphi\left(1 - \omega(t)\right)^{-\varphi-1}$. In ODEs like this one, $\omega$ depends on $t$ alone, so $L_1[\omega(t)]$ is a function of $t$ alone, not a pointwise field over a domain. Moreover, it does not contain any derivatives of $w$, so the closure of note * is not needed here. $L_1[\omega(t)]$ is real, so the growth coefficient is:

$$\lambda(t) = Re(L_1[\omega(t)]) = L_1[\omega(t)] \tag{33}$$

*Strict positivity holds on the truncated domain since $M, \sigma, \chi, \varphi > 0$ and $1 - \omega(t) > 0$ for $t \le t_{end}$ ; concretely, all four DS GTD-111 constants above are positive.*

Choosing $\Phi$ and $Q$ in this case is also trivial. With one mechanism $\Lambda = \lambda$, so $\Phi = \Phi_+ = \Phi_{net} = \Phi_{\Sigma} = |\lambda| = \lambda$, since positivity is ensured, thus $L[\omega(t)] = \Phi = L_1[\omega(t)]$. Since the equation has no spatial variable ($\omega$ depends on $t$ only), the spatial reduction $Q$ is not needed, and we get:

$$\boxed{\mathrm{rate}(t) = M\sigma^{\chi}\varphi\left(1 - \omega(t)\right)^{-\varphi-1}} \tag{34}$$

σ enters $\mathcal{N}(\omega)$ multiplicatively, as a fixed material-loading parameter, not as an additive forcing term, so this equation does not exercise the forcing-invisibility property of Section 3.2. What it does isolate is the runaway-stiffness behavior: since $\mathcal{N}(\omega)$ and $\mathrm{rate}(t)$ share the factor $(1-\omega)^{-\varphi}$, but $\mathrm{rate}(t)$ carries one additional power of $(1-\omega)^{-1}$ times $\varphi$, $\mathrm{rate}(t)$ grows strictly faster than $\dot{\omega}$ as $\omega \to 1$. Thus $rate(t)$ anticipates coming stiffness ahead of raw velocity for this monotone-runaway equation.

We finally take $\rho(t) = rate(t)$.

## 5.2 Burgers' Equation

The viscous Burgers' equation is a one-dimensional model for the interplay between nonlinear steepening and viscous diffusion of a wave, and it is the simplest equation in which that interplay appears [70]. It originates in Burgers' mathematical models of turbulence [71], and it follows from the Navier–Stokes equation for the velocity field once the pressure gradient is dropped [6]. Accordingly, it appears in fluid mechanics, gas dynamics, traffic flow, and nonlinear acoustics. The equation can be solved exactly: the Cole-Hopf transformation reduces it to the linear heat equation [70,72,73], so exact reference solutions are available [74]. Together with its use in the original PINN study [6] and in the PINNacle benchmark [15], this makes it a standard test problem. The governing equation, in the benchmark configuration of Raissi et al. [6], is

$$\frac{\partial u}{\partial t} + u\frac{\partial u}{\partial x} = \nu\frac{\partial^2 u}{\partial x^2}, \qquad x \in \Omega = [-1,1], t \in [0,1], \qquad u(x,0) = -\sin(\pi x), \quad u(\pm 1, t) = 0 \tag{35}$$

where $\nu > 0$ is the kinematic viscosity, which for the benchmark configuration is $\nu = \frac{0.01}{\pi}$, and the homogeneous Dirichlet conditions hold at both ends of the domain. In this configuration the solution develops a sharp internal layer around $t \approx 0.4$ [6].

Burgers is the first benchmark with a spatial domain, so the state is now a field $u(x,t)$ and the general recipe is applied in full: the decomposition of the operator, the growth coefficients $\lambda_i(x,t)$, the combination rule $\Phi$ and the spatial reduction $Q$ over $\Omega = [-1,1]$. Write $\mathcal{N}[u] = \nu\frac{\partial^2 u}{\partial x^2} - u\frac{\partial u}{\partial x}$, so that the equation is of the form $\frac{\partial u}{\partial t} = \mathcal{N}[u]$.
The operator $\mathcal{N}$ is decomposed by mechanism as $\mathcal{N} = \mathcal{N}_1 + \mathcal{N}_2$, with:

$$\mathcal{N}_1[u] = -u\frac{\partial u}{\partial x}, \quad \mathcal{N}_2[u] = \nu\frac{\partial^2 u}{\partial x^2} \tag{36}$$

the nonlinear advection and the diffusion.
Perturb $u \to u + w$. Expanding the product in $\mathcal{N}_1$ term by term,

$$\mathcal{N}_1[u+w] = -(u+w)\left(\frac{\partial u}{\partial x} + \frac{\partial w}{\partial x}\right) = \underbrace{-u\frac{\partial u}{\partial x}}_{\mathcal{N}_1[u]} + \left(-u\frac{\partial w}{\partial x} - w\frac{\partial u}{\partial x}\right)\underbrace{-w\frac{\partial w}{\partial x}}_{R_1(w^2)}, \tag{37}$$

and then

$$\mathcal{N}_2[u+w] = \underbrace{\nu\frac{\partial^2 u}{\partial x^2}}_{\mathcal{N}_2[u]} + \nu\frac{\partial^2 w}{\partial x^2} \tag{38}$$

Evaluating the linear-in-w terms at the pilot-training state $u = U(x,t)$,

$$L_1[U(x,t)]w = -U\frac{\partial w}{\partial x} - w\frac{\partial U}{\partial x}, \qquad L_2[U(x,t)]w = \nu\frac{\partial^2 w}{\partial x^2} \tag{39}$$

Both contain derivatives of w, so neither has a pointwise magnitude. Per Section 3.2 each is tested on the plane wave $w = e^{ikx}$, with the coefficients $U$ and $\frac{\partial U}{\partial x}$, which vary in $x$, frozen over the wavelength of the perturbation. Since $\frac{\partial^q w}{\partial x^q} = (ik)^q w$, we get

$$L_1[U(x,t)]w = \left(-ikU - \frac{\partial U}{\partial x}\right)w, \quad L_2[U(x,t)]w = -\nu k^2 w \tag{40}$$

The terms multiplying $w$ are the response of $\mathcal{N}_1$ and $\mathcal{N}_2$. The response to $\mathcal{N}_1$ has an imaginary part $-kU$, coming from the piece $-U\frac{\partial w}{\partial x}$. This piece has the form $c\frac{\partial w}{\partial x}$ of pure transport, which means it translates the perturbation wave, without changing its size (pure transport property, Section 3.2). Extracting the real part of the responses gives:

$$\lambda_1(x,t) = Re(L_1[U(x,t)]) = -\frac{\partial U}{\partial x}, \quad \lambda_2(x,t) = Re(L_2[U(x,t)]) = -\nu\left|\frac{\partial U}{\partial x}\right|^2, \tag{41}$$

where the second uses the closure $k = \left|\frac{\partial U}{\partial x}\right|$ defined in Section 3.2.

$\lambda_1$ is positive when $\frac{\partial U}{\partial x} < 0$ and negative when $\frac{\partial U}{\partial x} > 0$, while the second growth coefficient $\lambda_2 \leq 0$ everywhere (Figure 3). Therefore, the two growth coefficients oppose each other in the compressive region, where $\lambda_1 > 0$ and $\lambda_2 < 0$. This opposition is the physical content of the benchmark: steepening and diffusion compete, and diffusion eventually halts the steepening. To capture the effect of each mechanism, we take $\Phi = \Phi_\Sigma$. The net rule $\Phi_{net}$ lets the two cancel, and $\Phi_+$ is zero when $\frac{\partial U}{\partial x} \geq 0$, which covers half of $\Omega$ in the initial condition $u(x,0) = -\sin(\pi x)$, where $\frac{\partial u}{\partial x} > 0$ for $|x| > \frac{1}{2}$.

Thus, with $\Phi_\Sigma$ we get:

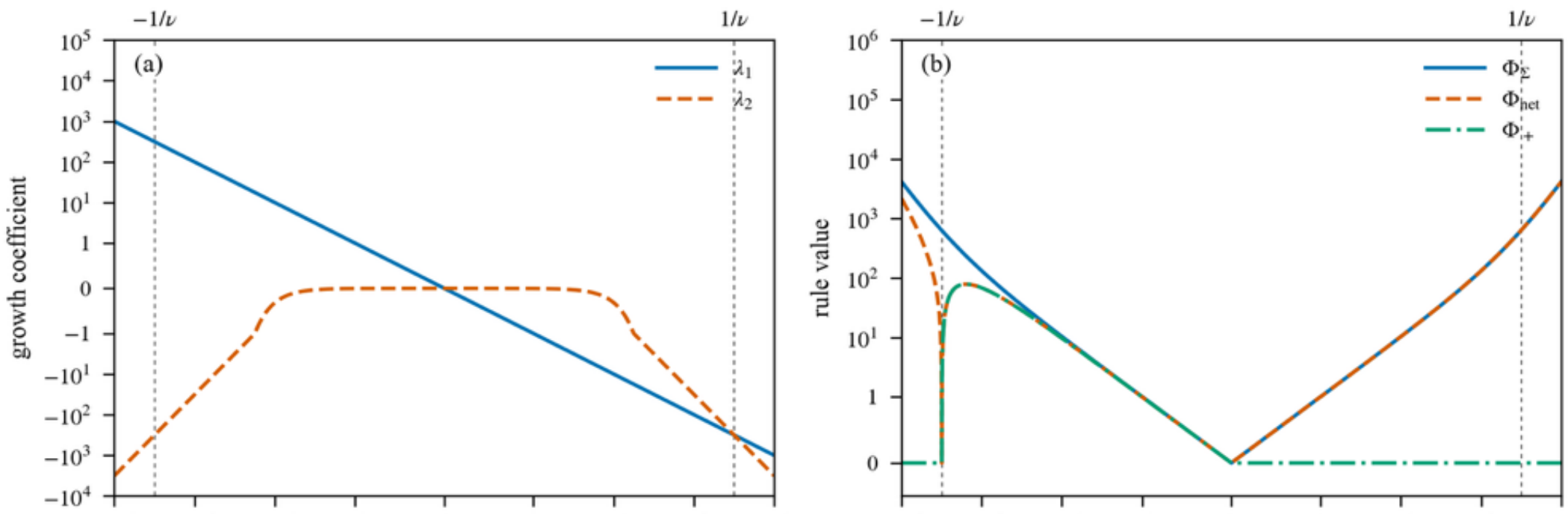

*Figure 3: Growth coefficients and combination rules for the Burgers equation as a function of ∂U/∂x. (a) λ_1 and λ_2. (b) Φ_Σ, Φ_net, Φ_+.*

$$|L[U(x,t)]| = |\lambda_1| + |\lambda_2| = \left|\frac{\partial U}{\partial x}\right| + \nu\left|\frac{\partial U}{\partial x}\right|^2 \tag{42}$$

At every time point $t$, $|L[U(x,t)]|$ is a field over $x$, while the warp density function $\rho(t)$ requires a real number. This is when the spatial reduction takes place. We take $Q = Q_{med}$, the median over $\Omega$, which satisfies the requirements on $Q$ in Section 3.2. Thus,

$$rate(t) = Q_{med}[|L[U(\cdot,t)]|] = median_{x\in\Omega}\left(\left|\frac{\partial U}{\partial x}\right| + \nu\left|\frac{\partial U}{\partial x}\right|^2\right) \tag{43}$$

Since $|L[U(x,t)]|$ is an increasing function of $\left|\frac{\partial U}{\partial x}\right|$, the median passes through it, finally giving us,

$$rate(t) = G(t) + \nu G^2(t), with\ G(t) = median_{x\in\Omega}\left(\left|\frac{\partial U}{\partial x}\right|\right) \tag{44}$$

Positivity holds, thus we take $\rho(t) = rate(t)$.

The choice of the reduction method as noted above is not arbitrary. To further illustrate this choice we will need to look at the spatial structure of the field.

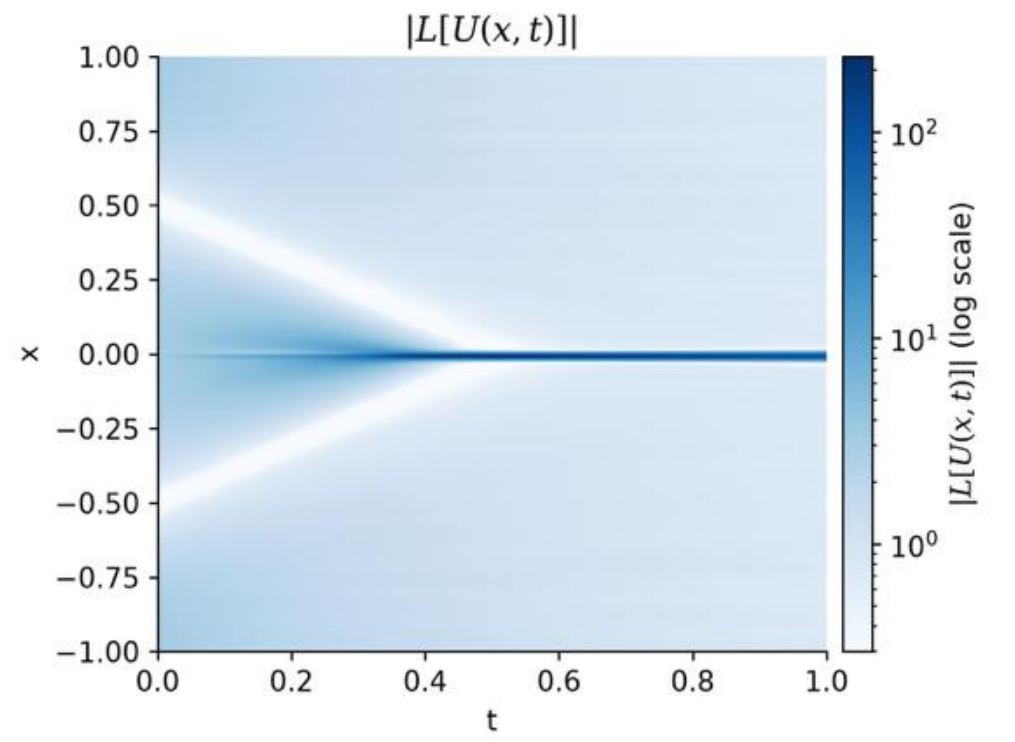

*Figure 4: Spatiotemporal map of |L[U(x,t)]| for the Burgers equation on the pilot-training state, for x ∈ [−1,1] and t ∈ [0,1] (logarithmic color scale).*

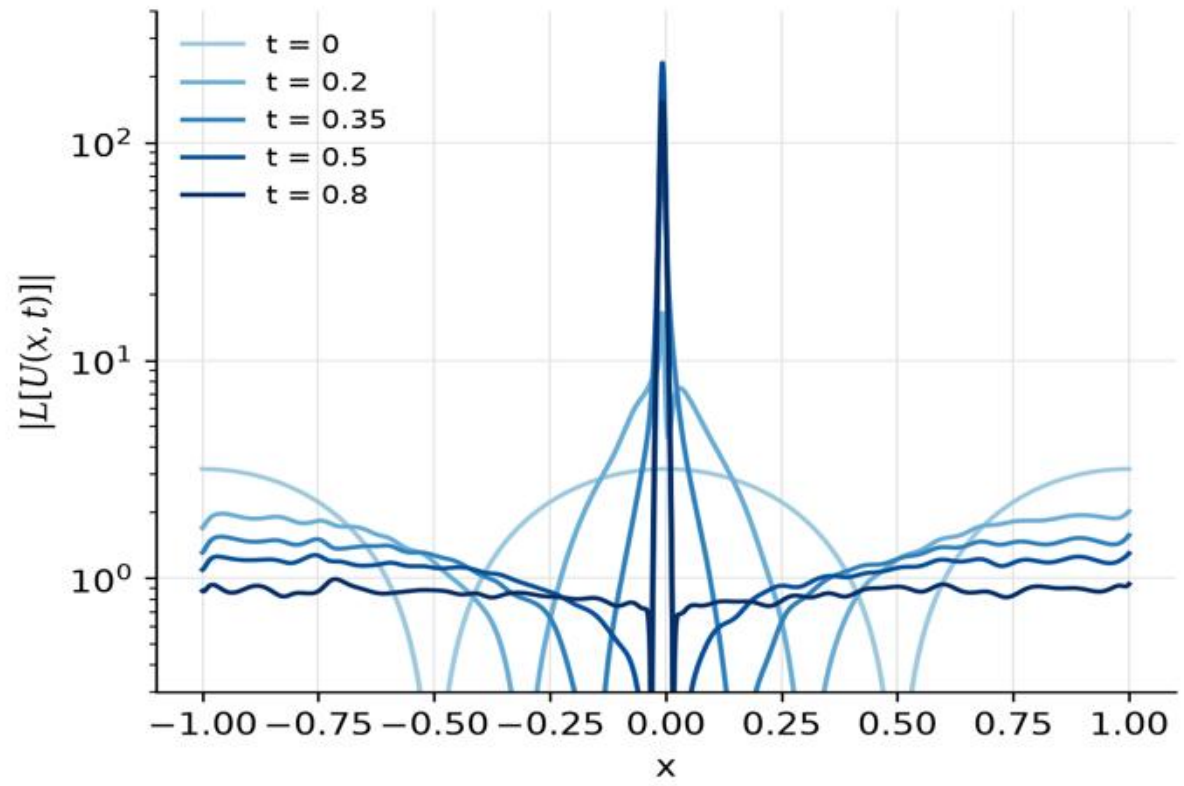

*Figure 5: Profiles of |L[U(x,t)]| over x at t = 0, 0.2, 0.35, 0.5 and 0.8 for the Burgers equation (logarithmic vertical axis).*

On the pilot-training state, $|L[U(x,t)]|$ is of order 1 to 3 except in a thin band at $x \approx 0$ (Figure 4), which appears from $t \approx 0.3$ and persists to $t = 1$; from $t \approx 0.35$ on, the profiles at individual times show a single narrow spike on a flat background (Figure 5). Among the reductions over $x$, the maximum follows the spike, rising from about 3 to about 230 near $t \approx 0.5$, and reports a single narrow location (Figure 6). The mean stays between about 1.5 and 2.3 for all $t$, because the growing contribution of the layer offsets the decay of the bulk, and so carries almost no time variation. The median decreases smoothly from about 2.3 to 0.75. From $t \approx 0.45$ on, the 10th–90th percentile band (Figure 6) is narrow, so most points take values close to the median; before $t \approx 0.3$ it is wider (about 0.5 to 4). The median thus describes a typical point of $\Omega$.

This choice follows from a simple criterion. Under the warp $\frac{\mathrm{d}s}{\mathrm{d}t} = \rho(t)$ with $\rho(t) = \mathrm{rate}(t)$, the equation in the computational clock is $\frac{\partial u}{\partial s} = \frac{\mathcal{N}[u]}{\mathrm{rate}(t)}$, so the pointwise rate field becomes $\frac{|L[U(x,t)]|}{\mathrm{rate}(t)}$, which should be of order one at a typical point of $\Omega$. Choosing $b = \mathrm{rate}(t)$ to minimize the deviation in orders of magnitude, $\int_\Omega \left|\ln\left(\frac{|L[U(x,t)]|}{b}\right)\right| \mathrm{d}x$, gives the median of $|L[U(x,t)]|$ over $x$, since the minimum is reached when equal measures of $\Omega$ lie above and below $b$. The mean instead minimizes the squared deviation on a linear scale, and the maximum is the smallest $b$ for which $\frac{|L[U(x,t)]|}{b} \leq 1$ everywhere. As the criterion weights all of $\Omega$ equally, a thin feature such as the shock layer barely moves the median. This argument explains why the median is a sensible scale, not that it trains best.

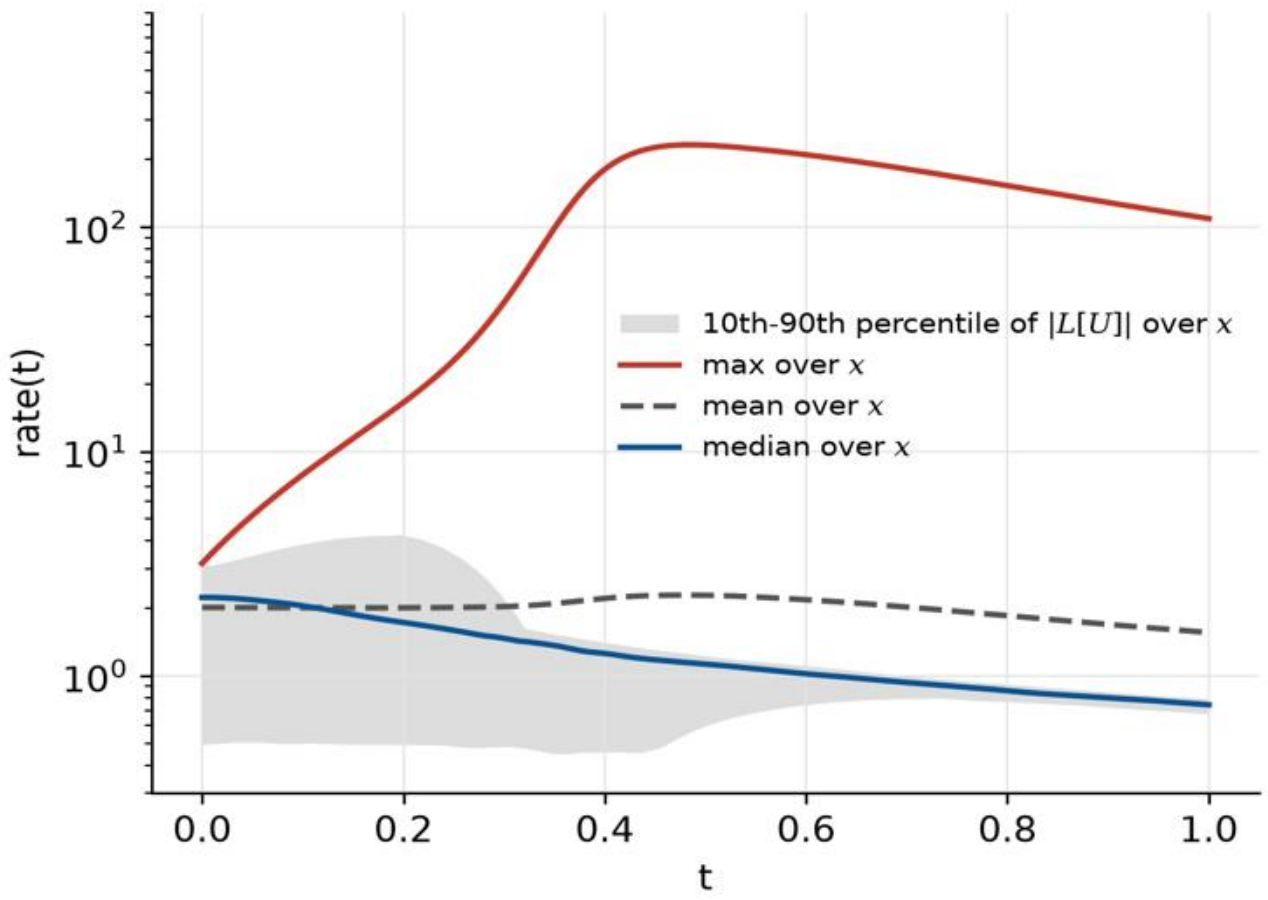

*Figure 6: Reductions of |L[U(x,t)]| over x (maximum, mean and median) as functions of t for the Burgers equation. The shaded band indicates the 10th–90th percentile of |L[U(x,t)]| over x (logarithmic vertical axis).*

## 5.3 Allen-Cahn

The Allen-Cahn equation models phase separation in a binary alloy, introduced by Allen and Cahn [75] for the motion of antiphase boundaries. The field $u$ is an order parameter with two stable phases, $u = \pm 1$. The reaction term drives $u$ toward the nearest phase, and the diffusion term smooths the thin interface between the two. The equation is the gradient flow of a Ginzburg-Landau free energy, and Allen and Cahn showed that the interface it describes moves with a velocity proportional to its mean curvature. This makes it the basic model of phase-field methods, which are also used for multiphase flow, crystal growth and image segmentation. It has been a standard PINN benchmark since Raissi, Perdikaris and Karniadakis [6], and is a case where vanilla PINNs are known to struggle.

We use the standard configuration:

$$\frac{\partial u}{\partial t} = \nu \frac{\partial^2 u}{\partial x^2} + 5u - 5u^3, x \in \Omega = [-1,1], t \in [0,1], \nu = 10^{-4}, \quad u(x,0) = x^2 \cos(\pi x) \tag{45}$$

with periodic boundary conditions $u(-1,t) = u(1,t)$ and $\frac{\partial u(-1,t)}{\partial x} = \frac{\partial u(1,t)}{\partial x}$.

We take:

$$\mathcal{N}[u] = \nu \frac{\partial^2 u}{\partial x^2} + 5u - 5u^3 \tag{46}$$

We now write the operator as a sum of three terms: $\mathcal{N}_1, \mathcal{N}_2, aqd\ \mathcal{N}_3$, where

$$\mathcal{N}_1[u] = \nu \frac{\partial^2 u}{\partial x^2}, \quad \mathcal{N}_2[u] = 5u, \quad \mathcal{N}_3[u] = -5u^3 \tag{47}$$

are the diffusion, the linear reaction and the cubic reaction, respectively. Perturbing $u \to u + w$

$$\mathcal{N}_1[u+w] = \underbrace{\nu \frac{\partial^2 u}{\partial x^2}}_{\mathcal{N}_1[u]} + \nu \frac{\partial^2 w}{\partial x^2}, \qquad \mathcal{N}_2[u+w] = \underbrace{5u}_{\mathcal{N}_2[u]} + 5w, \quad \mathcal{N}_3[u+w] = \underbrace{-5u^3}_{\mathcal{N}_3[u]} - 15u^2 w \underbrace{-15uw^2 - 5w^3}_{R_3(w^2)} \tag{48}$$

Retaining the linear-in-$w$ terms and evaluating at the pilot-training state $U(x,t)$ gives

$$L_1[U(x,t)]w = \nu \frac{\partial^2 w}{\partial x^2}, \quad L_2[U(x,t)]w = 5w, L_3[U(x,t)]w = -15U^2 w \tag{49}$$

Only the diffusion term contains a derivative of $w$ , so it is tested on the plane wave $w = e^{ikx}$. Since $\frac{\partial^q w}{\partial x^q} = (ik)^q w$, we get $L_1[U(x,t)]w = -\nu k^2 w \xrightarrow{k=\left|\frac{\partial U}{\partial x}\right|} -\nu \left|\frac{\partial U}{\partial x}\right|^2 w$. Extracting the real part of the responses we get the coefficients

$$\lambda_1 = Re(L_1[U(x,t)]) = -\nu \left|\frac{\partial U}{\partial x}\right|^2, \quad \lambda_2 = Re(L_2[U(x,t)]) = 5, \quad \lambda_3 = Re(L_3[U(x,t)]) = -15U^2 \tag{50}$$

so,

$$\Lambda(x,t) = \sum_i \lambda_i = \lambda_1 + \lambda_2 + \lambda_3 = 5 - 15U^2 - \nu \left|\frac{\partial U}{\partial x}\right|^2 \tag{51}$$

The next step is choosing $\Phi$ and $Q$. To do this, we need to first examine the sign structure of $\Lambda(x,t)$. The two reaction coefficients sum to $\lambda_2 + \lambda_3 = 5 - 15U^2$, a quantity that is positive for $U^2 < \frac{1}{3}$ (reaction amplifies the perturbation), and negative when $U^2 > \frac{1}{3}$ (reaction damps the perturbation). The diffusion coefficient $\lambda_1$ is always non-positive, so it only damps the perturbation. Consequently $\Lambda > 0$ implies $U^2 < \frac{1}{3}$. The equation thus has states on which $\Lambda$ is positive and states on which it is negative, and the combination rules of Section 3.3 differ in how they treat this sign structure.

Choice of $\Phi$: We apply the criterion of Section 3.3. Where the reaction damps, $U^2 > \frac{1}{3}$, the state relaxes towards the stable equilibria $u = \pm 1$ of the reaction, and an error there decays without changing the state it relaxes to. Since $\nu$ is small, the linearized operator acts almost pointwise in $x$, so these regions are not coupled to those where $\Lambda > 0$, and their error is not carried into them. The solution is decided where $\Lambda > 0$, near $u = 0$, where an error is amplified. $\Phi_+$ is the rate at which error introduced by the residual is amplified, and it is zero wherever the equation damps, $\Lambda \leq 0$. The other two rules assign a positive rate to the relaxing states: $\Phi_\Sigma = 5 + 15U^2 + \nu\left(\frac{\partial U}{\partial x}\right)^2 \geq \lambda_2 = 5$ never falls below the linear reaction rate and is at least 10 where the reaction damps, and $\Phi_{net} = |\Lambda|$ grows with $U^2$ as the damping strengthens. The clip is applied pointwise, before the reduction. We take $\Phi = \Phi_+$.

Choice of $Q$: Since $\lambda_1 \leq 0$ and $\lambda_3 \leq 0$, the field $\Phi_+$ satisfies $0 \leq \Phi_+ \leq \lambda_2 = 5$, so $Q_{\max} \leq 5$. $Q_{med}$ is zero whenever $\Lambda \leq 0$ on at least half of $\Omega$, and otherwise it ignores the size of the amplification on the rest. A reduction should reflect where the error lies. For a perturbation $w$ of the linearized equation on the periodic domain, integration by parts, with the derivative of the perturbation taken at the closure scale of note *, gives

$$\frac{1}{2}\frac{d}{dt}\int_\Omega w^2\,dx \approx \int_\Omega (5 - 15U^2)\,w^2 dx - \nu\int_\Omega \left|\frac{\partial U}{\partial x}\right|^2 w^2 dx, \tag{52}$$

so the growth rate of the $L^2$ norm is a weighted average of the local coefficient, with weights set by the distribution of $w^2$. This distribution is not known when the warp is built. The mean is the weighting that assumes nothing about it, each point of $\Omega$ contributing in proportion to its share of $\Omega$. $Q_{\max}$ is the $L^\infty$ counterpart and attributes all of the error to the single most amplifying point. We take $Q = Q_{mean}$, so:

$$rate(t) = \frac{1}{2}\int_{-1}^{1} \max\left(0{,}5 - 15U^2 - \nu\left|\frac{\partial U}{\partial x}\right|^2\right) dx \tag{53}$$

Positivity cannot be guaranteed by construction, thus we finally take as the warp density function:

$$\rho(t) = rate(t) + \varepsilon \tag{54}$$

## 5.4 Frank-Kamenetskii Thermal Explosion

A reactive solid that releases heat through a thermally activated exothermic reaction can ignite without an external source when the heat it generates cannot be conducted away fast enough [76–78]. The heat generation rate depends nonlinearly on temperature, exponentially for Arrhenius kinetics, whereas the heat loss is essentially linear [79]. Semenov's theory applies when the temperature in the body is uniform, and Frank-Kamenetskii's when heat is transported by conduction alone [80]. In the latter case, a steady state exists only when the Frank-Kamenetskii parameter does not exceed a critical value that depends on the geometry (0.88 for a slab) [80]. Above it, the temperature starting from a uniform initial state blows up in finite time [81–83]. Yuan et al. [84] cite this theory as the foundation of their model of self-heating ignition in coal layers. It has been applied to forest litter [85] and to stacked lithium-ion cells [86], whose critical ambient temperature falls from 165.5 °C for one cell to 153 °C for four.

Consider an infinite slab of half-thickness a whose faces are held at the ambient temperature $\Theta_0$. For $\frac{\mathcal{R}\Theta_0}{E} \ll 1$, with $E$ the activation energy and $\mathcal{R}$ the gas constant, the exponent of the Arrhenius source is linearized about $\Theta_0$. With $u = \frac{E(\Theta - \Theta_0)}{(\mathcal{R}{\Theta_0}^2)}$ , $x$ measured in units of a and t in units of $\frac{a^2}{\alpha}$, with α the thermal diffusivity, the heat balance becomes

$$\frac{\partial u}{\partial t} = \frac{\partial^2 u}{\partial x^2} + \delta e^u, \qquad x \in [-1,1], t \in [0, t_{end}],\ u(x,0) = 0, u(-1,t) = u(1,t) = 0 \tag{55}$$

where $a$ is the half-width of the slab and $\delta = \left(\frac{E}{\mathcal{R}\Theta_0^2}\right)\left(\frac{HZa^2}{\kappa}\right)e^{-\frac{E}{\mathcal{R}\Theta_0}}$ is the Frank-Kamenetskii parameter. Here H is the heat of reaction per unit volume, Z the pre-exponential factor and κ the thermal conductivity. The steady problem has a solution only for $\delta \le \delta_c = 0.8785$ (closed form in [87]). For $\delta > \delta_c$ the temperature at $x = 0$ diverges at a finite time $t_b$, and the system blows up.

The coefficients are taken from a lithium-ion cell (Sanyo UF103450P, $LiCoO_2$, 30% state of charge) [86]. The rate diverges as $t \to t_b$, so, as for Kachanov-Rabotnov, the domain is truncated at $t_{end} = (1 - \psi)t_b$, with $\psi = 3 \cdot 10^{-3}$. Since the exponent is linearized about $\Theta_0$, the problem is the Frank-Kamenetskii model with coefficients derived from a material, not a quantitative description of cell ignition.

We take $\mathcal{N}[u] = \frac{\partial^2 u}{\partial x^2} + \delta e^u$. We split it into two terms:

$$\mathcal{N}_1[u] = \frac{\partial^2 u}{\partial x^2}, \qquad \mathcal{N}_2[u] = \delta e^u \tag{56}$$

where $\mathcal{N} = \mathcal{N}_1 + \mathcal{N}_2$

Perturbing $u \to u + w$ we get

$$\mathcal{N}_1[u+w] = \frac{\partial^2 (u+w)}{\partial x^2} = \underbrace{\frac{\partial^2 u}{\partial x^2}}_{\mathcal{N}_1[u]} + \frac{\partial^2 w}{\partial x^2}, \qquad \mathcal{N}_2[u+w] = \delta e^{u+w} = \delta e^u e^w \tag{57}$$

Expanding the exponential $e^w$ in its Taylor series [69]

$$e^w = \sum_{l=0}^{\infty} \frac{w^l}{l!} = 1 + w + \frac{w^2}{2} + O(w^3) \tag{58}$$

Thus $\mathcal{N}_2[u+w]$ becomes

$$\mathcal{N}_2[u+w] = \delta e^u\left(1 + w + R_2(w^2)\right) = \underbrace{\delta e^u}_{\mathcal{N}_2[u]} + w\delta e^u + R_2(w^2) \tag{59}$$

Retaining the linear-in-$w$ terms and evaluating at the pilot-training state $U(x,t)$ gives

$$L_1[U(x,t)]w = \frac{\partial^2 w}{\partial x^2}, \quad L_2[U(x,t)]w = \delta e^U w \tag{60}$$

The derivative of $w$ is tested on the plane wave $w = e^{ikx}$. Since $\frac{\partial^q w}{\partial x^q} = (ik)^q w$, we get $L_1[U(x,t)]w = -k^2 w \xrightarrow{k=\left|\frac{\partial U}{\partial x}\right|} -\left|\frac{\partial U}{\partial x}\right|^2 w$. There are no imaginary parts, so we get the coefficients

$$\lambda_1 = Re(L_1[U(x,t)]) = -\left|\frac{\partial U}{\partial x}\right|^2, \quad \lambda_2 = Re(L_2[U(x,t)]) = \delta e^U \tag{61}$$

thus

$$\Lambda(x,t) = \sum_i \lambda_i = \lambda_1 + \lambda_2 = -\left|\frac{\partial U}{\partial x}\right|^2 + \delta e^U \tag{62}$$

The next step is choosing the right combination rule $\Phi$ and spatial reduction $Q$. The diffusion coefficient $\lambda_1 = -\left|\frac{\partial U}{\partial x}\right|^2$ only damps the perturbation, while the source coefficient $\lambda_2 = \delta e^U$ only amplifies it (Figure 7a), so $\Lambda > 0$ exactly where $\delta e^U > \left|\frac{\partial U}{\partial x}\right|^2$. Since $U$ is even in $x$, $\Lambda(0,t) = \delta e^{U(0,t)} > 0$ at the hot spot.

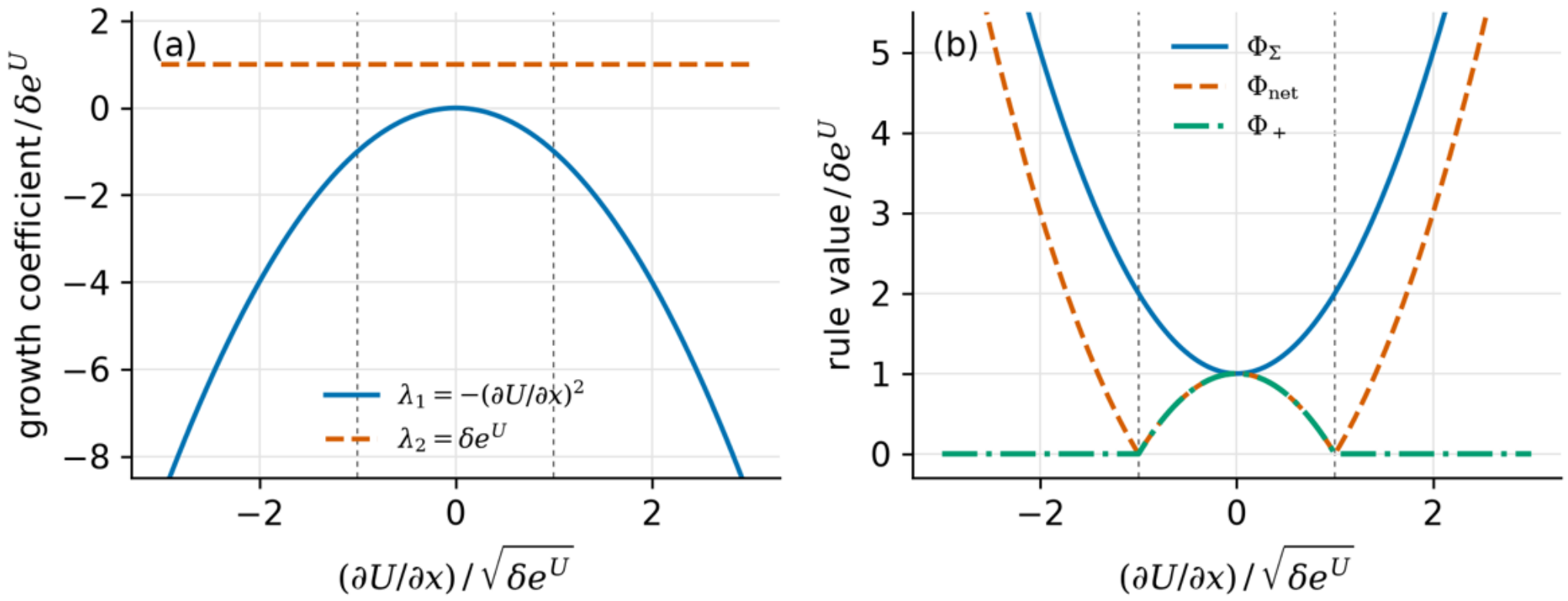

*Figure 7: Growth coefficients and combination rules for the Frank-Kamenetskii equation, normalized by $\delta e^U$, as functions of the normalized gradient $(\partial U/\partial x)/\sqrt{(\delta e^U)}$. (a) $\lambda_1$ and $\lambda_2$. (b) $\Phi_\Sigma$, $\Phi_{net}$, $\Phi_+$.*

Choice of $\Phi$: We apply the criterion of Section 3.3. The source amplifies, and diffusion damps where the gradient is steep, so $\Lambda > 0$ in the hot spot and $\Lambda < 0$ in the flanks, separated by the interface $\Lambda = 0$ (Figure 8). Since $\delta > \delta_c$, the equation has no steady state to which the flanks could relax: they keep heating as the hot spot grows and exchange heat with it through diffusion, which is of unit strength here, so an error in the flanks is carried into the hot spot, and their damping is a time scale that the warp must resolve. $\Phi_{\text{net}} = |\Lambda|$ is the size of the net response in both regions and vanishes only on the interface, where the two mechanisms cancel. $\Phi_+$ would discard the flanks, and $\Phi_\Sigma$ would count the two mechanisms even where they cancel (Figure 7b). Since the response is local in $x$ (frozen coefficients, Section 3.2), $\Phi_{\text{net}}$ is evaluated pointwise, before the reduction. We take $\Phi = \Phi_{\text{net}}$. The hot spot, where $\Lambda > 0$, shrinks to $|x| \approx 1/3$ by the final time, while $|\Lambda|$ at $x = 0$ diverges as blow-up is approached (Figure 8).

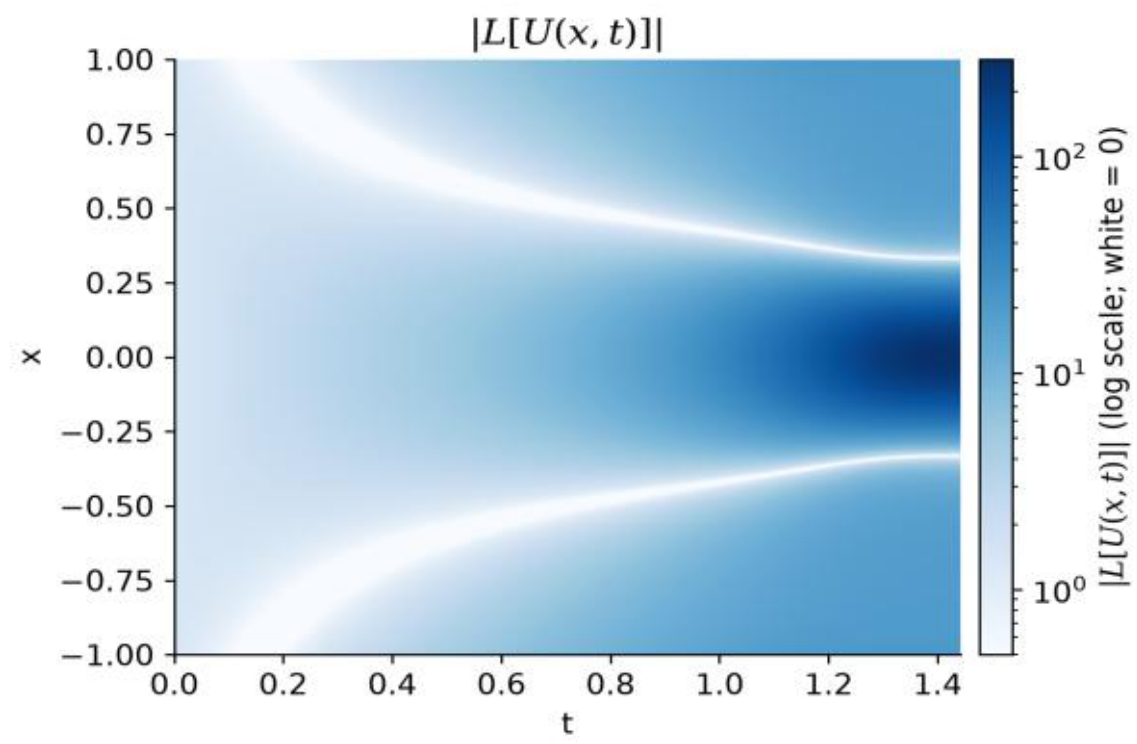

*Figure 8: Spatiotemporal map of $|L[U(x,t)]|$ for the Frank-Kamenetskii equation, for $x \in [-1,1]$ and $t \in [0, t_{end}]$ with margin $3 \cdot 10^{-3}$ (logarithmic color scale; white where $|L[U(x,t)]| = 0$).*

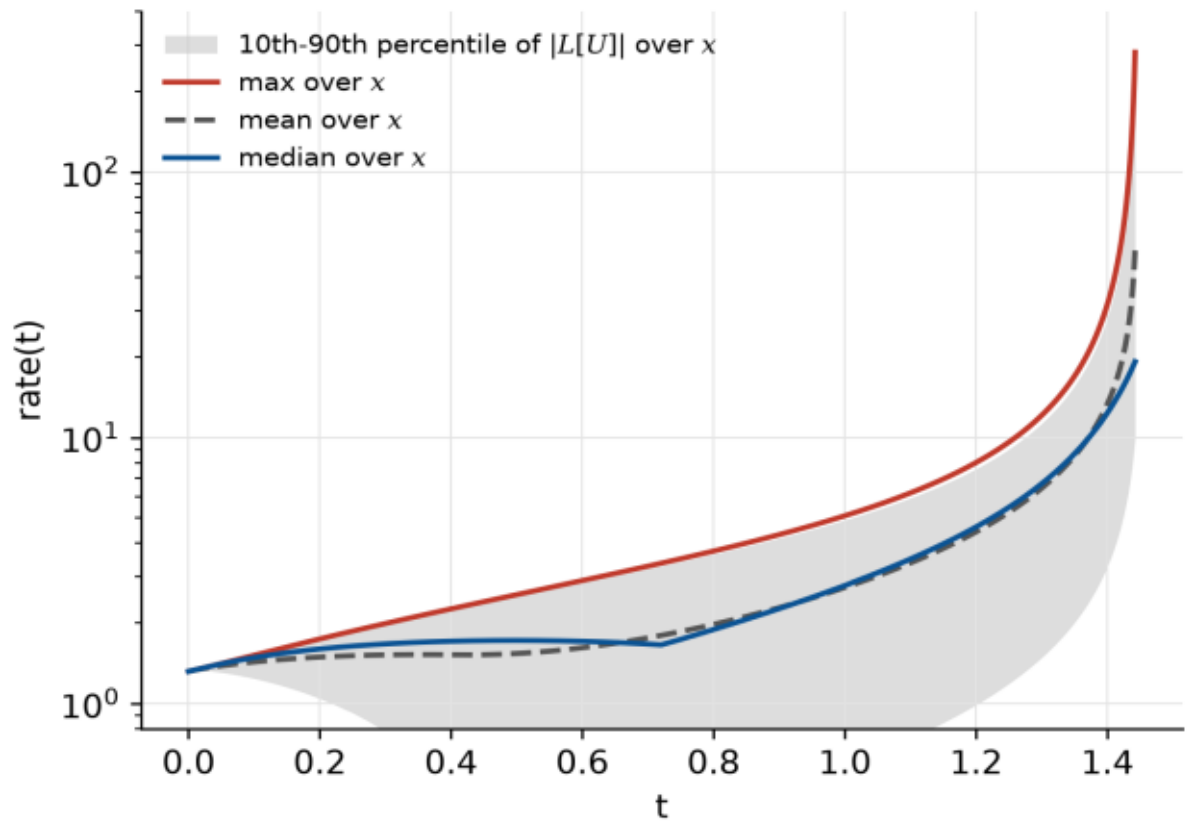

*Figure 9: Reductions of |L[U(x,t)]| over x (maximum, mean and median) as functions of t for the Frank-Kamenetskii equation. The shaded band indicates the 10th–90th percentile of |L[U(x,t)]| over x (logarithmic vertical axis).*

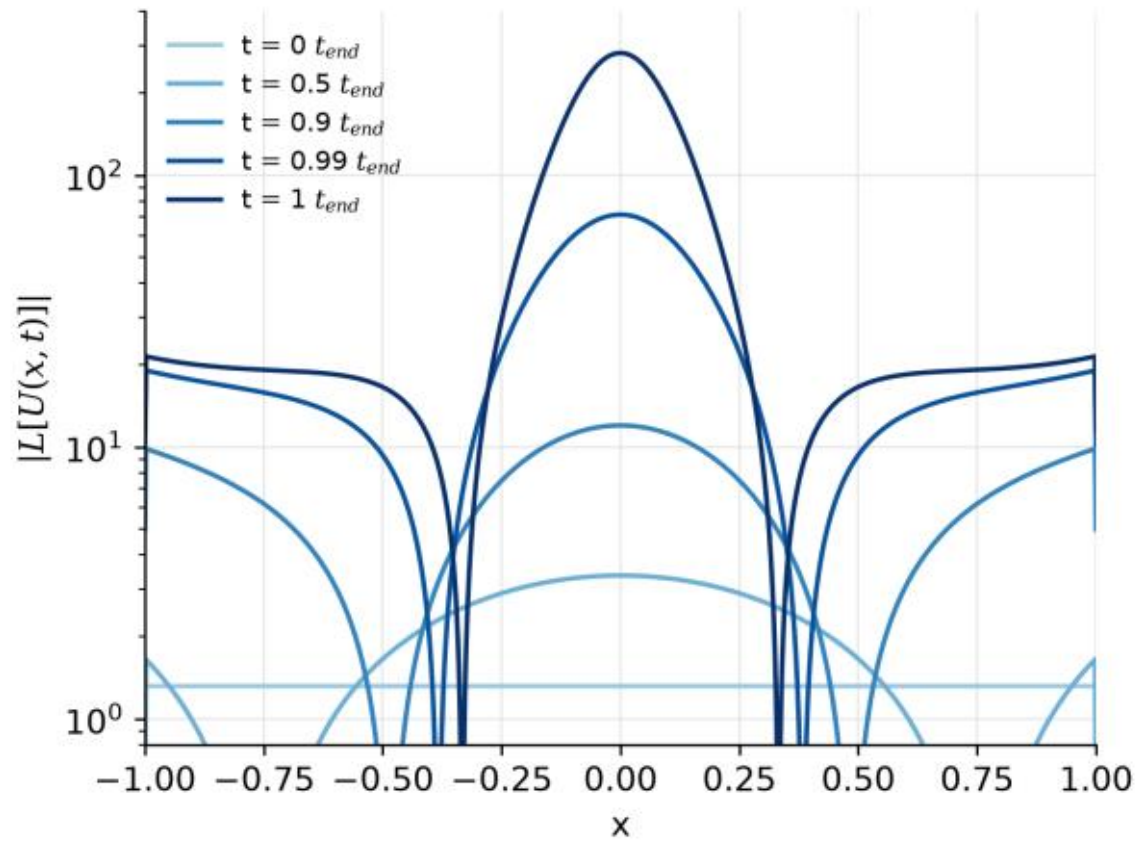

*Figure 10: Profiles of |L[U(x,t)]| over x at $t/t_{end}$ = 0, 0.5, 0.9, 0.99 and 1 for the Frank-Kamenetskii equation (logarithmic vertical axis).*

Choice of Q: For a perturbation $w$ with $w = 0$ on the boundary, $\frac{1}{2}\frac{d}{dt} \int w^2\, dx \approx \int \Lambda\, w^2\, dx$, so the growth rate of the $L^2$ norm is a weighted average of $\Lambda$ with unknown weights, and its magnitude is bounded by the same average of $|\Lambda|$. The mean assumes nothing about them. $Q_{\max}$ is the $L^\infty$ counterpart, and $Q_{med}$ ignores the size of the response in the hot spot, which covers less than half of the slab near blow-up (Figure 9). The profiles at individual times show the response diverging at the center of the slab (Figure 10). We take $Q = Q_{mean}$, so:

$$rate(t) = \frac{1}{2} \int_{-1}^{1} \left| \delta e^{U(x,t)} - \left|\frac{\partial U}{\partial x}\right|^2 \right| dx \tag{63}$$

Unlike the maximum, whose warp integral keeps growing, the mean stays bounded up to blow-up (Figure 11). Strict positivity holds by construction, since $\Lambda(0,t) > 0$ and $\Lambda$ is continuous in $x$. We therefore take $\rho(t) = rate(t)$.

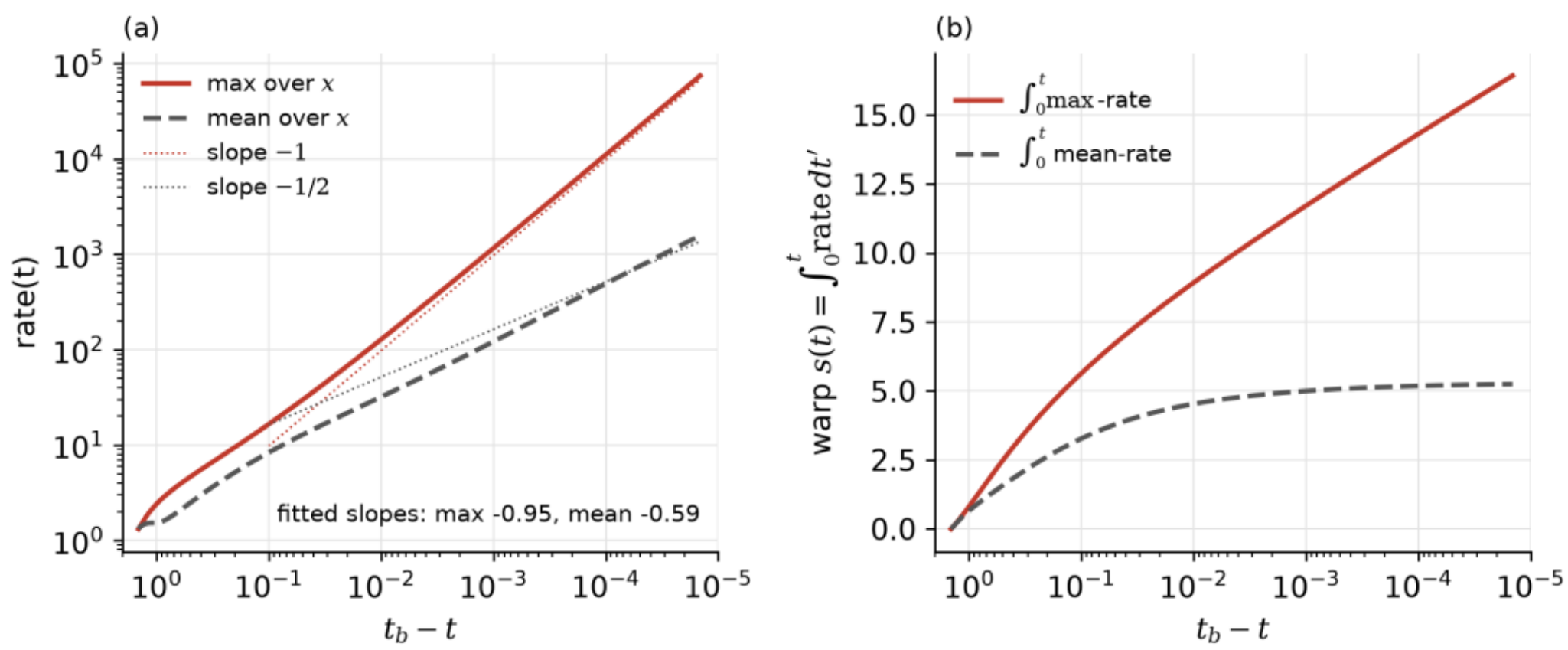

*Figure 11: Warp scaling near blow-up for the Frank-Kamenetskii equation, margin $10^{-5}$. (a) Maximum and mean of |L[U(x,t)]| over x against $t_b - t$, with the fitted slopes −0.95 and −0.59. (b) The warp $s(t) = \int_0^t rate(t')\, dt'$ for both reductions.*

# 6. Results & Discussion

The goal of this section is to demonstrate the effectiveness of the proposed method through numerical results on the four case studies introduced in Section 5. We place particular emphasis on systems that blow up in finite time, namely Kachanov-Rabotnov (Section 5.1) and Frank-Kamenetskii (Section 5.4), where our method predicts the behavior close to the blow-up time and captures their stiff dynamics.

In our implementation we apply the general pipeline of Section 3.1 with the hyperbolic tangent (tanh) activation function, unless otherwise stated. Our setups use some or all of the practical considerations of Section 4. All networks are trained with the Adam and L-BFGS optimizers, alone or in sequence. Adam [88] is a first-order gradient method that adapts the step size of each parameter from running averages of its gradient. It is robust far from a minimum, so it brings the network into a good region of the loss landscape. L-BFGS [89] is a quasi-Newton method that builds an approximation of the curvature from its most recent gradients and chooses its step with a line search. It converges much faster close to a minimum, so it refines the solution to the accuracy that stiff problems need. We compare against six existing methods: sequence-to-sequence learning and curriculum regularization [16], bc-PINN [18], causal training [14], TSONN [34] and a Plain PINN following the formulation of Raissi et al. [6]. Since these methods do not use a pilot training for the time reparameterization, we give them extra training epochs to match its cost, measured in floating-point operations, so that the comparison is fair.

Accuracy is measured by the relative $L^2$ error between the network prediction and the exact or reference solution, reported as the median over ten seeds unless stated otherwise.

All hyperparameter selections, computational costs, implementation details, validation metrics and analytical results and graphs can be found in the Appendix. The code can be found at

## 6.1 Kachanov-Rabotnov

The Kachanov-Rabotnov damage model of Section 5.1 is the simplest of our problems: a scalar ODE whose solution is known in closed form. Its solution blows up at the rupture time $t_f$, and the truncated horizon $t_{\text{end}} = (1 - \psi)t_f$ controls how close to rupture the network is trained. We study $\psi = 10^{-3}$ and $10^{-4}$; smaller $\psi$ moves $t_{\text{end}}$ closer to rupture and therefore produces a stiffer final stretch.

We begin with the least stiff case, $\psi = 10^{-3}$. The rate warp reaches a relative $L^2$ error of 0.04%, 0.21% and 0.15% for $N_t = 256$, 36 and 16, respectively (Table 1). This is 8 to 22 times lower than the error of the best competing method, which is sequence-to-sequence learning at $N_t = 256$ (0.75%) and TSONN at $N_t = 36$ and $N_t = 16$ (1.69% and 3.28%). All methods, including the Plain PINN, train successfully, so the differences reflect accuracy and not training failure.

The alternative reparameterization, arc-length, applied to PINNs reaches errors of 0.71%, 1.05% and 0.31%. It outperforms every competing method except sequence-to-sequence learning at $N_t = 256$, where the two are level, which indicates that time reparameterization is effective on PINNs in general. It does not, however, match the rate warp, being 18, 5.1 and 2.0 times less accurate at $N_t = 256$, 36 and 16. For this problem, this supports an equation-informed reparameterization over a purely kinematic one.

*Table 1: Accuracy of the rate warp (ours) against six existing methods on the Kachanov-Rabotnov damage model with truncation margin $\psi = 10^{-3}$, for (a) $N_t = 256$, (b) $N_t = 36$ and (c) $N_t = 16$ collocation points. Each entry is the median relative $L^2$ error over ten seeds (lower is better); in brackets, the factor by which the error of the method exceeds that of the rate warp.*

| (a) $N_t = 256$ | |
|---|---|
| Method | Median error |
| **Rate (ours)** | **0.0395%** |
| Seq2seq | 0.753% (19×) |
| TSONN | 2.77% (70×) |
| bc-PINN | 3.23% (82×) |
| Causal | 3.69% (94×) |
| Curriculum | 6.65% (168×) |
| Plain PINN | 7.00% (177×) |

| (b) $N_t = 36$ | |
|---|---|
| Method | Median error |
| **Rate (ours)** | **0.206%** |
| TSONN | 1.69% (8.2×) |
| bc-PINN | 2.29% (11×) |
| Seq2seq | 3.15% (15×) |
| Causal | 3.29% (16×) |
| Curriculum | 3.67% (18×) |
| Plain PINN | 5.49% (27×) |

| (c) $N_t = 16$ | |
|---|---|
| Method | Median error |
| **Rate (ours)** | **0.151%** |
| TSONN | 3.28% (22×) |
| Curriculum | 3.52% (23×) |
| bc-PINN | 3.80% (25×) |
| Causal | 4.04% (27×) |
| Plain PINN | 7.14% (47×) |
| Seq2seq | 22.4% (148×) |

Figure 12 shows how the three sampling strategies distribute the collocation points for $N_t = 36$. The competing methods sample uniformly in time (within each segment, for the time-marching ones), so only 4 of the 36 points fall in the last 10% of the horizon, where the damage rate diverges and the equation is stiffest. The arc-length warp raises this to 11 points and the rate warp to 25, of which 13 lie in the final 1% alone. Conversely, the rate warp places only 4 points in the first half of the horizon, where the solution varies slowly, against 13 for arc-length and 18 for uniform sampling. The median error at $N_t = 36$ decreases with this concentration: 3.7% for the same network with uniform sampling, 1.05% for arc-length and 0.21% for the rate warp (the uniform arm even receives the extra epochs described above). This ordering is consistent with the density near rupture driving the accuracy, although Figure 12 shows a single configuration and does not by itself establish the cause.

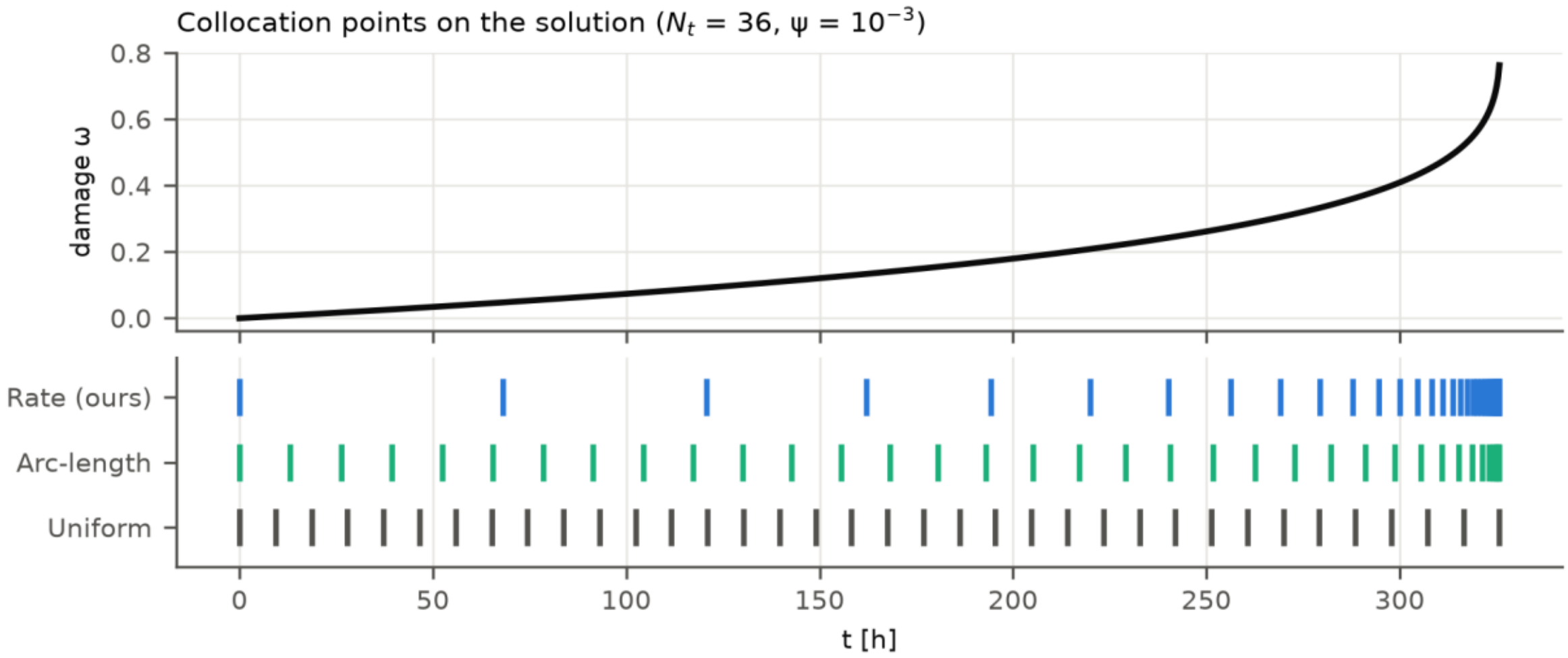

*Figure 12: Collocation points of the three sampling strategies on the Kachanov-Rabotnov damage solution for $N_t = 36$ and $\psi = 10^{-3}$. Uniform sampling spaces the points evenly in time, whereas the arc-length and rate warps concentrate them towards rupture.*

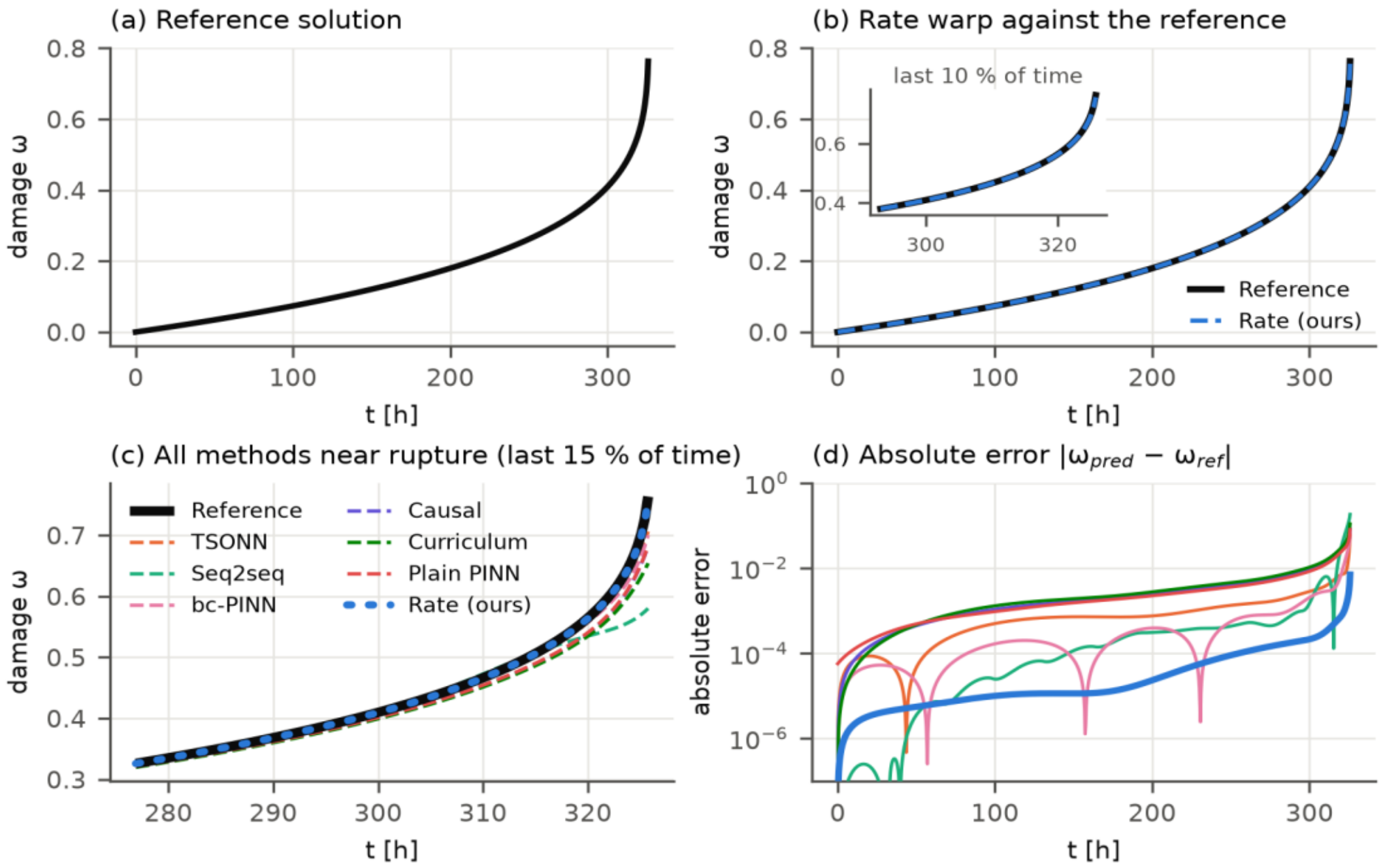

*Figure 13: Prediction of the damage ω for $N_t = 36$ and $\psi = 10^{-3}$ (one run, seed 90). (a) Reference solution. (b) Rate warp against the reference, with a zoom on the last 10% of time. (c) All methods over the last 15% of time; the rate warp (dotted) follows the reference, while the competing methods fall below it towards rupture. (d) Absolute error over the whole time domain.*

Figure 13 shows the predicted damage for a single run at $N_t = 36$, scored against the closed-form solution. All methods reproduce the reference over most of the horizon, and the deviation concentrates near rupture: every method has its largest error at the truncation point, and every method falls below the reference there. The error of the rate warp at that point is $7.1\times10^{-3}$, against $5.7\times10^{-2}$ for TSONN, $6.3\times10^{-2}$ for bc-PINN, $8.1\times10^{-2}$ for Causal and the Plain PINN, $1.1\times10^{-1}$ for Curriculum and $1.8\times10^{-1}$ for Seq2seq, which is 8 to 26 times larger (Figure 13d). Averaged over the last 10% of the horizon, the rate warp's error is $6.0\times10^{-4}$, against $5.6\times10^{-3}$ to $2.1\times10^{-2}$ for the competitors, and its median error over the whole horizon ($1.2\times10^{-5}$) is at least an order of magnitude below every competitor. This agrees with the denser sampling near rupture seen in Figure 12. As Figure 13 is a single run, the ten-seed medians of Table 1 remain the basis of the comparison.

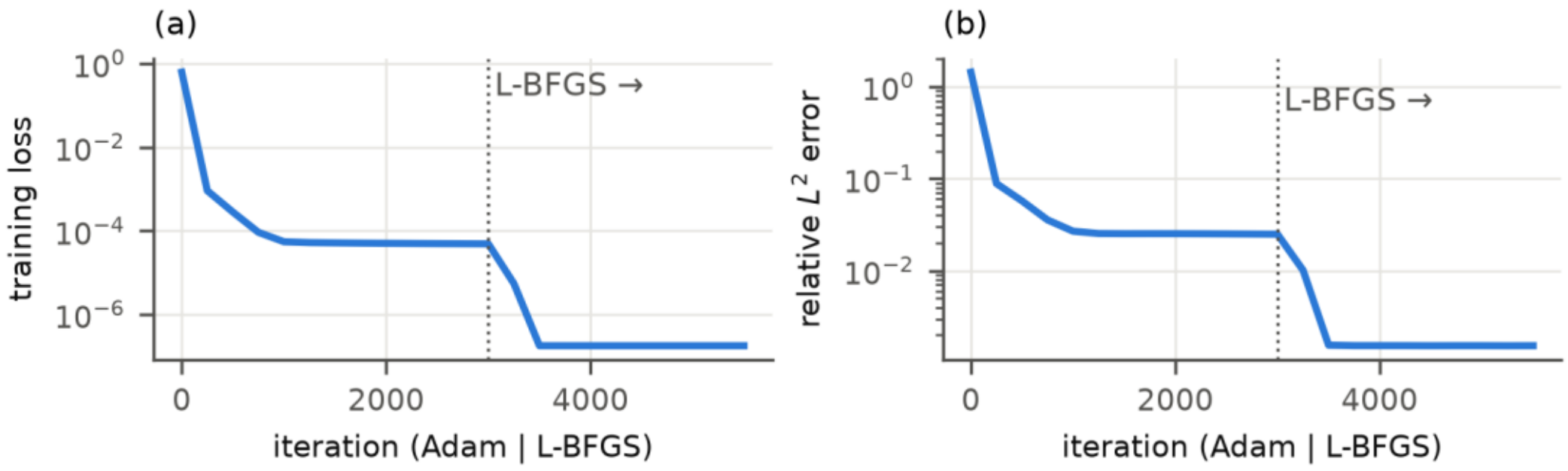

*Figure 14: Training of the rate warp for $N_t = 36$ and $\psi = 10^{-3}$ (one run): (a) training loss and (b) relative $L^2$ error against the iteration. The dotted line marks the switch from Adam to L-BFGS.*

We continue with the case $\psi = 10^{-4}$. The rate warp reaches a relative $L^2$ error of 0.053%, 0.088% and 0.39% for $N_t = 256$, 36 and 16, respectively (Table 2). This is 4.4 to 29 times lower than the error of the best competing method, which is sequence-to-sequence learning at $N_t = 256$ (1.52%) and TSONN at $N_t = 36$ and $N_t = 16$ (2.57% and 1.68%).

*Table 2: Accuracy of the rate warp (ours) against six existing methods on the Kachanov-Rabotnov damage model with truncation margin $\psi = 10^{-4}$, for (a) $N_t = 256$, (b) $N_t = 36$ and (c) $N_t = 16$ collocation points. Each entry is the median relative*

*$L^2$ error over ten seeds (lower is better); in brackets, the factor by which the method's error exceeds that of the rate warp. The rate oracle builds the warp from the exact rate instead of the pilot training.*

| (a) $N_t = 256$ | |
|---|---|
| **Method** | **Median error** |
| **Rate (ours)** | **0.0525%** |
| Rate oracle | 0.0183% |
| Seq2seq | 1.52% (29×) |
| bc-PINN | 2.72% (52×) |
| TSONN | 3.58% (68×) |
| Curriculum | 5.21% (99×) |
| Causal | 5.55% (106×) |
| Plain PINN | 6.90% (132×) |

| (b) $N_t = 36$ | |
|---|---|
| **Method** | **Median error** |
| **Rate (ours)** | **0.0879%** |
| Rate oracle | 0.0952% |
| TSONN | 2.57% (29×) |
| bc-PINN | 2.74% (31×) |
| Plain PINN | 5.09% (58×) |
| Causal | 5.20% (59×) |
| Curriculum | 5.36% (61×) |
| Seq2seq | 8.04% (91×) |

| (c) $N_t = 16$ | |
|---|---|
| **Method** | **Median error** |
| **Rate (ours)** | **0.386%** |
| Rate oracle | 0.377% |
| TSONN | 1.68% (4.4×) |
| Causal | 4.57% (12×) |
| Curriculum | 4.80% (12×) |
| Plain PINN | 5.55% (14×) |
| bc-PINN | 8.57% (22×) |
| Seq2seq | 23.7% (61×) |

The accuracy of the rate warp at this margin depends largely on the accuracy of the pilot training. The rate oracle, which builds the warp from the exact rate obtained from the reference solution instead of from the pilot training, reaches errors of 0.018%, 0.095% and 0.38%. The rate warp comes close to it (Figure 15) once the pilot training is refined by bootstrapping, applied twice here: each repetition runs a new full training on the warp built from the previous one, and its cost is credited to the competing methods as above. With it the rate warp matches the oracle at $N_t = 36$ and $N_t = 16$ and is within a factor of three of it at $N_t = 256$, while a single pilot training applied alone leaves the error 4 to 20 times higher.

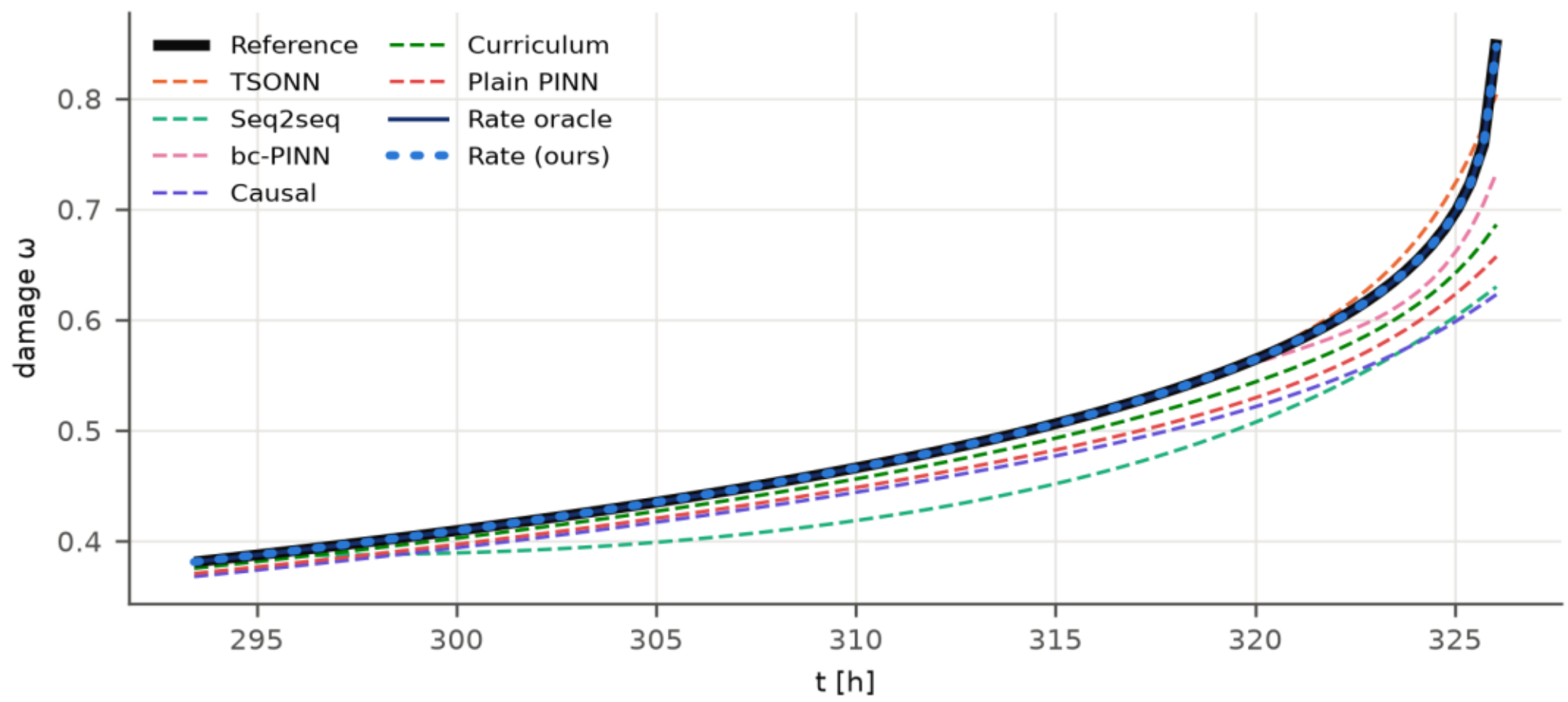

*Figure 15: Predicted damage ω over the last 10% of the time horizon for $N_t = 36$ and $\psi = 10^{-4}$ (one run): reference solution, rate warp (dotted), rate oracle and the competing methods.*

## 6.2 Burgers' Equation

Burgers' equation of Section 5.2 forms a shock at t ≈ 0.32, so the stiff stretch lies in the middle of the time horizon and not at its end.

*Table 3: Accuracy of the rate warp (ours) against six existing methods on Burgers' equation with 128 spatial points, for (a) $N_t = 4$ and (b) $N_t = 6$ time columns. Each entry is the median relative $L^2$ error over ten seeds (lower is better); in brackets, the factor by which the error of the method exceeds that of the rate warp.*

| (a) $N_t = 4$ | | (b) $N_t = 6$ | |
|---|---|---|---|
| **Method** | **Median error** | **Method** | **Median error** |
| **Rate (ours)** | **2.86%** | **Rate (ours)** | **3.02%** |
| TSONN | 3.45% (1.2×) | TSONN | 0.681% (0.23×) |
| Curriculum | 17.4% (6.1×) | Causal | 5.13% (1.7×) |
| Seq2seq | 36.1% (13×) | Curriculum | 5.57% (1.8×) |
| Causal | 65.6% (23×) | Seq2seq | 7.38% (2.4×) |
| Plain PINN | 87.8% (31×) | Plain PINN | 88.4% (29×) |
| bc-PINN | 92.0% (32×) | bc-PINN | 89.9% (30×) |

Table 3 shows the results. The rate warp reaches a relative $L^2$ error of 2.9% for $N_t = 4$ and 3.0% for $N_t = 6$. At $N_t = 4$ it has the lowest median of all methods, 1.2 times below TSONN (3.5%, not a significant difference), together with the lowest mean and worst seed, and it is 6 to 32 times more accurate than the other methods. At $N_t = 6$ TSONN is more accurate, 0.68% against 3.0%, although the rate warp has the lower mean (3.5% against 4.8%) and worst seed (6.6% against 22%), because TSONN fails on some of the seeds. The rate warp is 1.7 to 2.4 times more accurate than causal, curriculum and sequence-to-sequence learning at $N_t = 6$, and about 30 times more accurate than bc-PINN and the Plain PINN. Figure 16 shows the prediction of the rate warp for $N_t = 4$, and Figure 17 compares every method at t = 0.4, close to the formation of the shock; the training of the same run is shown in Figure 18a.

TSONN wins at $N_t = 6$ because of its resampling: it draws a fresh random set of collocation points, as many as the grid has, at each of its 46 and 39 pseudo-time steps (for 4 and 6 time columns), so it trains on about 40 times more distinct points than the fixed grid that every other method uses, which is what limits accuracy when only 4 to 6 time columns are available.

With 512 spatial points and 32 time columns the rate warp converges to 0.053% (Figure 18b). At this budget, however, the rate warp, TSONN, causal learning and sequence-to-sequence learning all lie between 0.053% and 0.072%, within the variation between seeds, and the remaining methods are about 3 to 6 times above them. We deem that a comparison of the methods at such a budget has no value.

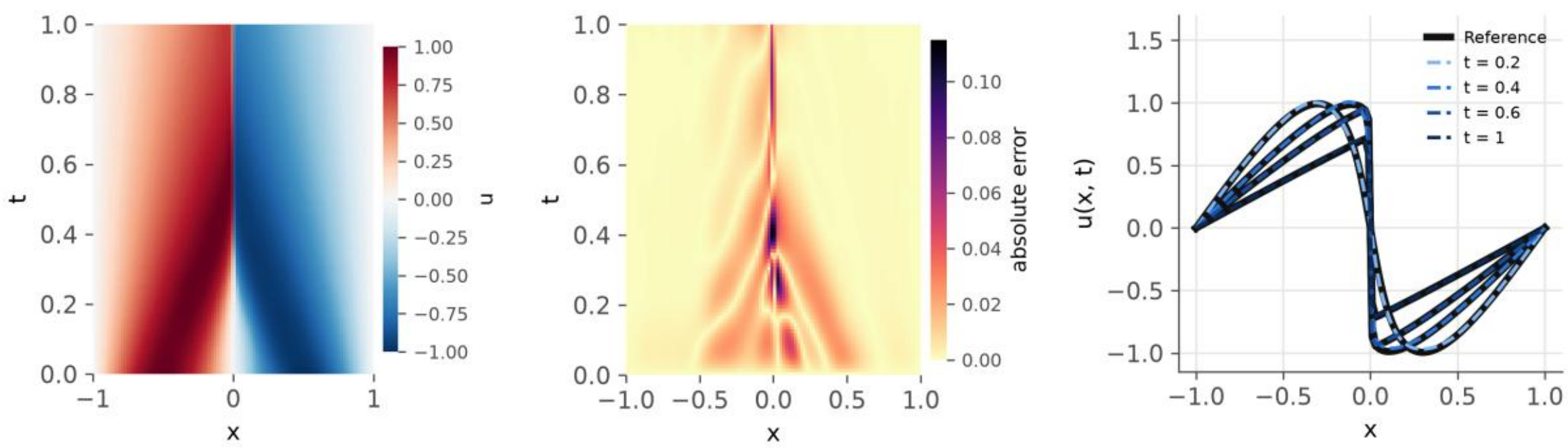

*(a)* *(b)* *(c)*

*Figure 16: Rate warp for Burgers' equation with $N_t = 4$. (a) Prediction. (b) Absolute error of the prediction. (c) Rate warp against the reference at four times.*

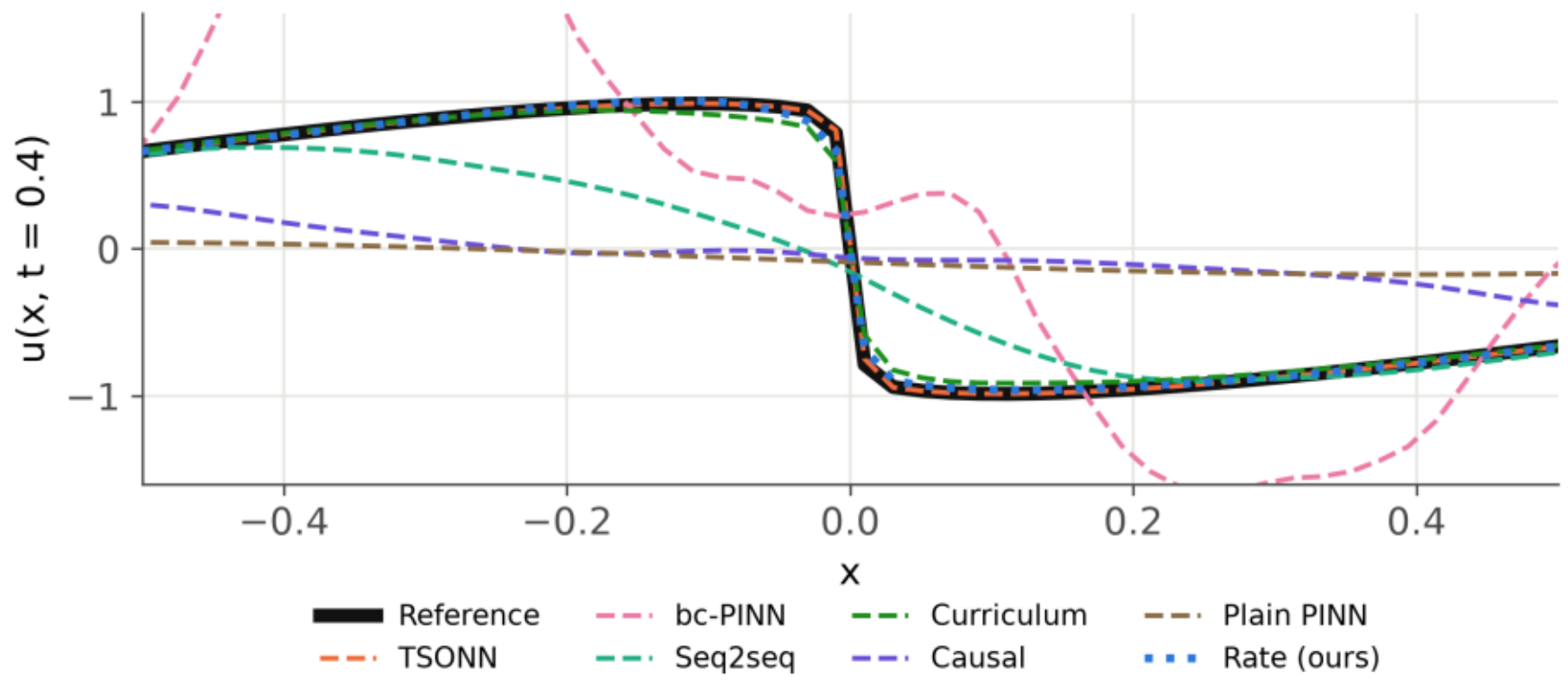

*Figure 17: Solution at t = 0.4, close to the formation of the shock, for $N_t = 4$.*

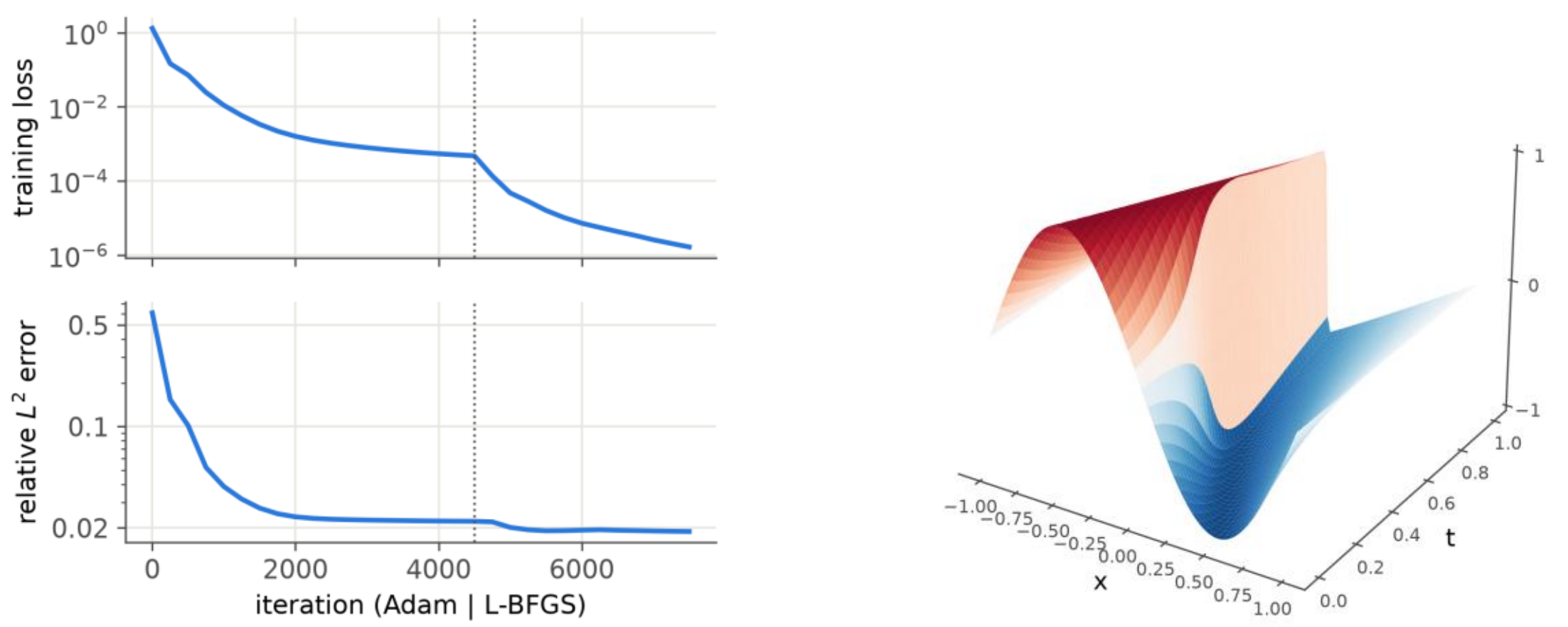

(a)

(b)

*Figure 18: (a) Training of the rate warp for $N_t = 4$ (same run as in the previous figure): loss and error against the iteration. (b) Prediction of the rate warp for 512 spatial points and 32 time columns.*

## 6.3 Allen-Cahn

The Allen-Cahn equation of Section 5.3 is challenging for a different reason than the blow-up problems. The function u = 0 is an exact but unstable equilibrium of the equation, and on the discretized problem it fits the loss better than the true solution: the runs that collapse to it end at a lower training loss than the runs that find the correct solution, so the loss cannot be used to tell them apart. The trap is decided in a thin layer between the initial condition and the first collocation times. We study $N_t = 24$, $N_t = 20$ and $N_t = 16$ time columns, with 10,000 L-BFGS iterations for every method. Here the full training and the competing methods use L-BFGS only, without an Adam phase; only the pilot training starts with Adam (Appendix D.3). Periodicity in x is imposed exactly by a periodic (Fourier) embedding of the input, which every method uses except the Plain PINN, which takes the raw coordinate and a periodic penalty.

Bootstrapping is what makes the rate warp work on this equation. The rate is estimated from a first training, the pilot training, which is a plain PINN on the uniform grid, exactly the setting in which Allen-Cahn PINNs fail: all ten pilot trainings end in the trivial solution (51% to 54% error) and give an almost flat rate (Figure 19b). We therefore bootstrap: a new training is carried out on the warp built from the pilot rate, the rate is recomputed from its solution, and this is repeated once more. The first round leaves the trivial solution (9.1% error, Figure 19a), the second reaches 2.9%, and the resulting final rate is within 12% of the exact rate, which is never used in training.

The residual is weighted by the normalized rate (Section 4.2), which gives the largest weight to the earliest time columns, where the sign of the solution is decided and the trap is set (Figure 19c, top). The weighting matters on top of the warp: the residual of the reparameterized equation as it stands, which divides by the rate and so gives the early columns the smallest weight, converges to the trivial solution in the interior of the domain (error 75%), whereas the unweighted physical residual reaches 0.64% and the rate-weighted residual 0.23% on the same warp (Figure 19c, bottom; one run per form).

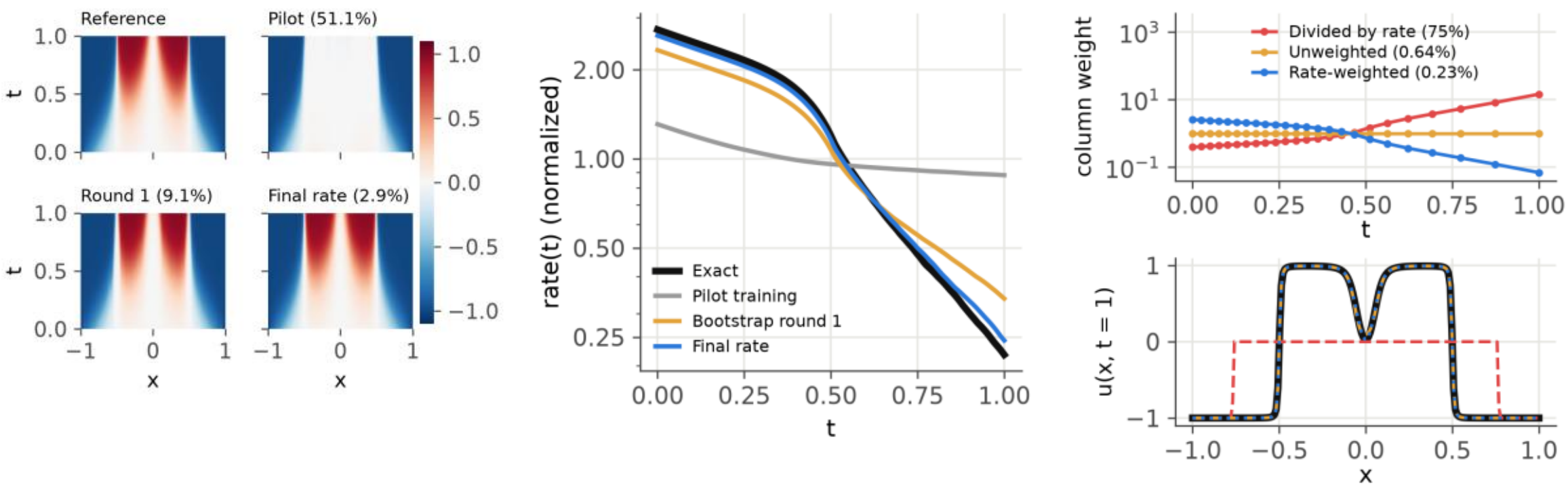

*(a)* *(b)* *(c)*

*Figure 19: Bootstrapping and residual weighting for Allen-Cahn. (a) Solutions of the pilot training, of the first bootstrap round and of the training that gives the final rate, against the reference (relative $L^2$ error in brackets). (b) Rate at each stage of the bootstrap, normalized by its geometric mean, against the exact rate computed from the reference solution. (c) Residual weighting for $N_t = 24$ (one run per form): column weights (top) and solution at $t = 1$ (bottom), with the error in brackets.*

*Table 4: Accuracy of the rate warp (ours) against six existing methods on the Allen-Cahn equation, for (a) $N_t = 24$, (b) $N_t = 20$ and (c) $N_t = 16$ time columns. Each entry is the median relative $L^2$ error over ten seeds (lower is better); in brackets, the factor by which the error of the method exceeds that of the rate warp.*

| (a) $N_t = 24$ | |
|---|---|
| Method | Median error |
| **Rate (ours)** | **0.245%** |
| TSONN | 0.310% (1.3×) |
| bc-PINN | 0.355% (1.4×) |
| Seq2seq | 1.26% (5.2×) |
| Curriculum | 8.96% (37×) |
| Causal | 54.0% (220×) |
| Plain PINN | 61.5% (251×) |

| (b) $N_t = 20$ | |
|---|---|
| Method | Median error |
| **Rate (ours)** | **0.290%** |
| TSONN | 0.360% (1.2×) |
| bc-PINN | 0.365% (1.3×) |
| Seq2seq | 1.66% (5.7×) |
| Curriculum | 9.24% (32×) |
| Causal | 54.4% (188×) |
| Plain PINN | 61.3% (211×) |

| (c) $N_t = 16$ | |
|---|---|
| Method | Median error |
| **Rate (ours)** | **0.300%** |
| bc-PINN | 0.340% (1.1×) |
| TSONN | 0.415% (1.4×) |
| Seq2seq | 1.60% (5.3×) |
| Curriculum | 9.03% (30×) |
| Causal | 55.2% (184×) |
| Plain PINN | 61.1% (204×) |

Table 4 shows the results. The rate warp reaches a relative $L^2$ error of 0.25%, 0.29% and 0.30% for $N_t = 24$, $N_t = 20$ and $N_t = 16$, respectively. This is 1.1 to 1.3 times lower than the error of the best competing method, which is TSONN at $N_t = 24$ and $N_t = 20$ (0.31% and 0.36%) and bc-PINN at $N_t = 16$ (0.34%). The other methods are further away. Sequence-to-sequence learning is 5 to 6 times less accurate and curriculum learning 30 to 37 times, and the Causal and Plain PINNs never leave the trivial solution (errors above 50%, where a run counts as failed when its error exceeds 15%). The training of the rate warp and its prediction over the whole domain are shown in Figure 20b and c, its solution at the final time against every method in Figure 21, and its solution at five times in Figure 22.

The arc-length reparameterization does not help here. It places almost the same columns as uniform sampling (Figure 20a), and the network falls into the trivial solution on 8, 9 and 10 of the ten seeds at $N_t = 24$, $N_t = 20$ and $N_t = 16$. Without any warp the same network falls into it on 2, 7 and 6 seeds.

The rate warp moves five of the 24 columns into the first half of the horizon (17, against 12 for uniform sampling and for arc-length) and never falls into it.

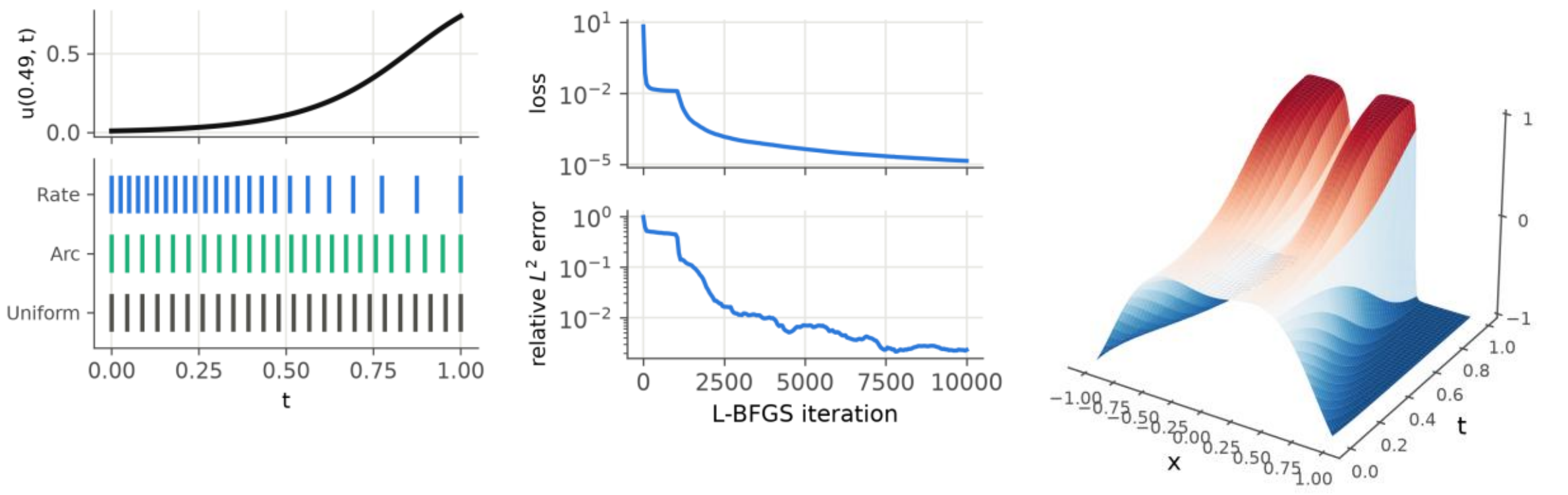

*(a)* *(b)* *(c)*

*Figure 20: Rate warp for Allen-Cahn. (a) Time columns of the three strategies for $N_t$ = 24 (top: reference at x = 0.49). (b) Training of the rate warp. (c) Prediction of the rate warp for $N_t$ = 16 over the whole domain.*

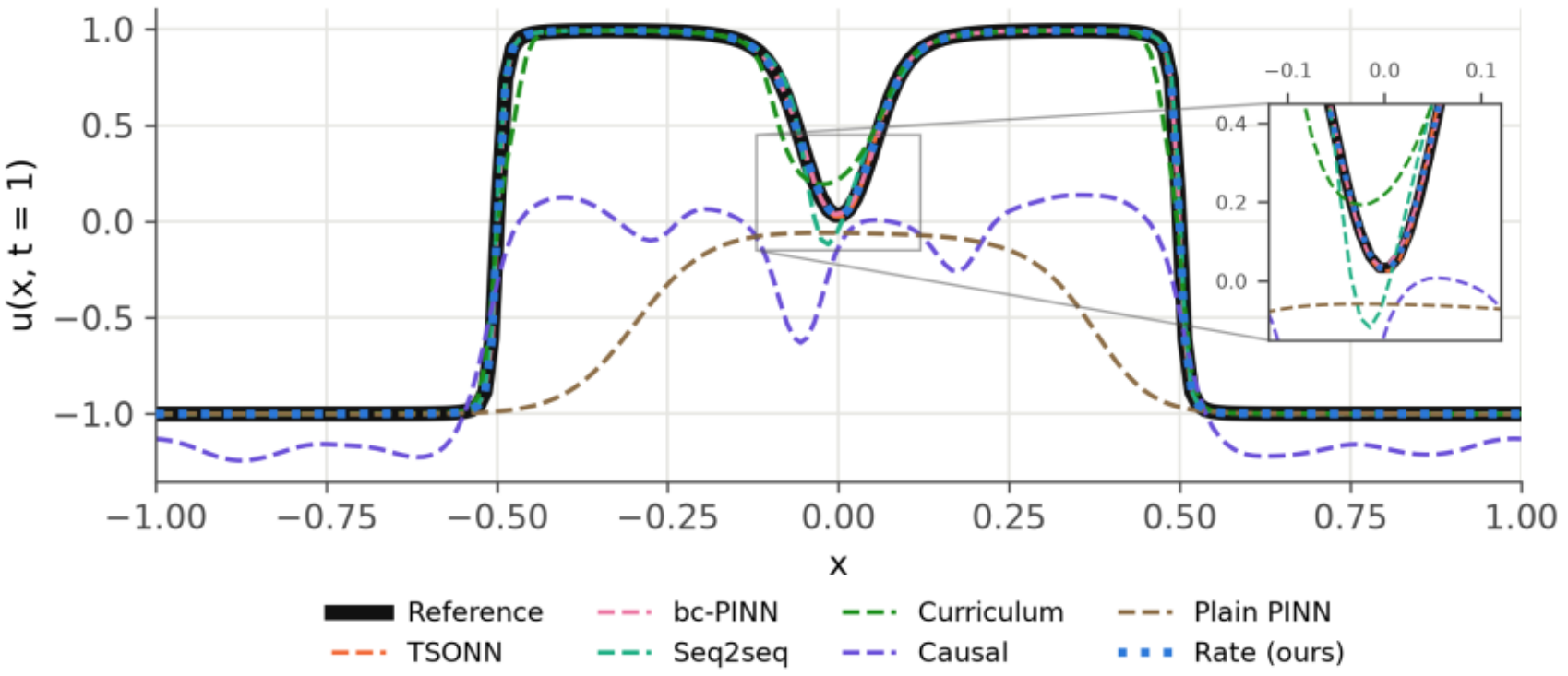

*Figure 21: Solution at the final time t = 1 for $N_t$ = 24; the inset zooms on the central dip.*

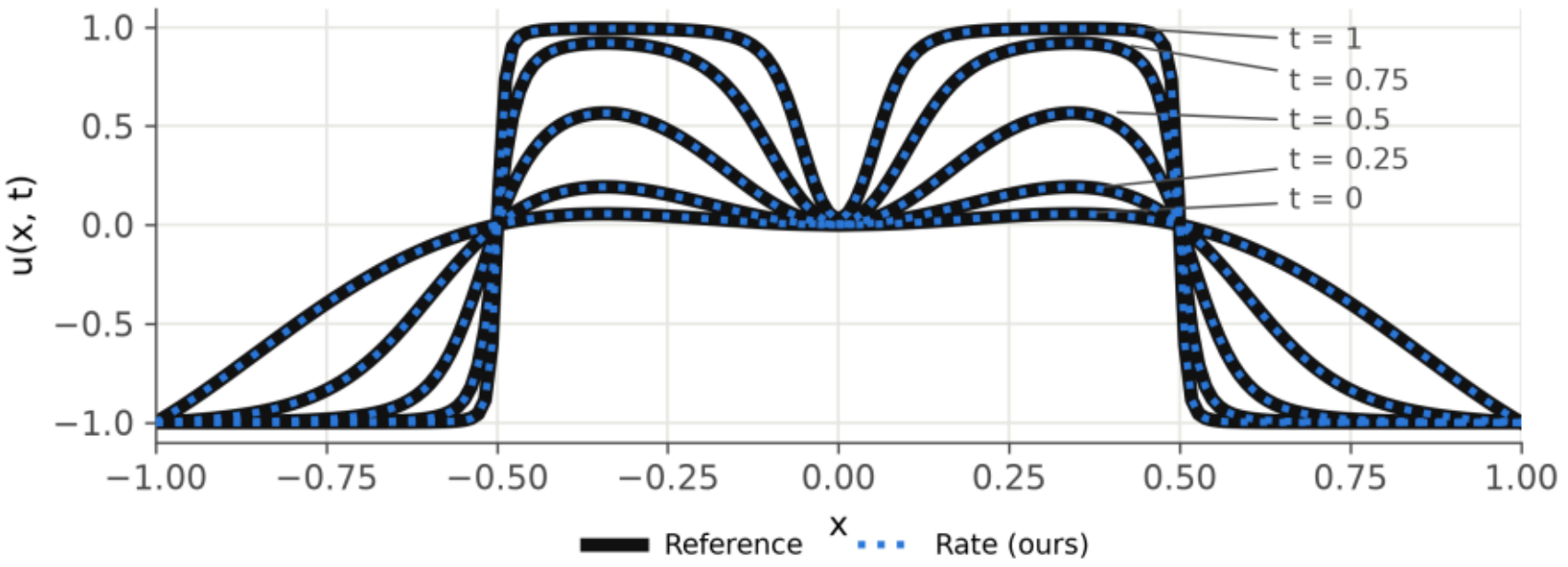

*Figure 22: Rate warp against the reference at t = 0, 0.25, 0.5, 0.75 and 1.*

## 6.4 Frank-Kamenetskii

The Frank-Kamenetskii equation of Section 5.4 is our second blow-up problem. At the center of the slab the temperature rises from zero to 5.4 over the horizon, and it diverges at $t_b = 1.447$, so the stretch before $t_b$ decides the accuracy. We take the truncation margin $\psi = 3 \cdot 10^{-3}$, which ends the horizon at $u(0, t_{end}) = 5.36$, with 256 spatial points and $N_t = 80$ time columns. The full training and the competing methods use L-BFGS only; the pilot training uses Adam (Appendix D.4).

Besides the relative $L^2$ error we report the peak error, the relative error of the maximum of the solution at $t_{end}$, which measures whether the network follows the blow-up. It is the harder of the two metrics: for every method in Table 5 it is four to five times larger than the relative $L^2$ error, because the error concentrates at the hot spot in the last time instants.

*Table 5: Accuracy of the rate warp (ours) against six existing methods on the Frank-Kamenetskii equation with truncation margin $\psi = 3 \cdot 10^{-3}$ and $N_t = 80$ time columns, for (a) the relative $L^2$ error and (b) the peak error at the final time. Each entry is the median over ten seeds (lower is better); in brackets, the factor by which the error of the method exceeds that of the rate warp. The rate oracle builds the warp from the exact rate instead of the pilot training.*

| (a) Relative $L^2$ error | | (b) Peak error | |
|---|---|---|---|
| Method | Median error | Method | Median error |
| **Rate (ours)** | **0.341%** | **Rate (ours)** | **1.81%** |
| Rate oracle | 0.209% | Rate oracle | 1.11% |
| Seq2seq | 1.80% (5.3×) | Seq2seq | 8.52% (4.7×) |
| TSONN | 2.94% (8.6×) | TSONN | 13.5% (7.5×) |
| bc-PINN | 3.19% (9.3×) | bc-PINN | 14.3% (7.9×) |
| Causal | 4.43% (13×) | Causal | 18.8% (10×) |
| Plain PINN | 4.56% (13×) | Plain PINN | 19.3% (11×) |
| Curriculum | 5.42% (16×) | Curriculum | 22.1% (12×) |

Table 5 shows the results. The rate warp reaches a relative $L^2$ error of 0.34% and a peak error of 1.81%, and no seed exceeds 5.8% in peak error. The competing methods are 5.3 to 16 times less accurate in relative $L^2$ error and 4.7 to 12 times in peak error. Sequence-to-sequence learning is the closest, with 1.80% and 8.5%, while TSONN, bc-PINN, causal learning, curriculum learning and the Plain PINN have peak errors between 13% and 22%. Every method underestimates the peak in the median, that is, its solution blows up too late. The differences between the rate warp and each competing method are significant for both metrics (Holm-corrected one-sided Mann-Whitney tests, $p < 0.05$). Figure 23 shows the prediction of the rate warp and its error, and Figure 24 compares every method at the final time, where all of them stay below the peak.

With the exact rate computed from the reference solution, which the method never uses, the median errors fall to 0.21% and 1.11%, so the pilot training costs a factor of 1.6 in the median. The two have the same mean peak error (2.2%), and the worst seed of the oracle (8.0%) is worse than that of the rate warp (5.8%).

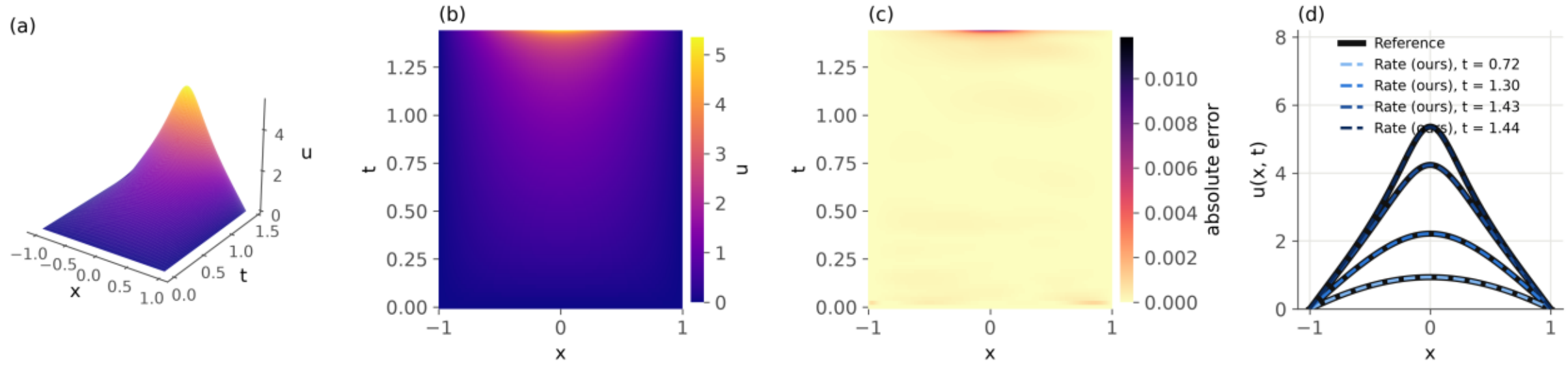

*Figure 23: Rate warp for the Frank-Kamenetskii equation with ψ = 3·10⁻³: (a) predicted solution; (b) the same as a map in the (x, t) plane; (c) absolute error against the reference; (d) rate warp (dashed) against the reference (black) at four times.*

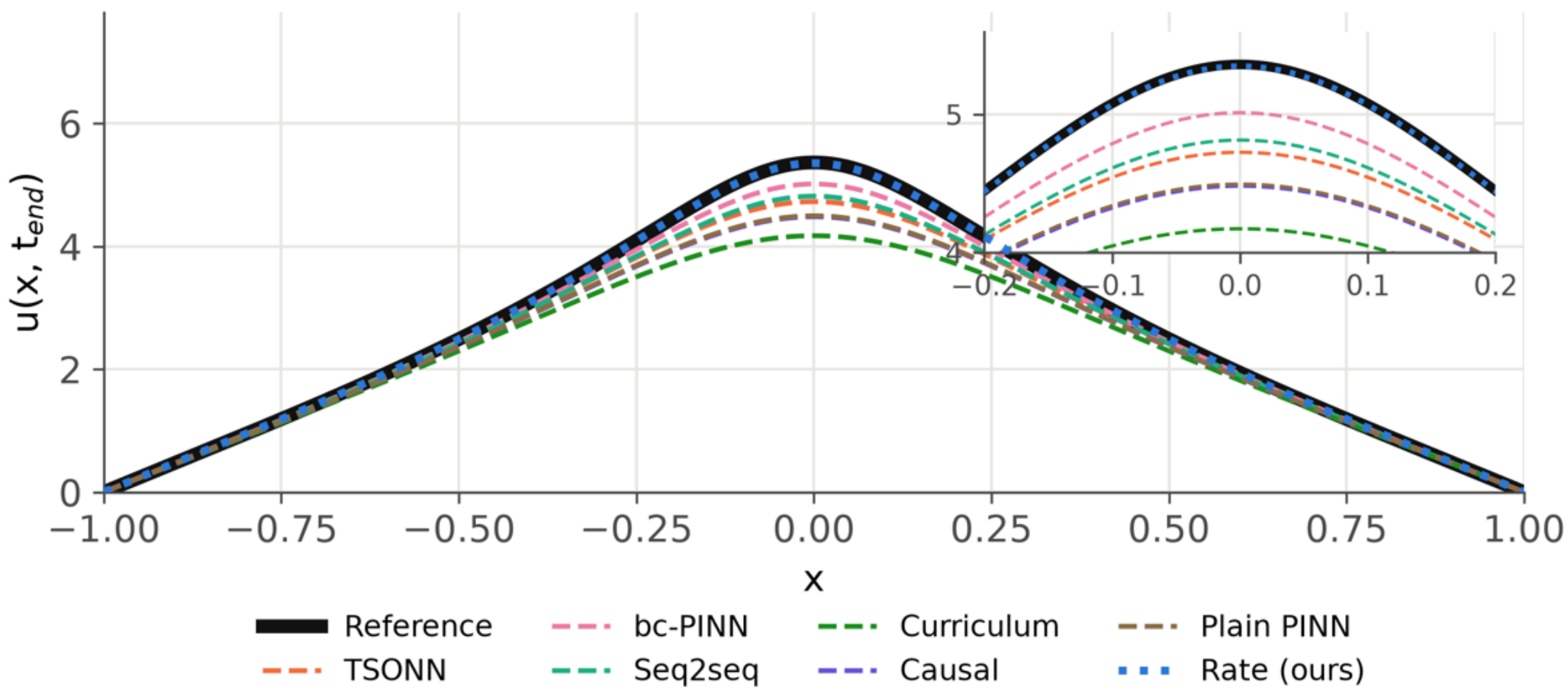

*Figure 24: Solution at the final time $t_{end}$ for ψ = 3·10⁻³. The inset zooms on the peak.*

Closer to blow-up, at $\psi = 10^{-5}$ ($u(0, t_{end}) = 10.94$), no method resolves the peak (Figure 25). The rate warp reaches 7.9% in relative $L^2$ error and 47% in peak error, against 8.5% to 13% and 49% to 61% for the competing methods. The rate warp is more accurate than every method that does not warp time, and with the exact rate the errors are 6.9% and 44%.

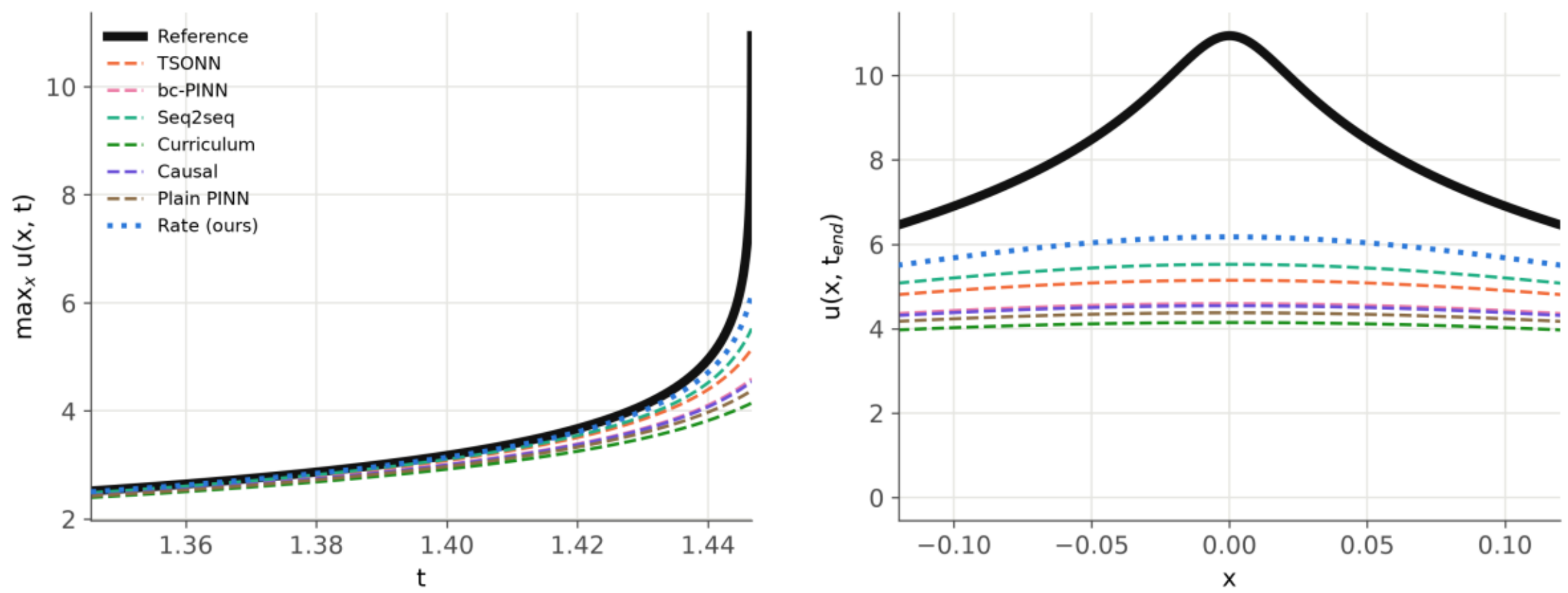

*(a)* *(b)*

*Figure 25: Rate warp for the Frank-Kamenetskii equation with ψ = 10⁻⁵ (one run, the best seed of the rate warp). (a) Largest value of the solution over x as a function of time. (b) Solution at the final time $t_{end}$; the reference peaks at 10.94.*

## 6.5 Discussion

The results show that time reparameterization is an effective way to tackle stiff and fast-varying dynamics in PINNs. On Kachanov-Rabotnov even the purely kinematic arc-length warp improves on most of the alternative methods, but on Allen-Cahn it places almost the same time columns as uniform sampling and falls into the trivial solution on most seeds. Deriving the reparameterization from the governing equation instead of from the solution's motion is what makes the difference, because it anticipates the stiffness before the solution visibly moves and separates amplification from transport and forcing; arc-length, which follows the solution's speed, does not help where the solution stays almost still while an instability builds. The gains are largest when the stiff stretch is localized and few time columns are available, and modest on smoother problems.

These gains trace back to where the collocation points are placed. In the new time coordinate the equation has an approximately uniform local time scale, so the stiffness is equalized and the points concentrate where the equation amplifies a perturbation. The best residual weighting depends on the problem. On Allen-Cahn and Burgers' equation, where the sampling already favors the high-rate times, dividing the residual by the rate works against the warp, while the physical residual preserves it; extra weight on the high-rate times helps on Allen-Cahn, where the trap is set in the earliest columns, but not on Burgers' equation. On the two blow-up problems the residual of the reparameterized equation, which divides by the rate, is used. The warp also tolerates an inexact pilot training as long as the pilot identifies where the dynamics are stiff, and the comparison with the exact rate shows only a modest gap; when the pilot is qualitatively wrong, bootstrapping recovers it. In such cases the training loss is a poor guide, since a collapsed solution can reach a lower loss than the correct one, in line with the remark of Turinici [21] that the loss does not reliably indicate the error of the final solution.

Because the rate warp changes only where the collocation points are, and leaves the order and weighting of the loss untouched, it can complement the alternative methods rather than replace them. It can be paired with causal training and time-marching, which Wang et al. [14] present as enhancements to one another, and with residual-based adaptive sampling [60], to tackle harder problems. The latter redistributes points a posteriori from the residual, whereas the rate warp places them a priori from the structure of the equation, before any residual is evaluated.

Two limitations qualify this picture. The method is hard to implement: being specific to each equation, the rate has to be derived anew for every equation, by linearizing the operator, choosing the combination rule and the spatial reduction, and working out the growth coefficients term by term. Automatic differentiation could in principle supply the linearization, which we have not explored. The method also depends on a closure of the wavenumber for even-order derivative terms, taken from the gradient of the pilot solution; it is an order-of-magnitude choice that sees only the scale of the solution.

Looking ahead, the natural next step is to ask whether the equation-informed reparameterization carries over to the spatial domain, and ultimately whether it can be combined with the time reparameterization to place collocation points where they are needed most in space and time. The construction already points in this direction: the space-time field of the response contains this information, and reducing it over time instead of space would give a spatial warp, in the way moving-mesh methods use a monitor function to place points [43].

**Acknowledgements.** This work was carried out within the INSANE research lab of IIT NCSR Demokritos. Although all ideas, analyses, methods, results and conclusions presented remain solely those of the authors, Claude AI (Anthropic) was used for proofreading specific paragraphs of this manuscript.

# Appendix

## A. Nomenclature

Table 6 summarizes the main symbols used throughout the paper; symbols specific to one equation are defined in Section 5.

*Table 6: Main symbols used throughout the paper.*

| Symbol | Meaning |
|---|---|
| $\Omega, x, t$ | Spatial domain, space, physical time |
| $\mathcal{N}$ | Governing differential operator |
| $\rho(t)$ | Warp density function |
| $s(t)$ | Warp function that maps each physical time point to a warped time point |
| $U(x,t)$ | Current solution state from the pilot training |
| $w$ | Perturbation of the state |
| $\lambda_i(x,t)$ | Growth coefficient of mechanism i |
| $I$ | Number of mechanisms in the decomposition of the operator |
| $\Lambda$ | Net growth coefficient, the sum of the $\lambda_i$ |
| $\Phi, Q$ | Combination rule and spatial reduction |
| $rate(t)$ | Equation-informed rate that sets $\rho$ |
| $\varepsilon$ | Positivity constant added to the rate |
| $\beta$ | Warp-strength exponent |
| $\zeta$ | Residual-weighting exponent |
| $n$ | Number of bootstrap rounds |
| $N_t, N_x$ | Time levels and spatial points |
| $\psi$ | Truncation margin before rupture or blow-up |

## B. Hyper-Parameters

Table 7 lists the network size, the collocation points and the rate-warp settings used in each case study.

*Table 7: Network size, collocation points and rate-warp settings of each case study. Training rounds counts the pilot training, the bootstrap rounds and the full training.*

| Case | Subcase | Depth | Width | $N_t$ | $N_x$ | $\beta$ | $\zeta$ | Training rounds |
|---|---|---|---|---|---|---|---|---|
| Kachanov-Rabotnov | $\psi = 10^{-3}$ | 2 | 16 | 256 / 36 / 16 | – | 1.5 / 1.1 / 1.3 | −2 | 2 |
| | $\psi = 10^{-4}$ | 2 | 16 | 256 / 36 / 16 | – | 1.3 / 1.25 / 1.0 | −2 | 4 |
| Burgers | – | 3 | 20 | 4 | 128 | 0.85 | 1 | 2 |
| | | | | 6 | 128 | | | |
| | | | | 32 | 512 | | | |
| Allen-Cahn | – | 4 | 64 | 24 / 20 / 16 | 256 | 0.76 / 1.01 / 0.76 | 2 | 4 |
| Frank-Kamenetskii | $\psi = 3 \cdot 10^{-3}$ | 4 | 64 | 80 | 256 | 1 | −2 | 2 |
| | $\psi = 10^{-5}$ | 4 | 64 | 80 | 256 | 1 | −2 | 2 |

## C. Computational Cost

Table 8 reports the full-training budget and wall time of the rate method, the cost of its pilot training and bootstrap rounds, and the extra epochs credited to the competing methods to match that cost.

*Table 8: Cost of the rate method and epochs credited to the competitors. Max. iterations refers to the full training; wall time is the total training time of one chain (pilot training, bootstrap rounds and full training) for one seed on an A100, in float64 except for Allen-Cahn, which runs in float32; cost is the number of floating-point operations of the pilot training and bootstrap rounds,*

*charged once. Competitor budget is the number of total competitors' epochs divided by the max iterations of the rate method; slash-separated values follow the $N_t$ order of Table 7. The Burgers subcases are $N_x \times N_t$.*

| Case | Subcase | Max. iterations (Adam + L-BFGS) | Wall time (s) | Pilot / bootstrap cost (FLOPs) | Competitor budget (× rate) |
|---|---|---|---|---|---|
| Kachanov-Rabotnov | $\psi = 10^{-3}$ | 3000 + 2000 | 110 / 112 / 114 | $2.9 \cdot 10^{8}$ | ×1.05 / 1.33 / 1.62 |
| | $\psi = 10^{-4}$ | 3000 + 2000 | 206 / 201 / 208 | $6.1 \cdot 10^{9}$ | ×2.1 / 7.9 / 14 |
| Burgers | 128 × 4 | 4500 + 3000 | 299 | $9.2 \cdot 10^{10}$ | ×1.85 |
| | 128 × 6 | | 299 | | ×1.57 |
| | 512 × 32 | | 304 | | ×1.03 |
| Allen-Cahn | – | 0 + 10 000 | 517 / 508 / 512 | $9.0 \cdot 10^{11}$ | ×1.04 / 1.05 / 1.06 |
| Frank-Kamenetskii | $\psi = 3 \cdot 10^{-3}$ | 0 + 10 000 | 613 | $1.5 \cdot 10^{10}$ | ×1.00 |
| | $\psi = 10^{-5}$ | 0 + 10 000 | 599 | $7.7 \cdot 10^{9}$ | ×1.00 |

# D. Case-Study Details

All competing methods are implemented on the same benchmarks as the rate warp, with the same network size, collocation grid, learning rate and seeds, and at least the same total training budget, extended by extra epochs for the cost of the pilot training and the bootstrap rounds. Where the authors released code for the equation (bc-PINN [18], TSONN [34], and Causal weighting [14] on Allen-Cahn), we follow it and keep their method-specific settings. Sequence-to-sequence (seq2seq) and curriculum training [16] are reproduced from the paper, whose released code contains neither. Where a choice was open, such as the initial-condition treatment of seq2seq, the rollback rule of TSONN or the causality parameter of causal weighting, we kept the variant that performed best on development seeds, not on the seeds of the final comparison, which favors the competitors.

## D.1 Kachanov-Rabotnov

The Kachanov-Rabotnov model is separable, so the reference solution is exact: $(1 - \omega)^{\varphi+1} = 1 - (\varphi + 1)M\sigma^{\chi}t$, with rupture time $t_f = \frac{1}{(\varphi+1)M\sigma^{\chi}}$, which equals 326.1 h for the constants of Section 5.1. The error is the relative $L^2$ error of $\omega$ over 1001 uniformly spaced times in [0, $t_{end}$], with $t_{end} = (1 - \psi)t_f$, and the rate oracle builds its warp from this exact solution. Table 9 lists the setup for each $\psi$.

*Table 9: Setup of the rate warp for each margin. Bootstrap rounds give the collocation points and the Adam + L-BFGS budget of each round; the final β is listed for $N_t$ = 256 / 36 / 16.*

| $\psi$ | Pilot training | Bootstrap rounds | $\beta$ of the rounds | Final $\beta$ |
|---|---|---|---|---|
| $\psi = 10^{-3}$ | width 8, depth 2, 64 points, Adam 1000 + L-BFGS 2000 | none | – | 1.5 / 1.1 / 1.3 |
| $\psi = 10^{-4}$ | width 8, depth 2, 64 points, Adam 1000 + L-BFGS 2000 | 1: 64 points, Adam 1000 + L-BFGS 600; 2: 256 points, Adam 3000 + L-BFGS 2000 | 1.75 | 1.3 / 1.25 / 1.0 |

For $\psi = 10^{-4}$ the pilot training is refined by two bootstrap rounds; with a single round, the rate warp reaches 0.66%, 0.88% and 0.97% for 256, 36 and 16 time columns, against 0.053%, 0.088% and 0.39% with two rounds.

Figure 26 shows how the rate and the solution improve through the bootstrap rounds at $\psi = 10^{-4}$ (a, b), the median error against $N_t$ (c) and the error during training (d).

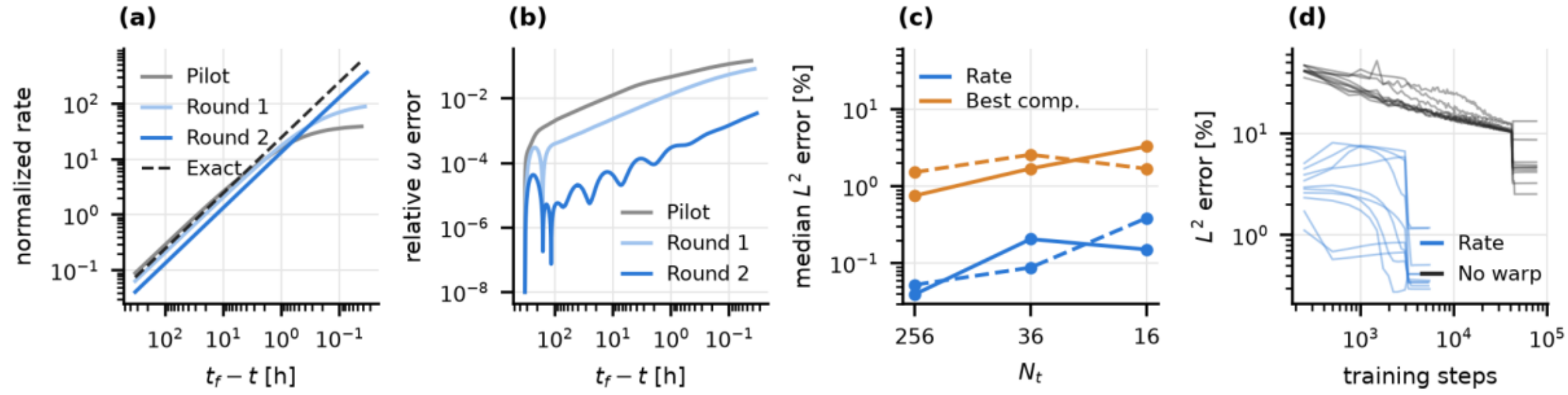

*Figure 26: Additional Kachanov-Rabotnov results. (a) Rate at each stage of the bootstrap chain for $\psi = 10^{-4}$, normalized by its geometric mean, against the exact rate. (b) Error of $\omega$ at each stage, relative to $\omega$ at the end of the horizon. (c) Median relative $L^2$ error against $N_t$ for the rate warp and the best competitor, with $\psi = 10^{-3}$ (solid) and $\psi = 10^{-4}$ (dashed). (d) Relative $L^2$ error during training for $\psi = 10^{-4}$ and $N_t = 16$, ten seeds per method; the no-warp network has the credited budget.*

Figure 27 shows the absolute error of $\omega$ against time for the rate warp and the unwarped network (a), for three values of $\beta$ (b), and the error against $\beta$ (c). Increasing $\beta$ puts more time columns near the rupture time.

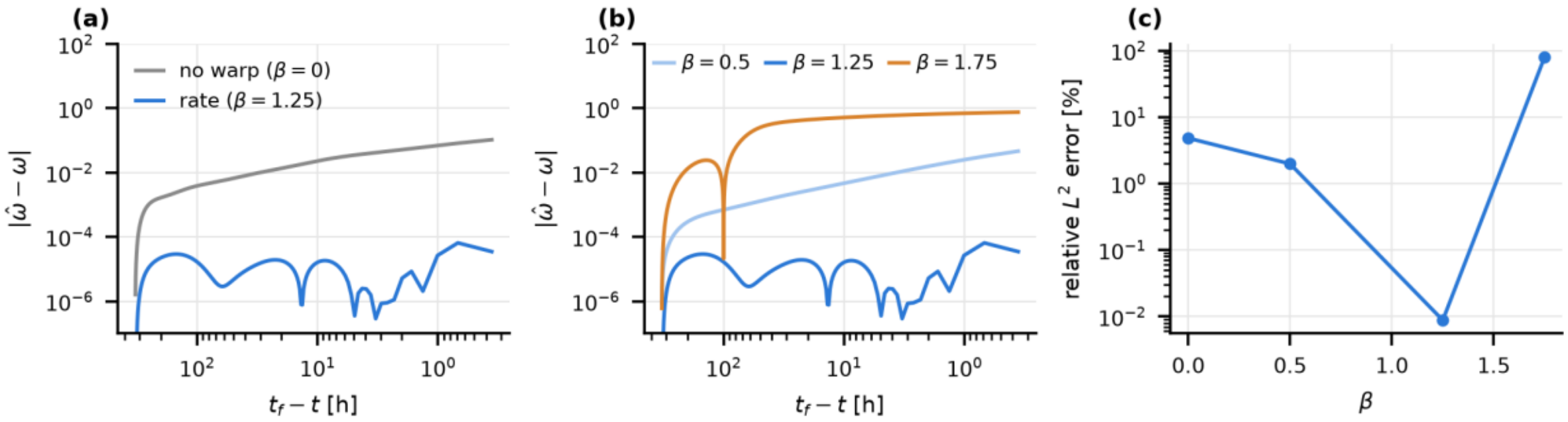

*Figure 27: Absolute error and the effect of $\beta$ for Kachanov-Rabotnov with $\psi = 10^{-4}$ and $N_t = 36$. (a) Absolute error of $\omega$ against the time to rupture $t_f - t$ for the rate warp ($\beta = 1.25$) and the unwarped network ($\beta = 0$). (b) The same for $\beta = 0.5$, 1.25 and 1.75. (c) Relative $L^2$ error against $\beta$.*

## D.2 Burgers

The reference solution of Burgers' equation of Section 5.2 is computed by the method of lines: central differences on 3201 equispaced points in $x$, integrated in time with a BDF scheme (relative tolerance $10^{-7}$, absolute tolerance $10^{-9}$) and interpolated by cubic splines onto a uniform 100 × 100 grid over $\Omega \times [0, 1]$. It lies within 0.01–0.04% of the converged solution. The error is the relative $L^2$ error over this grid. The initial and boundary conditions are imposed exactly by the ansatz $u = -sin(\pi x) + \tau(1 - x^2)N_\theta$, where $\tau$ is the time input of the network and $N_\theta$ receives $x$, the sine and cosine of πx, 2πx, 4πx and 8πx and $\tau$. Table 10 lists the setup of each stage. Burgers uses the pilot training only: its rate, the median over $x$ of the local rate of Section 5.2, enters the full training directly, without bootstrap rounds, and the time columns are placed at the midpoints of uniform intervals of the warped time.

*Table 10: Setup of the rate warp for Burgers. The network is given as hidden layers × width and the points as x × t columns; the full training is listed for the cells of Table 7.*

| Stage | Network | Points | Optimizer | $\beta$ | $\zeta$ |
|---|---|---|---|---|---|
| Pilot training | 3 × 20 | 128 × 16 | Adam 1000 + L-BFGS 600 | – | – |
| Full training | 3 × 20 | $N_x \times N_t$ | Adam 4500 + L-BFGS 3000 | 0.85 | 1 |

Figure 28 compares the prediction of the rate warp with the reference at four times. Figure 29 shows the reference solution (a), the absolute error of the prediction (b) and the rate of the pilot training against the exact rate (c). Figure 30 shows the absolute error against time for the rate warp and the unwarped network (a), for three values of $\beta$ (b), and the error against $\beta$ (c). Figure 31 shows the absolute error and the training error for three residual weightings of Section 4.2 ($\zeta = -2, 0, 1$).

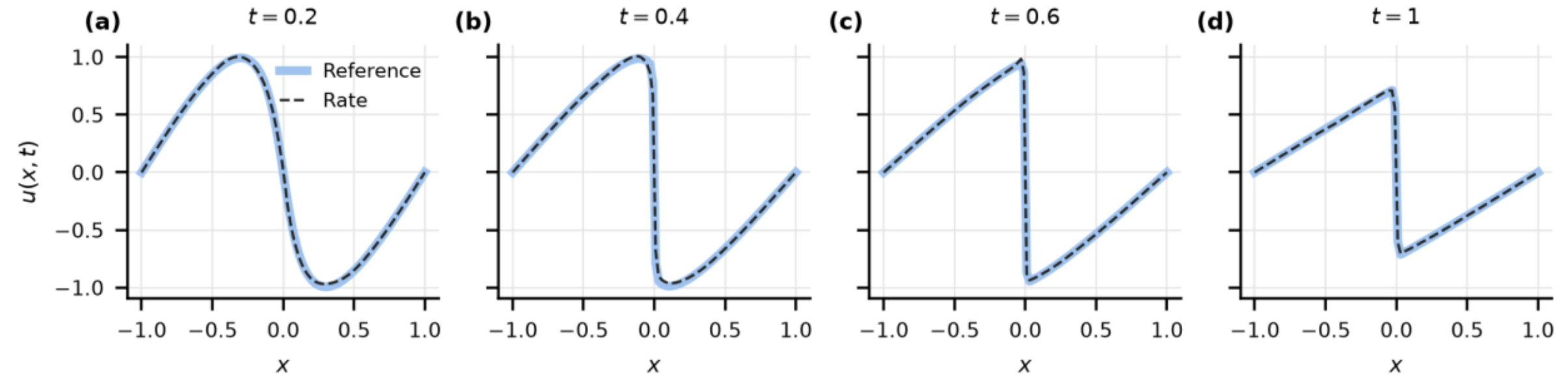

*Figure 28: Burgers solution of the rate warp (dashed) against the reference (thick) at $t$ = 0.2, 0.4, 0.6 and 1, $N_t$ = 4, $N_x$ = 128.*

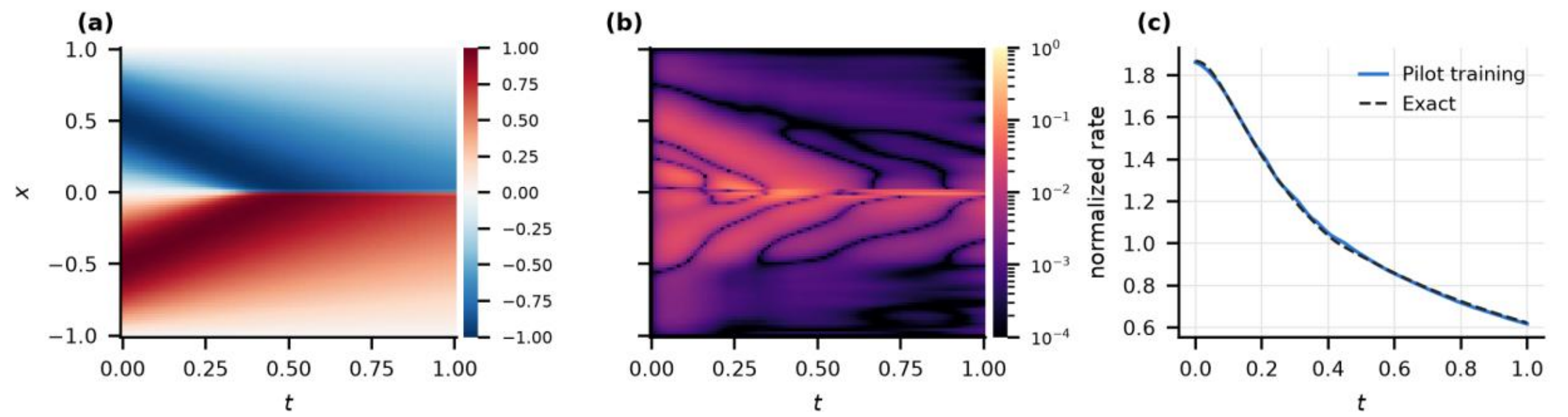

*Figure 29: Burgers results for $N_x$ = 128 and $N_t$ = 4. (a) Reference solution. (b) Absolute error of the rate warp. (c) Rate of the pilot training against the exact rate, both normalized by their geometric mean.*

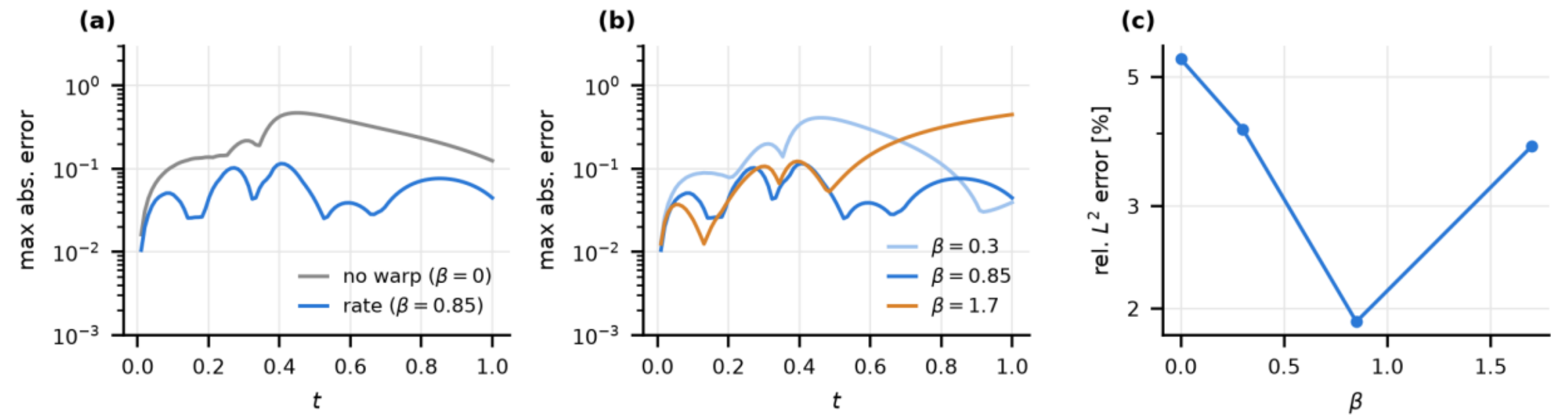

*Figure 30: Absolute error and the effect of $\beta$ for Burgers. (a) Maximum absolute error over $x$ against time for the rate warp ($\beta$ = 0.85) and the unwarped network ($\beta$ = 0). (b) The same for $\beta$ = 0.3, 0.85 and 1.7. (c) Relative $L^2$ error against $\beta$.*

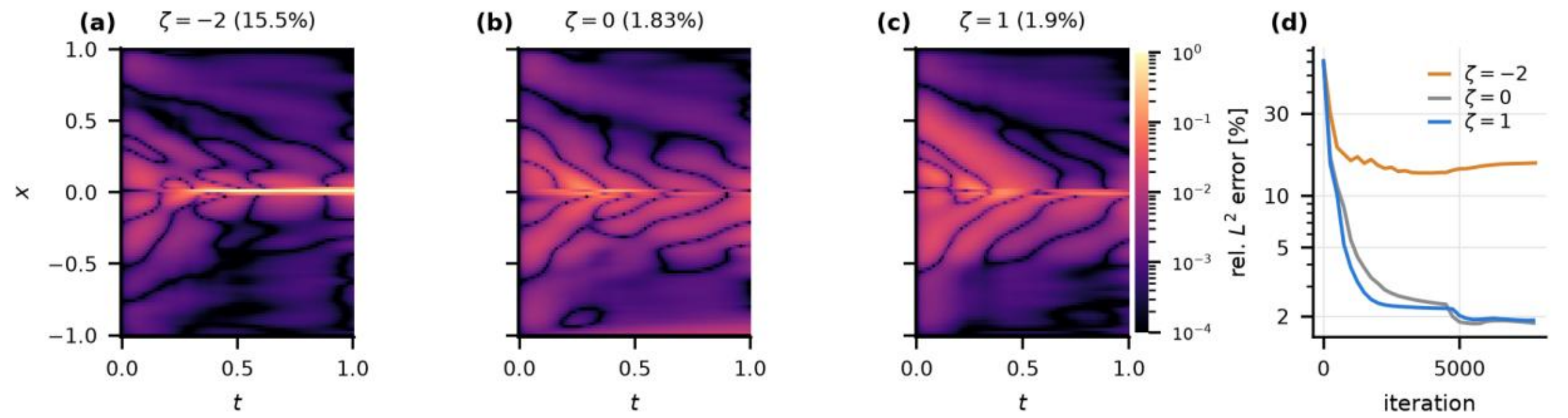

*Figure 31: Residual weighting on the same warp for Burgers. (a–c) Absolute error for $\zeta = -2$ (the residual divided by the rate), $\zeta = 0$ (the unweighted physical residual) and $\zeta = 1$ (the physical residual weighted by the rate), with the relative $L^2$ error in the title. (d) Relative $L^2$ error during training.*

## D.3 Allen-Cahn

The reference solution of the Allen-Cahn equation of Section 5.3 is computed with a pseudo-spectral method with 512 Fourier modes in $x$, integrated in time with a BDF scheme (relative tolerance $10^{-7}$, absolute tolerance $10^{-9}$) and interpolated by cubic splines onto a uniform 200 × 200 grid over $\Omega \times [0, 1]$. The error is the relative $L^2$ error over this grid. Table 11 lists the setup of each stage. All stages use the periodic embedding of Section 2.1 with $m = 10$; the bootstrap rounds additionally impose the sign of the solution with the factor $tanh(12\ cos(\pi x))$, and the rounds and the full training use the rate-weighted residual of Section 4.2.

*Table 11: Setup of the rate warp for Allen-Cahn. The network is given as hidden layers × width and the points as x × t columns; the final β is listed for $N_t$ = 24 / 20 / 16.*

| Stage | Network | Points | Optimizer | $\beta$ |
|---|---|---|---|---|
| Pilot training | 2 × 20 | 250 × 32 | Adam 1000 + L-BFGS 600 | – |
| Round 1 | 4 × 32 | 60 × 8 | L-BFGS 1000 | 0.1 |
| Round 2 | 4 × 32 | 128 × 16 | L-BFGS 2500 | 0.1 |
| Full training | 4 × 64 | 256 × $N_t$ | L-BFGS 10 000 | 0.76 / 1.01 / 0.76 |

Figure 32 compares the prediction of the rate warp with the reference at four times. Figure 33 shows the reference solution (a), the absolute error of the prediction (b), the error after each stage of the bootstrap chain (c) and the effect of $\beta$ (d). There is a balanced value of $\beta$: if it is too low the warp is negligible and the training is the same as without a warp, and if it is too high important regions are left without collocation points. Figure 34 shows the absolute error and the training error for the three residual weightings of Section 4.2 ($\zeta = -2, 0, 2$).

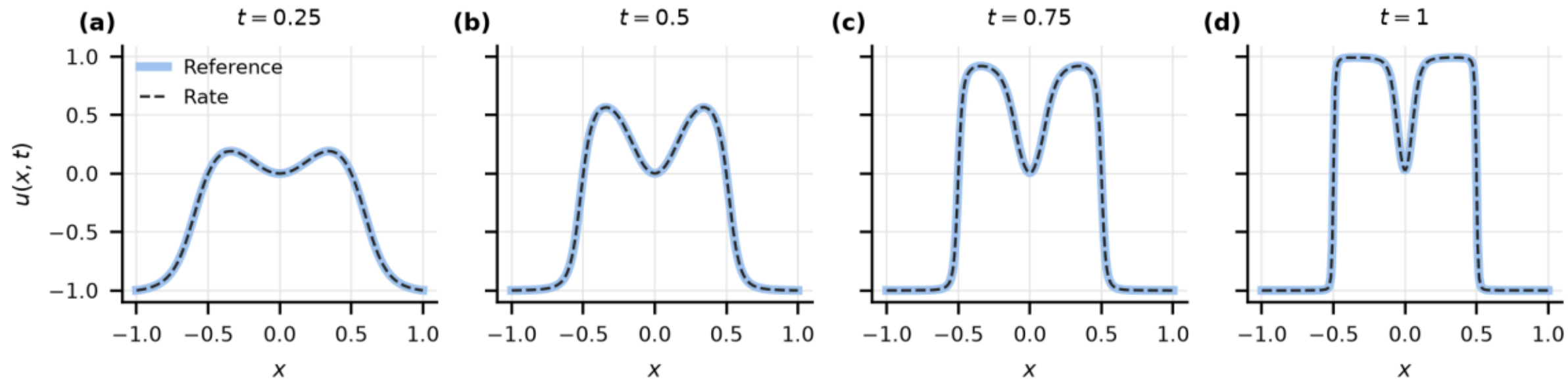

*Figure 32: Allen-Cahn solution of the rate warp (dashed) against the reference (thick) at t = 0.25, 0.5, 0.75 and 1, $N_t$ = 24.*

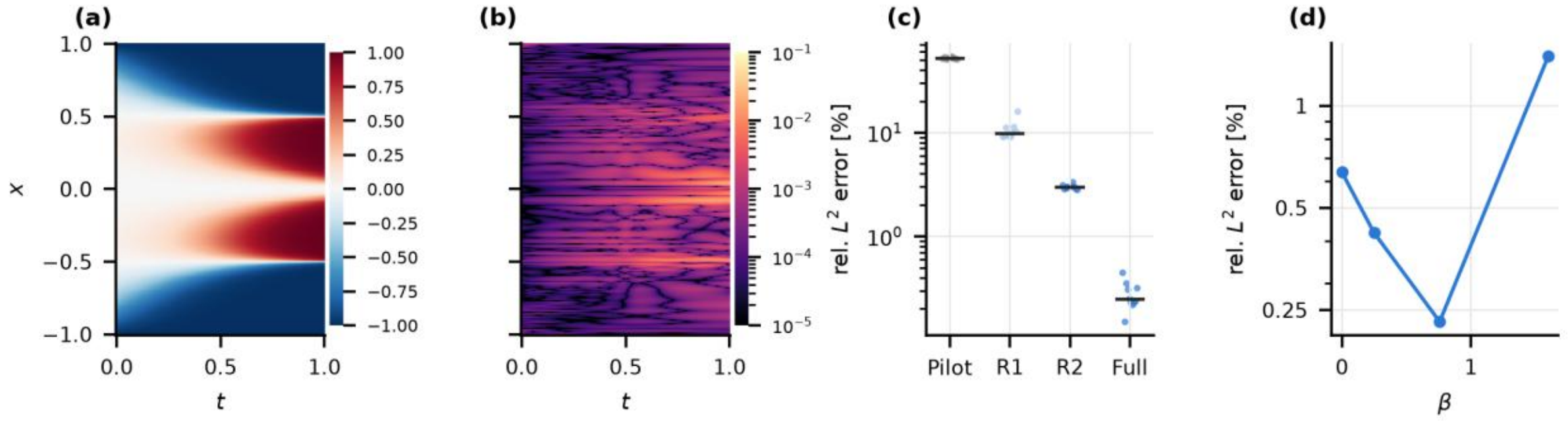

*Figure 33: Allen-Cahn results for $N_t$ = 24. (a) Reference solution. (b) Absolute error of the rate warp. (c) Relative $L^2$ error after the pilot training, Round 1, Round 2 and the full training; dots are the ten seeds and bars their medians. (d) Relative $L^2$ error against β; β = 0 is the unwarped network and β = 0.76 is the value used.*

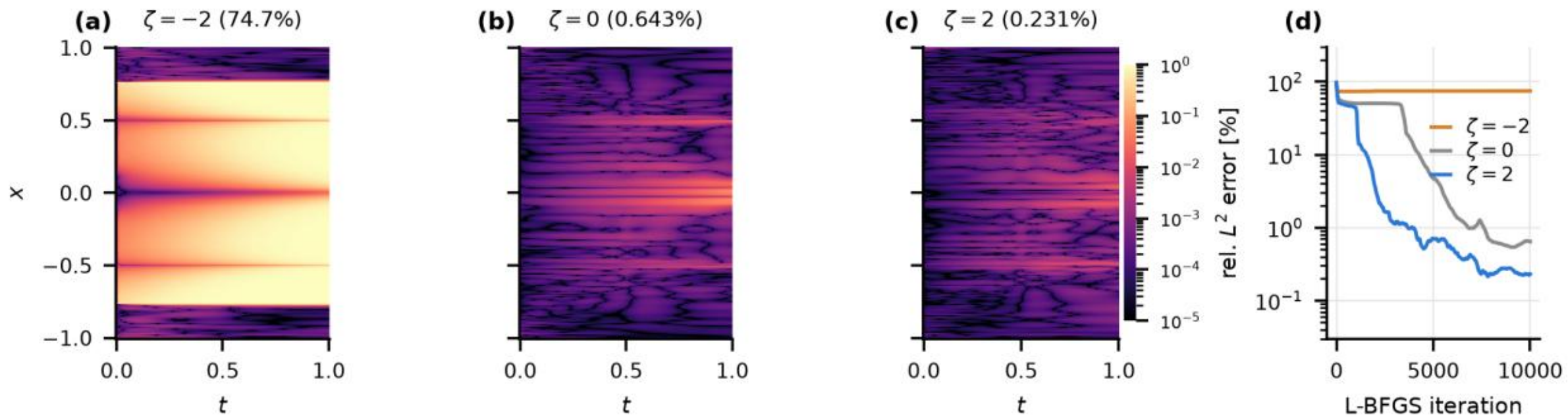

*Figure 34: Residual weighting on the same warp, $N_t$ = 24. (a–c) Absolute error for ζ = −2 (the residual divided by the rate), ζ = 0 (the unweighted physical residual) and ζ = 2 (the rate-weighted residual), with the relative $L^2$ error in the title. (d) Relative $L^2$ error during training.*

## D.4 Frank-Kamenetskii

The reference solution of the Frank-Kamenetskii equation of Section 5.4 uses the symmetry of the solution in $x$: it is computed on $[0, 1]$ with second-order finite differences (spacing $2.5 \cdot 10^{-4}$) and a BDF scheme (relative tolerance $10^{-11}$) and stored on 1001 points in $x$ and 1501 times clustered toward $t_{end}$. The blow-up time $t_b = 1.4467$ is obtained by integrating until $max\ u$ reaches 14 and adding the remaining time, which is of order $e^{-14}/\delta$. The error is the relative $L^2$ error over 2000 points in $x$ and 200 uniformly spaced times in $[0, t_{end}]$, with $t_{end} = (1 - \psi)t_b$; the peak error is the relative error of $max_x\ u$ at $t_{end}$. Table 12 lists the setup of each stage. Frank-Kamenetskii uses the pilot training only: its rate, the mean over $x$ of the magnitude $|\Lambda|$ of the net growth coefficient of Section 5.4, enters the full training directly, without bootstrap rounds. All stages run in float64.

*Table 12: Setup of the rate warp for Frank-Kamenetskii. The network is given as hidden layers × width and the points as x × t columns; slash-separated values follow $\psi = 3{\cdot}10^{-3} / 10^{-5}$.*

| Stage | Network | Points | Optimizer | $\beta$ | $\zeta$ |
|---|---|---|---|---|---|
| Pilot training | 3 × 16 | 32 × 32 | Adam 1000 / 500 | – | – |
| Full training | 4 × 64 | 256 × 80 | L-BFGS 10 000 | 1 | −2 |

Figure 35 compares the prediction of the rate warp with the reference at four times, given as fractions of $t_{end}$. Figure 36 shows the reference solution (a), the prediction (b) and the absolute error (c) for $\psi = 3 \cdot 10^{-3}$.

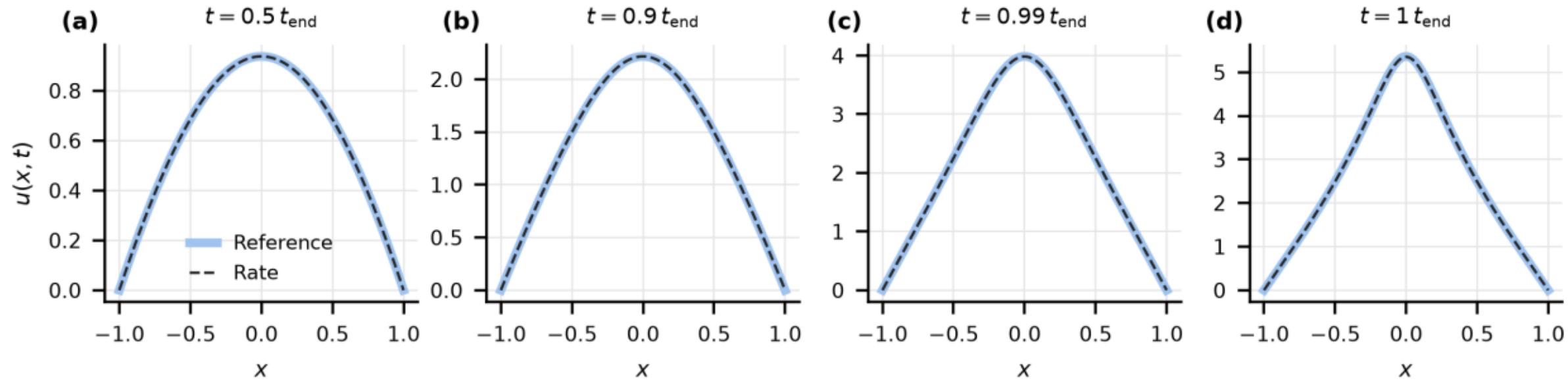

*Figure 35: Frank-Kamenetskii solution of the rate warp (dashed) against the reference (thick) at $t/t_{end}$ = 0.5, 0.9, 0.99 and 1, with $\psi = 3{\cdot}10^{-3}$.*

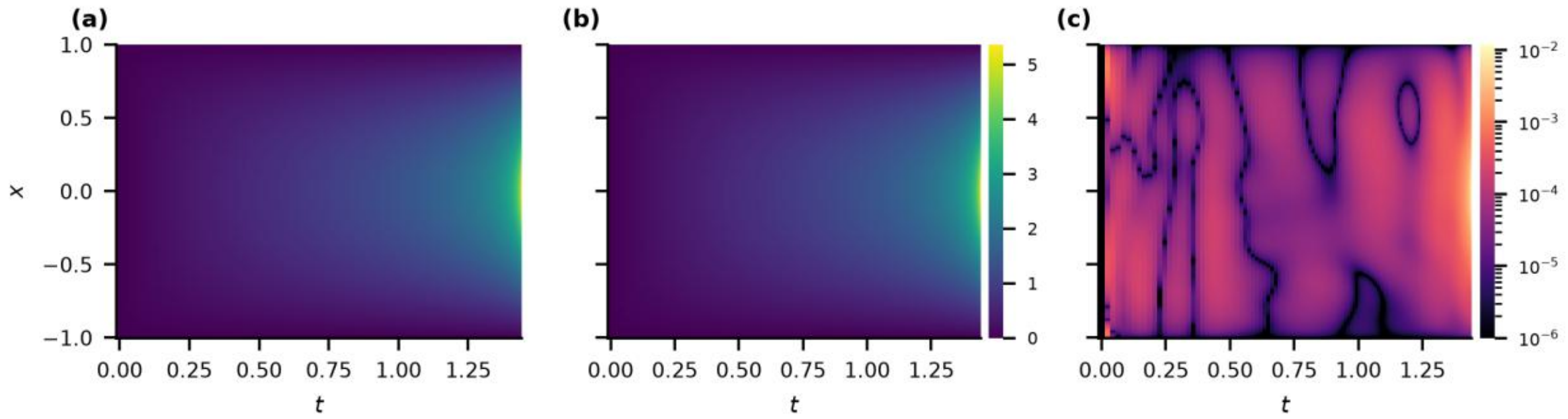

*Figure 36: Frank-Kamenetskii results for $\psi = 3{\cdot}10^{-3}$. (a) Reference solution. (b) Prediction of the rate warp. (c) Absolute error of the rate warp.*